\documentclass{article}

\usepackage{stmaryrd,graphicx,cite,yfonts}

\usepackage{amsmath,amssymb}

\DeclareMathOperator{\pr}{pr}
\DeclareMathOperator{\Sl}{sl}
\DeclareMathSymbol{\C}{\mathalpha}{AMSb}{"43}
\DeclareMathSymbol{\R}{\mathalpha}{AMSb}{"52}

\numberwithin{equation}{section}
\def\ss{\relax{\supset\kern-13pt +}}

\def\ss{\relax{\supset\kern-13pt +}}
 \author{{\bf Decio Levi},\\ Dipartimento di Ingegneria Elettronica, \\Universit{\`a} degli Studi Roma Tre and \\
I.N.F.N. Sezione di Roma III,\\
Via della Vasca Navale 84, 00146 Roma, Italy\\
email: levi@fis.uniroma3.it\\
{\bf Pavel Winternitz}, \\Centre de  recherches math{\'e}matiques, \\Universit{\'e} de Montr{\'e}al,\\
C.P. 6128, succ.\ Centre-ville,\\ Montr{\'e}al QC H3C 3J7, Canada \\
email:wintern@CRM.UMontreal.CA}

 \def\ra {\rightarrow}

 \def\neq {\not\equiv}

 \def\be   {\begin{equation}}   
 \def\ee   {\end{equation}}
 \def\ba   {\begin{array}}     
 \def\ea   {\end{array}}
 \def\bea  {\begin{eqnarray}}   
 \def\eea  {\end{eqnarray}}
 \def\bean {\begin{eqnarray*}}  
 \def\eean {\end{eqnarray*}}

\newcommand{\bref}[1]{\textnormal{\cite{#1}}}

\newcommand{\fref}[1]{\textnormal{(\ref{#1})}}

\newcommand\ds{\displaystyle}

\title{Continuous Symmetries of Difference Equations}

\begin{document}
 \maketitle
 \begin{abstract}
 Lie group theory was originally created more than 100 years ago as a tool for solving ordinary and partial differential equations. In this  article we review the results of a much more recent program: the use of Lie groups to study difference equations. We show that the mismatch between continuous symmetries and discrete equations can be resolved in at least two manners.  One is to use generalized symmetries acting on solutions of difference equations, but leaving the lattice invariant. The other is to restrict to point symmetries, but to allow them to also transform the lattice.  
 \end{abstract}
 
\tableofcontents
 
\section{Introduction} \label{i}

The symmetry theory of differential equations is well understood. it goes back to the classical work of Sophus Lie and is reviewed in numerous modern books and articles \cite{ref6,ref2,ref5,ref4,ref1,ref3,ref7,ref8,h2000,s2000,s1990,c2002,s1996,b2002,i1993,i1994,appr3}. As a matter of fact, Lie group theory is the most general and useful tool we have for obtaining exact analytic solutions of large classes of differential equations, specially nonlinear ones.

The application of Lie group theory to discrete equations is much more recent and a vigorous development of the theory only started in the 1990-ties \cite{l2003, w1999,rw2004,rsw1997,pw2004,tw1999,d1,l1,l2,l3,l4,l5,l6,l7,l8,l9,l11,l12,MF,b16,b18,s2001,sk2000,S2,b12,158,ref45,ref38b,lrdiscrete,olmo1,olmo2,ref41,ref28,ref28e,ref28d,ref28c,ref28b,refb,ref38,ref40,ref39,ref43,ref42,ref47,ref44,ref29,Proc.WNM,lT,ly1999,maeda91,maeda81,maeda82,ly2000,ly2001,ref30,ref22,hl,ref34,ref35,ref32,ref36,ref24,ref23,ref37,refc,LRR,y2004,5,6,ref33,ref48,ref49,refa,ref31,ref19,ref20,ref21,ref50,ref17,ref18,ref25,refe,ref26,ref27,refd}.

The purpose of this article is to provide a review of the progress made. 

In this whole field of research one uses group theory to do for difference equations what has been done for differential ones. This includes generating new solutions from old ones, identifying equations that can be transformed into each other, performing symmetry reduction, and identifying integrable equations.

When adapting the group theoretical approach from differential equations to difference ones, we must answer three basic questions:
\begin{enumerate}
\item What do we mean by symmetries?
\item How do we find the symmetries of a difference system?
\item What do we do with the symmetries, once we know them?
\end{enumerate}

Let us first briefly review the situation for differential equations.

Let us consider a general system of differential equations
\begin{equation}\label{1.1}
E_a(x, u, u_x, u_{2x}, \dots u_{nx}) = 0, \quad
x \in {\mathbb R}^p, u \in {\mathbb R}^q, a = 1, \dots N,
\end{equation}
where $u_{nx}$ denotes all (partial) derivatives of $u$ of order $n$. 
The numbers
$p, q, n$ and $N$ are all nonnegative integers.

We are interested in the symmetry group \textgoth{G} of system (\ref{1.1}),
i.e. in
the local Lie group of local point transformations taking solutions of
eq.~(\ref{1.1}) into solutions of the same equation. 
Point transformations in the space $X \times U$
of independent and dependent variables have the form
\begin{equation}\label{1.2}
\tilde x = \Lambda_{\lambda}(x, u), \quad \tilde u = \Omega_{\lambda}(x, u),
\end{equation}
where $\lambda$ denotes the group parameters. Thus
\[
\Lambda_0(x, u) = x, \quad \Omega_0(x, u) = u, \nonumber
\]
and the inverse transformation $(\tilde x, \tilde u) \mapsto (x, u)$ exists, at
least locally.

The transformations (\ref{1.2}) of local coordinates in  $X \times U$ also
determine the transformations of functions $u = f(x)$ and of derivatives of
functions. A group \textgoth{G} of local point transformations of $X \times U$ will be a
symmetry group of system (\ref{1.1}) if the fact that $u(x)$ is a solution
implies that $\tilde u(\tilde x)$ is also a solution.

How does one find the symmetry group \textgoth{G}?
Instead of looking for ``global'' transformations as in eq.~(\ref{1.2}) one
looks for infinitesimal ones, i.e.
one looks for the Lie algebra \textgoth{g} that corresponds to
\textgoth{G}. A one-parameter group of infinitesimal point
transformations will have the form
\begin{eqnarray} \label{1.3}
&&\tilde x_i  =  x_i + \lambda\xi_i(x, u), \quad \tilde u_{\alpha}  =  u_{\alpha} + \lambda\phi_{\alpha}(x, u), \\ &&|\lambda| \ll 1
 \quad 1 \leq i
\leq p, \quad 1 \leq \alpha \leq q.\nonumber
\end{eqnarray}

The search for the symmetry algebra \textgoth{g} of a system of differential equations
is best formulated in terms of vector fields acting on the space $X \times U$
of independent and dependent variables. Indeed, consider the vector field
\begin{equation}\label{1.4}
{\hat X} = \sum^p_{i=1}\xi_i(x, u)\partial_{ x_i} + \sum^q_{\alpha=1}\phi_{\alpha}
(x, u)\partial_{ u_{\alpha}},
\end{equation}
where the coefficients $\xi_i$ and $\phi_{\alpha}$ are the same as in
eq.~(\ref{1.3}). If these functions are known, the vector field (\ref{1.4}) can
be integrated to obtain the finite transformations (\ref{1.2}). Indeed, all we
have to do is to integrate the equations
\begin{equation}\label{1.5}
\frac{d\tilde x_i}{d\lambda} = \xi_i(\tilde x, \tilde u), \quad \frac{d\tilde
u_{\alpha}}{d\lambda} = \phi_{\alpha}(\tilde x, \tilde u),
\end{equation}
subject to the initial conditions
\begin{equation}\label{1.6}
\tilde x_i\mid_{\lambda = 0} = x_i \quad \tilde u_{\alpha}\mid_{\lambda=0} =
u_{\alpha}.
\end{equation}
This provides us with a one-parameter group of local Lie point
transformations of the form (\ref{1.2}) where $\lambda$ is the group parameter.

The vector field (\ref{1.4}) tells us how the variables $x$ and $u$ transform.
We also need to know how derivatives like $u_x$, $u_{xx}$, $\dots$ transform.
This is given by the prolongation of the vector field ${\hat X}$.

We have
\begin{eqnarray}
\lefteqn{\pr {\hat X}  =  {\hat X} +
\sum_{\alpha}\biggl\{\sum_i\phi^{x_i}_{\alpha}\partial_{ u_{\alpha,x_i}} +\sum_{i,
k}\phi^{x_ix_k}_{\alpha}\partial_{ u_{\alpha,x_ix_k}}}\label{1.7}\\ 
&\hspace{2in} & +  \sum^{x_ix_kx_l}_{i, k, l}\phi_{\alpha}^{x_ix_kx_l}
\partial_{ u_{\alpha,x_ix_kx_l}}
+
\dots\biggr\},\nonumber
\end{eqnarray}
where the coefficients in the prolongation can be calculated recursively, using
the total derivative operator,
\begin{equation}\label{1.8}
D_{x_i} = \partial_{x_i}+ u_{\alpha, x_i}\partial_{u_{\alpha}}+u_{\alpha,
x_ax_i}\partial_{u_{\alpha, x_a}} + u_{\alpha, x_ax_bx_i}\partial_{u_{\alpha,
x_ax_b}} + \dots
\end{equation}
(a summation over repeated indices is to be understood).
The recursive formulas are
\begin{eqnarray}
&&\phi^{x_i}_{\alpha}  =  D_{x_i}\phi_{\alpha} - (D_{x_i}\xi_a)u_{\alpha, x_a},
\quad
\phi^{x_ix_k}_{\alpha}  =  D_{x_k}\phi^{x_i}_{\alpha} - (D_{x_k}\xi_a)
u_{\alpha, x_ix_a},\label{1.9}\\
&&\phi^{x_ix_kx_l}_{\alpha}  =  D_{x_l}\phi^{x_ix_k}_{\alpha} - (D_{x_l}\xi_a)
u_{\alpha, x_ix_kx_a},\nonumber
\end{eqnarray}
etc.

The invariance condition for system (\ref{1.1}) is expressed in
terms of the operator (\ref{1.7}) as
\begin{equation}\label{1.10}
\pr^{(n)}{\hat X}E_a\mid_{E_1=\dots = E_N = 0} = 0, \quad a = 1, \dots N,
\end{equation}
where $\pr^{(n)}{\hat X}$ is the prolongation (\ref{1.7}) calculated up to
order $n$ (where $n$ is the order of system (\ref{1.1})).

Eq. (\ref{1.10}) provides a system of linear partial differential equations for the functions $\xi_i(x,u)$ and $\phi_{\alpha}(x,u)$, in which the variables $x$ and $u$ figure as independent variables. By definition of point transformations the coefficients $\xi_i$ and $\phi_{\alpha}$ depend only on ($x_1, \ldots, x_p, u_1, \ldots, u_q$), not on any derivative of $u_{\alpha}$. The action of $pr^{(n)}{\hat X}$ in eq.  (\ref{1.10}) will, on the other hand, introduce terms in  (\ref{1.10}), involving the derivatives $\frac{\partial^k u}{\partial x_1^{k_1} \ldots \partial x_p^{k_p}}$, $k = k_1 + \ldots + k_p$, $1 \le k \le n$. We use equations  (\ref{1.1}) to eliminate $N$ (the number of equations) such derivatives. We then collect all linearly independent remaining expressions in the derivatives and set the coefficients of these expressions equal to zero. This provides the "determining equations": a set of linear partial differential equations for the functions $\xi_i(x,u)$ and $\phi_{\alpha}(x,u)$. The order of the system of determining equations is the same as the order of the studied system  (\ref{1.1}), however the determining system is linear, even if the system  (\ref{1.1}) is nonlinear. It is usually overdetermined and not difficult to solve. Computer programs using various symbolic languages exist that derive the determining system and solve it, or at least partially solve it \cite{ref9,ref10,ref11,s1996}.

The solution of the determining system may be trivial, i.e. $\xi_i = 0$, $\phi_{\alpha} = 0$. Then the symmetry approach is of no avail. Alternatively, the general solution may depend on a finite number $K$ of integration constants. The Lie algebra of the symmetry group, the "symmetry algebra", for short, is then $K$-dimensional and must be identified as an abstract Lie algebra \cite{ref13,ref15,ref12,ref14}. Finally the general solution of the determining equations may involve arbitrary functions and the symmetry algebra is infinite-dimensional.

So far we have considered only point transformations, as in eq.~(\ref{1.2}), in
which the new variables $\tilde x$ and $\tilde u$ depend only on the old ones,
$x$ and $u$. More general transformations are ``contact transformations'',
where $\tilde x$ and $\tilde u$ also depend on first derivatives of $u$ \cite{ref1,ref16,ref2,ref4,h2000,s2000}. A
still more general class of transformations are generalized transformations,
also called ``Lie-B{\"a}cklund'' transformations \cite{ref1,ref16,k2000}. For these, the transformations are
\be \label{1.11}
\tilde x  =  \Lambda_{\lambda}(x, u, u_x, u_{xx}, \dots), \quad
\tilde u  =  \Omega_{\lambda}(x, u, u_x, u_{xx}, \dots), 
\ee
involving derivatives up to an arbitrary, but finite
order. The coefficients $\xi_i$ and
$\phi_{\alpha}$ of the vector fields (\ref{1.4}) will then also depend on
derivatives of $u$.

When studying generalized symmetries, and sometimes also point symmetries, it
is convenient to use a different formalism, namely that of evolutionary vector
fields.
Let us first consider the case of Lie point symmetries, i.e. vector fields of
the form (\ref{1.4}) and their prolongations (\ref{1.7}). To each vector field
(\ref{1.4}) we can associate its evolutionary counterpart ${\hat X}_e$, defined as
\be 
{\hat X}_e =  Q_{\alpha}(x, u, u_x)\partial_{ u_{\alpha}},\label{1.16} \qquad 
Q_{\alpha}  =  \phi_{\alpha} - \xi_j  u_{\alpha, x_j}.
\ee
The prolongation of the evolutionary vector field (\ref{1.16}) is defined as
\begin{eqnarray}
\pr {\hat X}_e &=& Q_{\alpha}\partial_{u_a} + Q^{x_j}_{\alpha} \partial_{ u_{\alpha, x_j}} +
Q^{x_j x_k}_{\alpha}\partial_{ u_{\alpha, x_j x_k}} + \dots\label{1.18}\\
Q^{x_j}_{\alpha} &=& D_{x_j}Q_{\alpha}, \quad Q^{x_j x_k}_{\alpha} = D_{x_j}D_
{x_k}Q_{\alpha}, \dots\nonumber.
\end{eqnarray}
The functions $Q_{\alpha}$ are called the characteristics of the vector field.
Observe that ${\hat X}_e$ and $\pr {\hat X}_e$ do not act on the independent variables $x_j$.
For Lie point symmetries evolutionary and ordinary vector fields are entirely
equivalent and it is easy to pass from one to the other. Indeed,
eq.~(\ref{1.16}) gives the connection between the two.
The symmetry algorithms for calculating the symmetry algebra \textgoth{g} in terms of ordinary, or 
evolutionary vector fields, are also equivalent. Eq.~(\ref{1.10}) is simply
replaced by
\begin{equation}\label{1.19}
\pr^{(n)}{\hat X}_e E_a\mid_{E_1=\dots=E_N=0}, = 0, \quad a = 1, \dots N.
\end{equation}
The reason that eq.~(\ref{1.10}) and (\ref{1.19}) are equivalent is the
following:
\begin{equation}\label{1.20}
\pr^{(n)}{\hat X}_e = \pr^{(n)}X - \xi_iD_i.
\end{equation}
The total derivative $D_i$ is itself a generalized symmetry of eq.~(\ref{1.1}),
i.e.,
\begin{equation}\label{1.21a}
D_iE_a\mid_{E_1=E_2=\dots=E_N=0}, =0 \quad i = 1, \dots p, \quad a = 1, \dots N.
\end{equation}
Eqs.~(\ref{1.20}) and (\ref{1.21a}) prove that systems (\ref{1.10}) and
(\ref{1.19}) are equivalent. Eq.~(\ref{1.21a}) itself follows from the fact that
$D_iE_a=0$ is a differential consequence of eq.~(\ref{1.1}), hence every
solution of eq.~(\ref{1.1}) is also a solution of eq.~(\ref{1.21a}).

To find generalized symmetries of order $k$, we use eq.~(\ref{1.16}) but allow
the characteristics $Q_{\alpha}$ to depend on all derivatives  of $u$
up to order $k$. The prolongation is calculated using eq.~(\ref{1.18}). The
symmetry algorithm is again eq.~(\ref{1.19}).

A very useful property of evolutionary symmetries is that $Q_{\alpha}$ provide
compatible flows. This means that the system of equations
\begin{equation}\label{1.21b}
\frac{\partial u_{\alpha}}{\partial \lambda} = Q_{\alpha}
\end{equation}
is compatible with system (\ref{1.1}). In particular, group invariant
solutions, i.e., solutions invariant under a subgroup of \textgoth{G},
 are obtained as
fixed points
\begin{equation}\label{1.22}
Q_{\alpha} = 0.
\end{equation}
If $Q_{\alpha}$ is the characteristic of a point transformation, then
(\ref{1.22}) is a system of quasilinear first order partial differential
equations. They can be solved and their solutions 
can be substituted into (\ref{1.1}), yielding 
the invariant solutions explicitly.

Many different extensions of Lie's original method of
group invariant solutions exist. Among them we mention, first of all,
 {\it conditional symmetries} \cite{levi4,bluman,fuch,fuch1}. For differential equations, they were introduced under several different names \cite{bluman,levi4,olver2,CK} in order to obtain dimensional reductions of partial differential equations, beyond those obtained by using ordinary Lie symmetries.  

Another valuable extension is the concept of {\it partial
symmetries}.  They
correspond to the existence of a subset of solutions which, without
necessarily being invariant, are mapped into each other by 
 the transformation \cite{CGpar,CiK}. Further extensions are given by  {\it asymptotic symmetries} \cite{Gasy,GaM}, when extra symmetries are obtained in the asymptotic regime, or {approximate symmetries} \cite{appr1,appr2,appr3} where one considers the symmetries  of  approximate solutions of a system depending on a small parameter.

Let us now return to the problem at hand, namely symmetries of difference systems. We wish to study the continuous symmetries and use Lie algebra techniques. However, the equations are now discrete, i.e. they involve functions $u(x)$ that are themselves continuous, but evaluated , or sampled, at discrete points. Several different approaches to this problem have been developed and are discussed below.

In Section \ref{sec.nn2} we discuss  point symmetries of difference equations defined on fixed, nontransforming lattices \cite{ref28,ref28e,ref28d,ref28c,ref28b,ref22,ref24,ref23,ref19,ref20,ref21,ref25,ref26,ref27,lT,ref17,maeda81,maeda82,ref18}.
The symmetry transformations are assumed to have the form (\ref{1.2}). They must take solutions into solutions and the lattice into itself. This approach is fruitful mainly for differential-difference equations (D$\Delta$E's), where not only the dependent variables, but also some of the independent ones are continuous.

Section \ref{sec.n3} is devoted to
{\sl generalized point symmetries} on fixed lattices \cite{ref29,ref30,ref32,ref33,ref38b,ref31,olmo1,olmo2}.
The concept of symmetry is generalized in that transformations act simultaneously at several points of the lattice, possibly infinitely many ones. In the continuous limit they reduce to point transformations. This approach is fruitful mainly for linear equations, or equations that can be linearized by a transformation of variables.

In Section \ref{sec.n4} we consider  generalized symmetries on fixed lattices \cite{hl,ref34,ref35,ref36,ref37}
This approach is fruitful for discrete nonlinear integrable equations, i.e., nonlinear difference equations possessing a Lax pair. The symmetries are generalized ones, treated in the evolutionary formalism. In a continuous limit they reduce to point and generalized symmetries of integrable differential equations.

Point symmetries transforming solutions and lattices \cite{ref41,ref38,ref40,ref39,ref43,ref42,ref45,ref47,ref44,ref48,ref49,ref50} are considered in Section \ref{sec.n2}.
The transformations have the form (\ref{1.2}) and they act on solutions and on lattices simultaneously. The lattices themselves are given by difference equations and their form is dictated by the symmetries. Their main application is to discretize given differential equations while preserving their symmetries.

A brief conclusion with an overview of the results presented and pending issues is given in Section \ref{co}.

\section{Point Symmetries of Difference Equations Defined on Fixed, Nontransforming Lattices} \label{sec.nn2} 

The essentially continuous techniques  for finding Lie symmetries for differential equations can be extended in a natural way to the discrete case by acting just on the continuous variables ~\cite{ref17,maeda81,maeda82,ref19,ref20,ref18,ref26}, leaving the lattice invariant. Transformations of the lattice are considered only at the level of the group which itself is finite or discrete. Depending on the discrete equation we are considering we can have translations on the lattice by multiples of the lattice spacing and rotations through fixed angles, i.e., for example, by $\frac{\pi}{2}$ when the partial difference equation is invariant with respect to the interchange of the two lattice variables.

In Section \ref{sect1} we consider, as an example, the Lie point symmetries of the discrete time  Toda lattice \cite{ref37}.
In Section \ref{sect2} we review the steps necessary to obtain continuous symmetries
for differential--difference equations (D$\Delta$E's) and then in Section \ref{sect3} we carry
out the explicit calculations in the case of the Toda equation in 1+1
dimensions. In doing so we will present and compare the results
contained in our articles on the subject, which correspond to both the
\emph{intrinsic method} and the \emph{differential equation method}
proposed earlier, 
and those of the {\it global method} introduced by Quispel \emph{et
  al}.~\cite{ref26}.

Apart from the standard Toda lattice we will consider a Toda lattice
equation with variable coefficients and will show how, by analyzing its
symmetry group, one can find the Lie point transformation which maps it into
the standard Toda lattice equation with constant coefficients. 
In Section \ref{sect4}, we will present some results on the classification of
nonlinear D$\Delta$E's with nearest neighbour interactions.  Finally, symmetries of a two--dimensional Toda lattice are discussed in Section \ref{sect5}.

\subsection{Lie point symmetries of the discrete time Toda equation} \label{sect1}

The discrete time  Toda equation \cite{hirota2} is one of the integrable completely discrete partial differential equations ($\Delta \Delta$E) \cite{suris,hirota1,hirota3,hirota4,ref56,ns1997,nc1995,sr2000,s1997,a2001,o1978,lps1981,npc1992,ssyy2001} and is given by
\bea \label{a14}
\Delta_{Toda} & = & e^{u_{n,m} - u_{n,m+1}} - e^{u_{n,m+1} - u_{n,m+2}} - \\
\nonumber & - & \alpha^{2} ( 
e^{u_{n-1,m+2} - u_{n,m+1}} - e^{u_{n,m+1} - u_{n+1,m}} ) = 0.
\eea 
On the left hand side of 
eq. (\ref{a14}) we can easily obtain the second difference of the 
function $u_{n,m}$ with respect to the discrete-time $m$. Thus, defining 
\be \label{a15}
t = m \sigma_t; \qquad v_{n}(t) = u_{n,m}; \qquad \alpha = \sigma_t^{2}
\ee
we find that eq.(\ref{a14}) reduces to the continuous-time Toda  equation:
\be \label{a16}
\Delta_{Toda}^{(2)} = {v}_{n,tt} - e^{v_{n-1} - v_{n}} + e^{v_{n} - v_{n+1}} = 0,
\ee
when $\sigma_t \ra 0$ and $m \ra \infty$ in such a way that $t$ remains finite.
The  Toda equation \eqref{a16}
is probably the best known and most studied differential--difference equation. 
It plays, in the case of lattice equations, the same role as the Korteweg -
de Vries equation for partial differential equations\cite{1,1t}.  It was obtained by Toda \cite{t1} when trying to explain the Fermi, Pasta and Ulam results \cite{fpu} on the numerical experiments on the equipartition of energy in a nonlinear lattice of interacting oscillators. As shown below in Section \ref{B2} eq. (\ref{a16}) reduces, in the continuum limit, to the potential Korteweg-de Vries equation. It can be encountered in many applications from solid state physics to DNA biology, from molecular chain dynamics to chemistry \cite{rb}.

Eq.(\ref{a14}) can be obtained as the compatibility condition of the following over-determined  pair of linear difference equations:
\bea \label{a17a}
 \psi_{n-1,m} + (\alpha + \frac{1}{\alpha} -  \alpha e^{u_{n-1,m+1} - 
u_{n,m}}  
-\frac{e^{u_{n,m} - u_{n,n+1}}}{ \alpha}) \psi_{n,m} 
\\ \nonumber  + e^{u_{n,m} - 
u_{n+1,m}} \psi_{n+1,m} =  \lambda \psi_{n,m}
\\ \label{a17b}
 \psi_{n,m+1} =  \psi_{n,m} - \alpha e^{u_{n,m+1} - u_{n+1,m}} 
\psi_{n+1,m}
\eea
Let us consider the Lie point symmetries of the  discrete-time Toda lattice 
 (\ref{a14}), on a fixed nontrasforming bidimensional lattice characterized by two lattice spacings in the two directions $m$ and $n$, $\sigma_t$ and $\sigma_x$. 
  
 If the lattice is uniform and homogeneous in both variables, we can represent the lattice by the following set of equations:
\bea \label{b5a}
x_{n,m}  = n  \sigma_x,
\quad t_{n,m} = m  \sigma_t.
\eea
Eqs. (\ref{b5a}) from now on will be denoted  as $\Delta_{Lattice}=0$. 

A Lie point symmetry is defined by giving its infinitesimal 
generators, i.e. the vector field 
\bea \label{b6}
\hat X_{n,m} & = & \xi_{n,m} (x_{n,m}, t_{n,m}, u_{n,m}) \partial_{x_{n,m}} +  \\ \nonumber 
& + & \tau_{n,m} (x_{n,m}, t_{n,m}, u_{n,m}) \partial_{t_{n,m}} + \phi_{n,m} (x_{n,m}, t_{n,m}, u_{n,m}) \partial_{u_{n,m}},
\eea
and generates an infinitesimal transformation in the site $(n,m)$ of its 
coordinates and of the function $u(x_{n,m}, t_{n,m}) = u_{n,m}$. The action of (\ref{b6}) on 
the discrete--time Toda lattice equation (\ref{a14}) is obtained by 
prolonging (\ref{b6}) to all points of the lattice. The prolongation 
is obtained \cite{ref49,ref19,ref20}  by shifting (\ref{b6}) to these 
points
\be \label{b7}
pr \hat X = \hat X_{n,m} + \hat X_{n+1,m} + \hat X_{n,m+1} + \hat 
X_{n,m+2} + \hat X_{n-1,m+2}.
\ee
The invariance condition then reads:
\bea \label{b8}
pr \hat X \Delta_{Toda}|_{(\Delta_{Toda}=0, \Delta_{Lattice}=0)} =0, \quad
pr \hat X \Delta_{Lattice}|_{(\Delta_{Toda}=0, \Delta_{Lattice}=0)} =0
\eea
The action of (\ref{b7}) on the lattice equation (\ref{b5a}) gives $\xi_{n,m} = 0$, $\tau_{n,m} = 0$.  When we act with (\ref{b7}) on the Toda equation, we get
\bea \label{b9}
e^{u_{n,m} - u_{n,m+1}} [ \phi_{n,m} - \phi_{n,m+1}] - e^{u_{n,m+1} - u_{n,m+2}} [\phi_{n,m+1} - \phi_{n,m+2}] - \\
\nonumber - \alpha^{2} \{ 
e^{u_{n-1,m+2} - u_{n,m+1}}[\phi_{n-1,m+2} -  \phi_{n,m+1}] -\\ \nonumber - e^{u_{n,m+1} - u_{n+1,m}}[\phi_{n,m+1} - \phi_{n+1,m}] \} = 0
\eea
From the action of $\hat X$ on \eqref{b5a} the infinitesimal coefficients $\xi$ and $\tau$ are zero  and thus the variables $x$ and $t$ are invariant. If we differentiate eq. (\ref{b9}) twice  with respect to $u_{n,m+2}$ we get that $\phi_{n,m} = c_1 e^{u_{n,m}} + c_2$, where $c_1$ and $c_2$ are two integration constants (that can depend on  $x_{n,m}$ and $t_{n,m}$). Introducing this result into eq. (\ref{b9}) we get that $c_1$ must be equal to zero. Taking into account that, due to the form of the lattice, all points are independent we get that $c_2$ must be just a constant.

To sum up, the discrete time Toda equation \eqref{a14} considered on a fixed lattice has only a one-dimensional continuous symmetry group. It consists of the translation of the dependent variable $u$, i.e. $\tilde u_{n,m} = u_{n,m} + \lambda$ with $\lambda$ constant. This symmetry is obvious from the beginning as eq. (\ref{a14}) does not involve $u_{n,m}$ itself but only differences between values of $u$ at different points of the lattice. Other transformations that leave the lattice and solutions invariant will be discrete \cite{lrdiscrete}. In this case they are simply translations of $x$ and $t$ by integer multiples of the lattice spacing $\sigma_x$ and $\sigma_t$.

The same conclusion holds in the general case of difference equations on fixed lattices. Lie algebra techniques will provide transformations of the continuous dependent variables only, through the transformations can depend on the discrete independent variables.

In Section \ref{sec.n4} we will see that the situation is completely different when generalized symmetries are considered. In Section \ref{sec.n2} we will consider transforming lattices, which greatly increases the role of point symmetries.

\subsection{Lie point symmetries of D$\Delta$Es} \label{sect2}

Let us now consider the more interesting case of  D$\Delta$Es. For notational simplicity, let us restrict ourselves to scalar D$\Delta$Es for one
real function $u(n,t)$ depending on one lattice variable $n$ and
one continuous real variable, $t$. Moreover, we will only be  interested
in D$\Delta$Es containing up to second order derivatives, as those are the ones of
particular interest in applications to dynamical systems. We write such
equations as: 
\begin{multline}
\Delta _n^{(2)}\equiv \Delta \bigl(t,n,u_{n+k}\big| _{k=a_0}^{b_0},u_{n+k,t}\big|
_{k=a_1}^{b_1},u_{n+k,tt}\big| _{k=a_{2}}^{b_{2}}\bigr)=0 \\ 
a_j \le b_j \in \mathbb{Z},
\label{t1}
\end{multline}
with $u_n \equiv u_n(t)$.

The lattice is uniform, time independent and fixed, the continuous variable $t$ is the same at all points of the lattice. Thus to eq. (\ref{t1}) we add 
\be \label{t1b}
t_n = t_{n+1} = t
\ee

Examples of such equations which will be considered in the following are the
Toda lattice equation \eqref{a16}
and the inhomogeneous Toda lattice~\cite{lr}
\begin{multline}
\label{t3}
{\tilde \Delta}_n^{(2)}=w_{,tt}(n)-\frac 12 w_{,t}+\frac 14 
- \frac n2 +\biggl[\frac 14 (n-1)^2+1\biggr]e^{w(n-1)-w(n)}\\
\smash{-\biggl[\frac 14 n^2+1\biggr]}
e^{w(n)-w(n+1)} = 0.
\end{multline}

We are interested in Lie point transformations which leave the solution set
of eqs.~\eqref{t1}, \eqref{t1b}  invariant. They have the form:
\begin{equation}
\label{t4}
\tilde t=\Lambda _g\bigl(t,n,u_n(t)\bigr),\quad \tilde u_{\tilde n}(\tilde
t)=\Omega_g\bigl(t,n,u_n(t)\bigr),\quad\tilde n= n 
\end{equation}
where $g$ represents a set of continuous or discrete group parameters.

Continuous transformations of the form \eqref{t4} are generated by a Lie
algebra of vector fields of the form:
\begin{equation}
\label{t5}
\widehat X=\tau_n \bigl(t,u_n(t)\bigr)\partial _t+\phi_n
\bigl(t,u_n(t)\bigr)\partial _{u_n}  
\end{equation}
where $n$ is treated as a discrete variable and we have set
$\tilde n=n$, when considering continuous transformations.

Invariance of the condition (\ref{t1b}) implies that $\tau$ does not depend on $n$.

As in the case of purely differential equations, the following invariance condition
\be
\label{t6}
\pr^{(2)}\widehat X\Delta _n^{(2)}\big| _{\Delta _n^{(2)}=0}=0
\ee
must be true if $\widehat X$ is to belong to the Lie symmetry algebra
of $\Delta_n^{(2)}$. The symbol $\pr^{(2)}\widehat X$ denotes the second
prolongation of the vector field $\widehat X$, i.e. 
\begin{equation}
\begin{split}
\pr^{(2)}\widehat X=\tau \bigl(t,u_n\bigr)\partial_t &+\sum_{k=n-a}^{n+b}\phi_k
\bigl(t,u_k\bigr)\partial_{u_k} \\ &+\sum_{k=n-a_1}^{n+b_1}\phi_k
^t\bigl(t,u_k,u_{k,t}\bigr)\partial _{u_{k,t}}\\
&+\sum_{k=n-a_{2}}^{n+b_{2}}\phi_k
^{tt}\bigl(t,u_k,u_{k,t},u_{k,tt}\bigr)\partial_{u_{k,tt}} 
\end{split}\label{t7}
\end{equation}
with
\begin{align}
\label{t8}
\phi_k^t\bigl(t,u_k,u_{k,t}\bigr) &=D_t\phi_k
\bigl(t,u_k\bigr)-\bigl[D_t\tau_k \bigl(t,u_k\bigr)\bigr]u_{k,t} \\ 
\label{t9}
\phi_k ^{tt}\bigl(t,u_k,u_{k,t},u_{k,tt}\bigr) &=D_t\phi_k
^t\bigl(t,u_k,u_{k,t}\bigr)-\bigl[D_t\tau_k
\bigl(t,u_k\bigr)\bigr]u_{k,tt} 
\end{align}
where $D_t$ is the total derivative with respect to $t$.

Let us notice moreover that $\phi ^t$ and $\phi ^{tt}$
are the prolongation coefficients with respect to the continuous variable.
The prolongation  with respect to the discrete
variable is reflected in the summation over $k$.

Eq.~\eqref{t6} is to be viewed as just one equation with $n$ as a discrete
variable; thus we have a finite algorithm for obtaining the determining
equations, a usually overdetermined system of linear partial differential
equations for $\tau \bigl(t,u_n\bigr)$ and $\phi \bigl(n,t,u_n\bigr)$.

A different approach consists of considering eq.~\eqref{t1} as a system of
coupled differential equations for the functions $u_n(t)$. Thus, in general
we have infinitely many equations for infinitely many functions. In this
case the Ansatz for the vector field $\widehat X$ would be:
\begin{equation}
\label{t10}
\widehat X=\tau (t,\{u_j(t)\}_j)\partial _t+\sum_k\phi
_k(t,\{u_j(t)\}_j)\partial _{u_k(t)}
\end{equation}
where by $\{u_j(t)\}_j$ we mean the set of all $u_j(t)$ and $j$ and $k$ vary
a priori over an infinite range. Calculating the second prolongation $
\pr^{(2)}\widehat X$ in a standard manner ( see eqs. (\ref{1.7}), (\ref{1.9})) and imposing
\begin{equation}
\label{t11}
\pr^{(2)}\widehat X\Delta _n^{(2)}\big| _{\Delta _j^{(2)}=0}=0\quad
\forall n,j 
\end{equation}
we obtain, in general, an infinite system of determining equations for an
infinite number of functions.
Conceptually speaking, this second method, called 
the \emph{differential equation method} in Ref.~\cite{ref20}, may give rise
to a larger symmetry group than the \emph{intrinsic method} we introduced
before. In fact the intrinsic method yields purely point transformations,
while the differential equation method can yield generalized symmetries with
respect to the differences (but not the derivatives). 
In practice, it turns out that usually no
higher order symmetries with respect to the discrete variable exist; then
the two methods give the same result and the intrinsic method is
simpler.

A third approach  \cite{ref26,ref27} consists of
interpreting the variable $n$ as a continuous variable and consequently the
D$\Delta$E as a differential-delay equation. In such an approach $u_{n+k}(t)\equiv
\exp  [\frac{k ~\partial}{\partial n}] \{u_n(t)\} $ and consequently the D$\Delta$E is interpreted as a partial differential equation of infinite order. In such a case
formula~\eqref{t6} is meaningless as we are not able to construct the infinite order prolongation of a vector field $\widehat X$. The  Lie symmetries are obtained
by requiring that the solution set of the
equation $\Delta _n^{(2)}=0$ ~\eqref{a16} be invariant under the infinitesimal
transformation
\begin{equation}
\label{t13} 
\begin{aligned}
{\tilde t}&=t+\epsilon \tau_n \bigl(t,u_n(t)\bigr), \\ 
{\tilde n}&=n+\epsilon \nu_n \bigl(t,u_n(t)\bigr), \\ 
{\tilde u}_{\tilde n}({\tilde t})&=u_n(t)+\epsilon \phi_n \bigl(t,u_n(t)\bigr).
\end{aligned}
\end{equation}

To obtain conditional symmetries we add to eq.~\eqref{t1} a constraining 
equation which we choose in such a way that 
it is automatically annihilated on its solution set by the
prolongation of the vector field. 
 Such an equation is the invariant 
surface condition
\begin{equation}
\label{t14}
\Delta _n^{(1)}=\phi_n\bigl(t,u_n(t)\bigr)-\tau_n \bigl(t,u_n(t)\bigr)
u_{n,t}(t)=0. 
\end{equation}
In general eqs.~\eqref{t1} and~\eqref{t14} may  not be compatible;
if they are, then eq.~\eqref{t14} provides a reduction of 
the number of the independent variables by one. This is the essence of the
symmetry reduction by conditional symmetries.  Due to the fact that 
eq.~\eqref{t14} is written in terms of $\tau$ and $\phi$, which are the 
coefficients of the vector field $\widehat X$, the determining 
equations are nonlinear. The vector fields do not form an algebra, nor
even a vector space, 
since each vector field is adapted to a different
condition~\cite{bluman,levi4,CK}. 

\subsection{Symmetries of the Toda lattice} \label{sect3}

Let us now apply the techniques introduced in  Section \ref{sect2} to the
case of eq.~\eqref{a16}. In this case eq.~\eqref{t6} reduces to an overdetermined
system of determining equations obtained by equating to zero the
coefficients of $[v_{n,t}]^k$, $k=0$, 1, 2, 3 and of $v_{n\pm 1}$. They imply:
\begin{equation}
\label{t2.1} 
\tau =at+d,\quad  \phi =b+2an+ct, 
\qquad a,b,c,d \text{ real constants,} 
\end{equation}
corresponding to a four dimensional Lie algebra generated by the vector
fields:
\begin{equation}
\label{t2.2} 
\widehat D=t\partial _t+2n\partial _{v_n}, \quad \widehat
T=\partial _t, \quad  
\widehat W=t\partial _{v_n}, \quad \widehat U=\partial _{v_n}. 
\end{equation}

The group transformation which will leave eq.~\eqref{a16} invariant is hence:
\begin{equation}
\label{t2.3}
{\tilde v}_n({\tilde t})=v_n({\tilde t}e^{-\lambda _4/2}- \lambda _3)+\lambda _2({\tilde t}e^{-\lambda
  _4/2}-\lambda_3)+\lambda _4n+\lambda _1  
\end{equation}
where $\lambda _j$, $j=1$, 2, 3, 4, are real group parameters.

To the transformation \eqref{t2.3} we can add some discrete ones
\cite{5}: 
\begin{equation}
\label{t2.4}{\tilde n}=n+N \quad N\in \mathbb{Z} 
\end{equation}
and
\begin{equation}
\label{t2.5}
\bigl(t,v_n\bigr)\to \bigl(-t,v_n\bigr);\quad \bigl(t,v_n\bigr)\to
\bigl(t,-v_{-n}\bigr).
\end{equation}

We write the symmetry group of eq.~\eqref{a16} as:
\begin{equation}
\label{t2.6}
 \textgoth{G} = \textgoth{G}_D\triangleright  \textgoth{G}_C 
\end{equation}
where $ \textgoth{G}_D$ are the discrete transformations \eqref{t2.4}, \eqref{t2.5} and the
invariant subgroup $ \textgoth{G}_C$ corresponds to the transformation \eqref{t2.3}. A
complete classification of the one dimensional subgroups of $ \textgoth{G}$ can be
easily obtained~\cite{ref20}.

In fact, if we complement the Lie algebra \eqref{t2.2} by the vector field
\begin{equation}
\label{t2.7}
\widehat Z=\frac \partial {\partial n} 
\end{equation}
and require, at the end of the calculations, that the corresponding group
parameter be integer, the commutation relations become:
\begin{equation}
\label{t2.8}
[\widehat Z,\widehat D]=2\widehat U;\quad [\widehat D,\widehat T]=-\widehat T;\quad [\widehat
D,\widehat W]=\widehat W;\quad [\widehat T,\widehat W]=\widehat U.
\end{equation}

The one dimensional subalgebras are:
\begin{equation}
\label{t2.9} 
\begin{gathered}
\{\widehat Z+a\widehat D+b\widehat U\},\quad \{\widehat Z+a\widehat T+k\widehat W\},\quad \{\widehat
Z+\epsilon \widehat W\},\quad \{\widehat Z+a\widehat U\}, \\ 
\{\widehat T+c\widehat W\},\quad \{\widehat D+c\widehat U\},\quad \{\widehat U\},\quad \{\widehat
Z\},\quad \{\widehat W\}, \\ 
(a,b,c)\in \mathbb{R};  a\neq 0; k=0,1,-1; \epsilon =\pm 1 .
\end{gathered}
\end{equation}

Nontrivial solutions, corresponding to reductions with respect to 
\emph{continuous subgroups} $ \textgoth{G}_0\subset \textgoth{G}_C$, are obtained by considering
invariance of the Toda lattice under $\{\widehat T+c\widehat W\}$, or
$\{\widehat D+c\widehat U\}$. They are:
\begin{align}
\label{t2.10}
v_n(t) &=p-\frac 12ct^2-\sum_{j=1}^n\log (q-cj), \\
\label{t2.11}
v_n(t)&=p+2(n+c)\log (t)-\smash[t]{\sum_{j=0}^n}\log [q+(2c-1)j+j^2], 
\end{align}
where $p$ and $q$ are arbitrary integration parameters.

Reduction by the purely \emph{discrete subgroup}, $ \textgoth{G}_0\subset  \textgoth{G}_D$,
given in eq.~\eqref{t2.4} 
 implies the
invariance of eq.~\eqref{a16} under discrete translation of $n$ and makes it
possible to impose the periodicity condition
\begin{equation}
\label{t2.12}
u(n+N,t)=u(n,t).
\end{equation}
This reduces the D$\Delta$E~\eqref{a16} to an ordinary differential
equation (or a finite system of equations).
For example for $N=2$, we get
\begin{equation}
\label{t2.14}
v _{,tt}=-4\sinh v, 
\end{equation}
while for $N=3$, we get the  Tzitzeica
differential equation \cite{tz,tz1}: 
\begin{equation}
\label{t2.16}
v_{,tt}=e^{-2v}-e^v. 
\end{equation}

Let us now consider a subgroup $ \textgoth{G}_0\subset  \textgoth{G}$ that is not contained
in $ \textgoth{G}_C$, nor in $ \textgoth{G}_D$, i.e.\ a \emph{nonsplitting subgroup} of
$ \textgoth{G}$. A reduction corresponding to $\widehat Z+a\widehat
D+b\widehat U$ yields the equation
\begin{equation}
\label{t2.17}
F''(y)=e^{-b}[\exp (F(ye^a)-F(y)+a)-\exp(F(y)-F(ye^{-a})-a)] 
\end{equation}
where the symmetry variables $F(y)$ and $y$ are defined by:
\begin{equation}
\label{t2.18} 
y=te^{-an}, \quad
v_n(t)=an^2+bn+F(y).
\end{equation}
Using the subalgebra $\widehat Z+a\widehat T+k\widehat W$ we get
\begin{equation}
\label{t2.19}
F''(y)=e^{F(y+a)-F(y)}-e^{F(y)-F(y-a)}-\frac ka, 
\end{equation}
where
\begin{equation}
\label{t2.20} 
y=t-an,\quad v_n(t)=\frac k{2a}t^2+F(y).
\end{equation}

Eq.~\eqref{t2.17} can be called a ``differential dilation'' type
equation; it involves one independent variable $y$, but the function $F$ and
its derivatives are evaluated at the point $y$ and at the dilated
points $ye^a$ and $ye^{-a}$. Eq.~\eqref{t2.19} is a differential
delay equation 
which has interesting solutions, such as  the soliton and periodic
solutions of the Toda lattice (for $k=0$).

The other two nonsplitting subgroups give rise to linear delay equations
which can be solved explicitely.

This same calculation can also be
carried out for the inhomogeneous Toda lattice \eqref{t3}. The symmetry
algebra is
\begin{equation}
\label{t2.21} 
\begin{gathered}
\widetilde D=2\partial _{\tilde t}+\frac 12\partial _{w_n},\quad
\widetilde T=e^{-\tilde t/2}\biggl[\partial _{\tilde t}
-\biggl(w_n-\frac 12\biggr)\partial _{w_n} \biggr]\\ 
\widetilde W=2e^{\tilde t/2}\partial _{w_n},\quad \widetilde
U=\partial _{w_n}.
\end{gathered}
\end{equation}
These vector fields have the same commutation relations as those
of the Toda 
lattice \eqref{a16}. This is a necessary condition for the existence of a point
transformation between the two equations. In fact by comparing the two sets
of vector fields, we get the following transformation which transforms a solution $v_n(t)$ of equation \eqref{a16} into a solution $w_n(\tilde t)$ of eq.~\eqref{t3}
\begin{equation}
\label{t2.22} 
\begin{gathered}
\tilde t=2 \log \biggl(\frac t2\biggr), \\ 
w_n(\tilde t)=v_n(t)-2\biggl(n-\frac 12\biggr)\log (t)
+\sum_{j=0}^n\biggl[\frac 14(j-1)^2+1\biggr].  
\end{gathered}
\end{equation}

Let us now calculate the conditional symmetries of the Toda lattice. We assume that $\tau$ is not  zero and
 the determining equation reads:
\begin{equation}
\label{t2.23} 
\begin{split}
\phi_{n,tt}&+\phi_{n,t}\phi_{n,v_n}+2\phi_n\phi_{n,t v_n}
+\phi_n[\phi_{n,v_n}^2+\phi_n\phi_{n,v_n v_n)}] \\ &+
[2\phi_n-\phi_{n-1}-\phi_{n+1}]e^{v_n-v_{n+1}} +[\phi_n-\phi_{n-1}][\phi_{n,t}-\phi_n\phi_{n,v_n}]=0.
\end{split}
\end{equation}
This implies
\begin{equation}
\label{t2.24}
\phi_n\bigl(t,v_n(t)\bigr)= \alpha(t) + \beta(t) n
\end{equation}
with 
\begin{equation}
\label{t2.25}
\beta_{,tt}+\beta \beta_{,t}=0;\quad \alpha_{,tt}+\beta\alpha_{,t}=0.
\end{equation}
Solving eqs. (\ref{t2.25}) we get 
\begin{equation}
\label{t2.26}
\phi_n(t,v_n(t))=\begin{cases}
K_0+\biggl(2\dfrac {K_1}{K_3}+nK_3\biggr)\tanh
\biggl[\dfrac{K_3}2(t-t_0)\biggr], 
 & \text {for $K_3 \ne0$}\\[3pt]
\dfrac{K_0t+K_1+2n}{t-t_0}, & \text{for $K_3=0$.}
\end{cases}
\end{equation}
For $K_3=0$ we get the results given in
eq.~\eqref{t2.1}. 
For $K_3 \ne0$ an additional ``symmetry'' is given by 
$\phi=K_3n\tanh(K_3 t/2)$. This gives a new 
explicit solution of the Toda lattice:
\begin{equation} \label{t2.26a}
v_n(t)=u_0+2nK_3\log\biggl[\cosh\biggl(\frac{K_3}{2}t\biggr)\biggr].
\end{equation}
 
\subsection{Classification of differential equations on a lattice} \label{sect4}

Group theoretical methods can also be used to classify equations according
 to their symmetry groups. This has been done in the case of partial
differential equations\cite{gazeau} showing, for instance, that in the
class of variable 
 coefficient Korteweg-de~Vries equations, the Korteweg-de~Vries
 itself has the largest symmetry group. The same kind of 
results can also be obtained in the case of  differential-difference equations.

Let us consider a class of equations involving nearest neighbour interactions \cite{ref21}
\begin{equation}
\label{t2.27}
\Delta _n=u_{n,tt}(t)-F_n(t,u_{n-1}(t),u_n(t),u_{n+1}(t))=0,  
\end{equation}
where $F_n$ is nonlinear in $u_k(t)$ and coupled, i.e. such that $F_{n,u_k}\neq 0$
for some $k\neq n$.  

We consider  point symmetries only; the continuous transformations of
the form~\eqref{t4} 
 are again generated by a Lie algebra of vector fields of the
 form~\eqref{t5}.  
Taking into account the form of eq~\eqref{t2.27}, we have
\begin{equation}
\label{t2.28}
\widehat X = \tau(t)\partial_t + \biggl[\biggl(\frac12\tau_{,t} +
a_n\biggr) u_n + b_n(t)\biggr] \partial_{u_n} 
\end{equation}
with $ a_{n,t} = 0$. The determining equations reduce to 
\begin{multline}
\label{t2.29}
\frac 12 \tau_{,ttt} u_n + b_{n,tt} + (a_n-\frac 32 \tau_{,t}) F_n - \tau F_{n,t}
\\ -\sum_{k=0,\pm 1}\biggl[\biggl(\frac 12 \tau_{,t} +a_{n+k}\biggr)
u_{n+k} +b_{n+k}(t) \biggr] F_{n,u_{n+k}} = 0.
\end{multline}

Our aim is  to solve eq.~\eqref{t2.29} with respect to both the form of
the nonlinear equation,  
i.e.\ $F_{n}$, and the symmetry vector field $\widehat X$, i.e.\
$\bigl(\tau(t), a_n, b_n(t)\bigr)$.  
In other words, for every nonlinear interaction $F_n$ we wish to find the
 corresponding maximal symmetry group $\textgoth{G}$. Associated with any symmetry
 group $\textgoth{G}$  
there will be a whole class of nonlinear differential-difference
equations related  
to each other by point transformations. To simplify the results,
 we will just look for the simplest element of a given class of nonlinear 
differential-difference equations, associated to a certain symmetry group. 
To do so we introduce so called \emph{allowed transformations}, i.e.\
a set of transformations of the form 
\begin{equation}
\label{t2.30}
\tilde t = \tilde t(t),\quad
\tilde n = n\quad
u_n(t) = \Omega_n(\tilde u_n(\tilde t),t)
\end{equation}
that transform  equation~\eqref{t2.27} into a different one of the same type.
By a straightforward calculation we find that the only allowed
transformations \eqref{t2.30} are given by 
\begin{equation}
\label{t2.31}
\tilde t = \tilde t(t),\quad
\tilde n = n,\quad
u_n(t) = \frac{A_n}{\sqrt{\smash[b]{\tilde t_{,t}(t)}}} \tilde
u_n(\tilde t) + B_n(t)  
\end{equation}
with $B_n(t)$, $A_n$, $\tilde t(t)$ arbitrary functions of their arguments. 

Under an allowed transformation  equation \eqref{t2.27} is transformed into 
\begin{equation}
\label{t2.32}
\tilde u_{n,{\tilde t}{\tilde t}}(\tilde t) = \widetilde F_n\bigl(n,\tilde t, \tilde
u_{n+1}(\tilde t), \tilde u_n(\tilde t), \tilde u_{n-1}(\tilde
t)\bigr) 
\end{equation}
with
\begin{multline}
\label{t2.33}
\tilde F_n = \frac 1 {\tilde t_{,t}^{\frac{3}{2}} A_n}
\biggl\{F_n(n,t,\{u_{n+1}(t),u_n(t),u_{n-1}(t)\})-B_{n,tt} - \\
-\biggl[\frac34 \frac
{\tilde t_{,tt}^2}{\tilde t_{,t}^{\frac{5}{2}}}- 
\frac12\frac{\tilde t_{,ttt}}{\tilde t_{,t}^{\frac{3}{2}}}\biggr] A_n
\tilde u_n(\tilde t)\biggr\} 
\end{multline}
and a symmetry generator \eqref{t2.28} into
\begin{multline}
\label{t2.34}
\widehat{\widetilde X} = [\tau(t) \tilde t_{,t}]\partial_{\tilde t} + 
\biggl\{\biggr[\frac {\tau(t)}2 \frac {\tilde t_{,tt}}{\tilde t_{,t}}
+\frac 12 \tau_{,t}(t) + a_n\biggr]\tilde u_n  + \\ 
+ \frac {\tilde t_{,t}^{\frac{1}{2}}}{A_n} \biggl[ -\tau(t)
B_{n,t}(t)+B_n(t)\biggl(\frac12
\tau_{,t}(t)+a_n\biggr)+b_n(t)\biggr]\biggr\}\partial_{\tilde u_n}. 
\end{multline}

We see that, up to an allowed transformation, every one-dimensional
symmetry algebra 
associated to eq.~\eqref{t2.27}, can be represented  by one of the
following vector fields: 
\begin{equation}
\label{t2.35}
\widehat X_1 = \partial_t + a_n^1 u_n \partial_{u_n}\quad
\widehat X_2 = a_n^2 u_n \partial_{u_n}\quad
\widehat X_3 = b_n(t) \partial_{u_n}
\end{equation}
where $a_n^j$ with $j=1,2$ are two arbitrary functions of $n$ and
$b_n(t)$ is an arbitrary  
 function of $n$ and $t$.
The vector fields $\widehat X_j$, $j=1$, 2, 3 are the symmetry vectors
of the Lie point symmetries of the following nonlinear differential
difference equations: 
 \begin{equation}
\label{t2.36}
\begin{aligned}
\widehat X_1&: &\quad u_{n,tt} &= e^{a_n^1 t} f_n(\xi_{n+1}, \xi_n,
\xi_{n-1}), &\quad &\text{with } \xi_j = u_j e^{-a_j^1 t}\\
\widehat X_2&: & u_{n,tt} &= u_n f_n(t, \eta_{n+1}, \eta_{n-1}),
&&\text{with }\eta_j =\frac{(u_j)^{a_n^2}}{(u_n)^{a_j^2}}\\
\widehat X_3&: &u_{n,tt} &= \frac{b_{n,tt}}{b_n} u_n + f_n(t,
\zeta_{n+1}, \zeta_{n-1})  
&&\text{with } \zeta_j = u_j b_n(t)-u_n b_j(t).
\end{aligned}
\end{equation}
These equations are still quite general, as they are written in terms
of arbitrary  
functions depending on three continuous variables.  More specific
equations are obtained  
for larger symmetry groups \cite{ref21}.

The Toda equation (\ref{a16}) is included in a class of equations whose infinitesimal symmetry generators satisfy a four dimensional solvable symmetry algebra with a nonabelian nilradical. The interactions in this class are  given by
\be \label{t2.36a}
F_n(t,u_{n-1}(t),u_n(t),u_{n+1}(t)) = e^{-2\frac{u_{n+1}-u_n}{\gamma_{n+1}-\gamma_n}} f_n(\xi)
\ee
where $\xi = (\gamma_n(t) - \gamma_{n+1}(t))u_{n-1} + (\gamma_{n+1}(t) - \gamma_{n-1}(t))u_{n} + (\gamma_{n-1}(t) - \gamma_{n}(t))u_{n+1}$ and the function $\gamma_n(t)$ is such that $\gamma_{n+1}(t) \ne \gamma_n(t)$ and $\frac{\partial \gamma_n(t)}{\partial t} = 0$. The associated symmetry generators are:
\bea \label{t2.36b}
{\widehat X}_1 = \partial_{u_n}, \quad {\widehat X}_2 = \partial_t, \quad {\widehat X}_3 = t \partial_{u_n}, \quad {\widehat Y} = t \partial_t + \gamma_n(t) \partial_{u_n}.
\eea
The Toda equation (\ref{a16}) is obtained by choosing $\gamma_n(t) = 2 n$ and $f_n(\xi) = -1 + e^{\frac{1}{2} \xi}$. Among the equations of the class \eqref{t2.27}, the Toda equation does not have the largest group of point symmetries. 

A complete list of all equations of the type \eqref{t2.27} with nontrivial symmetry group is given in the original article \cite{ref21} with the additional assumption that the interaction and the vector fields depend continuously on $n$.  Here we just give two examples of interactions with symmetry groups with dimension seven.  The first one is solvable, nonnilpotent and its Lie algebra is given by
\begin{equation}
\label{t2.37}
\begin{gathered}
\widehat X_1 = \partial_{u_n},\quad
\widehat X_2 = (-1)^n \partial_{u_n},\quad
\widehat X_3 = t \partial_{u_n},\\
\widehat X_4 = (-1)^n t \partial_{u-n},\quad
\widehat X_5 = (-1)^n u_n \partial_{u_n},\quad
\widehat X_6 = \partial_t,\quad
\widehat X_7 = t \partial_t +2 u_n \partial_{u_n}.
\end{gathered}
\end{equation}
The invariant equation is
\begin{equation}
\label{t2.39}
u_{n,tt} = \frac{\gamma_n}{u_{n-1} - u_{n+1}}.
\end{equation}
This algebra was not included in Ref \cite{ref21} because of its nonanalytical dependence on $n$ (in ${\hat X}_2$, ${\hat X}_4$ and ${\hat X}_5$).

The second symmetry algebra is nonsolvable; it contains the simple Lie algebra $\Sl(2,\mathbb{R})$ as a 
subalgebra. A basis of this algebra is
\begin{equation}
\label{t2.38}
\begin{gathered}
\widehat X_1 = \partial_{u_n},\quad
\widehat X_2 = t \partial_{u_n},\quad
\widehat X_3 = b_n \partial_{u_n} \\
\widehat X_4 = b_n t \partial_{u_n},\quad
\widehat X_5 = \partial_t,\quad
\widehat X_6 =  t \partial_t +\frac 12 u_n \partial_{u_n},\quad
\widehat X_7 = t^2 \partial_t + t u_n \partial_{u_n}
\end{gathered}
\end{equation}
with $b_{n,t} = 0$, $b_{n+1}\ne b_n$.
The corresponding invariant nonlinear differential equation is:
\begin{equation}
\label{t2.40}
u_{n,tt} = \frac {\gamma_n}
{[(b_{n+1}-b_n)u_{n-1} + (b_{n-1}-b_{n+1})u_n + (b_n-b_{n-1})u_{n+1}]^{3}}
\end{equation}
where $\gamma_n$ and $b_n$ are arbitrary $n$-dependent constants.

In \cite{ly2004} the integrability conditions for equations belonging to the class  \eqref{t2.27} has been considered. It has been shown that any equation of this class which has local generalized symmetries can be reduced by point transformations of the form
\be \label{t2.40a}
{\tilde u}_n = \sigma_n(t, u_n), \quad {\tilde t} = \theta(t)
\ee
to either the Toda equation \eqref{a16} or to the potential Toda equation 
\be \label{t2.40b}
u_{n,tt} = e^{u_{n+1} - 2 u_n + u_{n-1}}.
\ee

\subsection{Symmetries of the Two Dimensional Toda equation} \label{sect5}

Let us now apply the techniques introduced in Section \ref{sect2} to the two-dimensional Toda system (TDTS) 
\be \label{t2.41}
\Delta_{TDTS}=u_{n,xt}- e^{u_{n-1} - u_n} + e^{u_n-u_{n+1}} = 0
\ee
where $u_n$ is also a function of $t$ and $x$. The TDTS was proposed and studied by
Mikhailov \cite{m78} and Fordy and Gibbons \cite{fg80} and is an integrable D$\Delta$E, having a Lax pair, infinitely
many conservation laws, B\"acklund transformations, soliton solutions, and all the usual attributes
of integrability \cite{AC,AS,cdI}.

The continuous symmetries for eq. \eqref{t2.41} are obtained by considering the infinitesimal symmetry generator
\be \label{t2.42}
{\hat X} = \xi_n(x, t, u_n) \partial_x + \tau_n(x, t, u_n) \partial_t + \phi_n(x, t, u_n) \partial_{u_n}.
\ee
From the determining equation \eqref{t6} we get
\be 
\tau_n = f(t), \quad \xi_n = h(x). \quad \phi_n = ( h_{,x} + f_{,t} ) ~n + g(t) + k(x),
\ee
where $f(t)$, $g(t)$, $h(x)$ and $k(x)$ are arbitrary $C^{\infty}$ functions. A basis for the symmetry algebra is given by
\bea \label{t2.43}
T(f) = f(t) \partial_t + n  f_{,t} \partial_{u_n}, \quad U(g) = g(t) \partial_{u_n}, \\ \nonumber
X(h) = h(x) \partial_x + n  h_{,x} \partial_{u_n}, \quad W(k) = k(x) \partial_{u_n},
\eea
where, to avoid redundancy, we must impose $k_{,x} \ne 0$.

The nonzero commutation relations are
\bea \label{t2.44}
&&[ T(f_1), T(f_2)] = T(f_1 f_{2,t} - f_{1,t} f_2), \quad [ T(f), U(g) ] = U(f g_{,t}), \\ \nonumber
&&[ X(h_1), X(h_2) ] = X (h_1 h_{2,x} - h_{1,x} h_2 ), \\ \nonumber
&& [ X(h), W(k) ] = {\Bigl \{} \begin{array}{ccc} W(h k_{,x}), & (h k_{,x})_{,x} \ne 0 & \\ c U(1), & (h k_{,x})_{,x} = 0, & h k_{,x} = c. \end{array}
\eea

Thus $\{T(f), U(g) \}$ form a Kac--Moody--Virasoro $\hat u(1)$ algebra, as do \;
\( \{X(h), \) \( W(k), h_{,x} \ne 0, U(1) \} \).  However the two $\hat u(1)$ algebras are not disjoint. This Kac--Moody--Virasoro character of the symmetry algebra is found also in the case of a ($ 2 + 1$)--dimensional Volterra equation. It is also characteristic of many other integrable equations involving three continuous variables, such as the Kadomtsev--Petviashvili or three--wave equations \cite{ow97,dklw85,dklw86,mw89,levi1,cw}. From the symmetry algebra we can construct the  group of symmetry transformations which leave the TDTS \eqref{t2.41} invariant and transform noninvariant solutions into new solutions. Moreover, we can use the subgroups to reduce the TDTS \eqref{t2.41} to equations in a lower dimensional space. 

By adding to \eqref{t2.41} the equation 
\be \label{t2.45}
\Delta_n^{(1)} = \xi_n(x, t, u_n) u_{n,x} + \tau_n(x, t, u_n) u_{n,t} - \phi_n(x, t, u_n) = 0 
\ee
we obtain the conditional symmetries. It is easy to show that conditional symmetries in the intrinsic method do not provide any  further symmetry reduction.    It is, however, worthwhile to notice that conditional symmetries in the differential equation method, when
\be \label{t2.46}
{\hat X} = \xi_n(x, t, u_j(x,t)) \partial_x + \tau_n(x, t, u_j(x,t)) \partial_t + \sum_k \phi_k(x, t, u_j(x,t)) \partial_{u_k},
\ee
contain the B\"acklund transformation of the TDTS. In fact, by choosing $\xi_n = 0$, $\tau_n = 1$, eq. \eqref{t2.46} reduces to $\Delta_n^{(1)} = u_{n,t} -  \phi_n(x, t, u_j)  = 0$ and the determining equations are solved by putting 
\be \label{t2.46b}
\phi_n(x, t, u_j) = f_{n,t}(x,t) + a [ e^{u_n - f_{n+1}} - e^{u_{n-1} - f_n} ]
\ee
 where $a$ is a real constant and $f_n(x,t) = {\tilde u}_n$ is a set of functions which solves the TDTS \eqref{t2.41}. The B\"acklund transformation for the TDTS can be indeed written as \cite{fg80}
\bea \label{t2.47}
u_{n,t}  - {\tilde u}_{n,t} = a  [ e^{u_n - {\tilde u}_{n+1}} - e^{u_{n-1} - {\tilde u}_n} ], \\ \nonumber
u_{n-1,x} - {\tilde u}_{n,x} = \frac{1}{a}  [ e^{{\tilde u}_{n-1} - u_{n-1}} - e^{{\tilde u}_{n} - u_n} ].
\eea

\section{Generalized Point Symmetries on a Fixed Uniform Lattice} \label{sec.n3}

We saw in Section \ref{sec.nn2} that point symmetries for purely difference equations on a fixed uniform lattice do not provide very powerful tools, though they work well for differential - difference equations.

Here we shall consider an approach that makes use of a certain type of generalized symmetries that act simultaneously at more than one point of the lattice. We call them "generalized point symmetries", because in the continuous limit they reduce to point symmetries.

This approach is directly applicable to linear difference equations, indirectly to nonlinear equations that are linearizable by a change of variables.

The underlying formalism is called "umbral calculus", or "finite operator calculus". Its modern development is mainly due to G.C. Rota and his collaborators \cite{ref52,ref53,onrota}. For a review article with an extensive list of references, see \cite{ref54}. Umbral calculus has been implicitly used in mathematical physics \cite{ref29,ref30,ref31,gorskii,olmo1,olmo2}. The only explicit use in physics that we know of is in \cite{dimakis,ref33,ref38b,rotafis}.

\subsection{Basic Concepts of Umbral Calculus} \label{sec.n31}

{\bf Definition 1.} 
{\it A } shift operator $T_{\sigma}$ {\it is a linear operator acting on polynomials or
formal power series $f(x)$ in the following manner}
\begin{equation}\label{5.1}
T_{\sigma}f(x) = f(x+\sigma), \quad x \in \mathbb R, \quad \sigma \in \mathbb R.
\end{equation}

{\it For functions of several variables we introduce shift operators in the same
manner}
\begin{eqnarray}
\lefteqn{T_{\sigma_i}f(x_1, \dots x_{i-1}, x_i, x_{i+1}\dots x_n)}\nonumber\\
&&\hspace{1in} = f(x_1, \dots, x_{i-1}, x_i+\sigma_i, x_{i+1}, \dots,
x_n).\label{5.2}
\end{eqnarray}

In this section we restrict the exposition to the case of one real variable $x
\in \mathbb R$. The extension to $n$ variables and other fields is obvious.
We will sometimes drop
the subscript on the shift operator $T$ when that does not give rise to
misinterpretations.

{\bf Definition 2.}
{\it An operator } $U$ {\it  is called a } delta operator {\it  if it satisfies the following
properties,}
\begin{enumerate}
\item[\textup{1)}]{\it  It is shift invariant;}
\be \label{5.3}
T_{\sigma} U = U T_{\sigma}, \quad \forall\sigma \in \mathbb R,
\ee
\item[\textup{2)}] 
\be \label{5.4}
U x = c~{\not =}~0, \quad c = \mbox{const},
\ee
\item[\textup{3)}] 
\begin{equation}\label{5.5}
U a = 0, \quad a = \mbox{const},
,\end{equation}
\end{enumerate}
{\it and the kernel of} $U$ {\it consists precisely of constants.}

Important properties of delta operators are:
\begin{enumerate}
\item[1.] For every delta operator $U$ there exists a unique series of basic
polynomials $\{P_n(x)\}$ satisfying
\begin{equation}\label{5.6}
P_0(x) = 1 \quad P_n(0) = 0, \quad n \geq 1, \quad UP_n(x) = nP_{n-1}(x).
\end{equation}

\item[2.] For every delta operator $U$ there exists a conjugate operator
$\beta$, such that
\begin{equation}\label{5.7}
[U, x\beta] = 1.
\end{equation}
The operator $\beta$ satisfies
\begin{equation}\label{5.8}
\beta = (U ^{\prime})^{-1}, \quad U ^{\prime} = [U, x].
\end{equation}
\end{enumerate}
Eq. (\ref{5.7}) can be interpreted as the Heisenberg relation between the delta operator $U$ and its conjugate $x\beta$.

We shall make use of two types of delta operators. The first is the ordinary (continuous) derivative operator, for which we have
\be \label{5.9}
U = \partial_x, \quad \beta = 1, \quad P_n(x) = x^n.
\ee
The second is a general \textit{ difference operator} which we define as 
\be \label{5.10}
U = \Delta = \frac{1}{\sigma} \sum_{k=l}^m a_k T^k_{\sigma}, \quad l,m \in {\mathbb Z}, \quad l < m,
\ee
where $a_k$ and $\sigma$ are real constants and $T_{\sigma}$ is a shift operator.  In order for $\Delta$ to be a delta operator, we must impose
\be \label{5.11}
\sum_{k=l}^m a_k = 0.
\ee
We shall also require that the continuous limit of $\Delta$ be $\partial_x$. This imposes a further condition, namely
\be \label{5.12}
\sum_{k=l}^m k a_k = 1.
\ee
Than eqs. (\ref{5.3}), \ldots, (\ref{5.5}) hold, with $c = 1$.

If eqs. (\ref{5.11}), (\ref{5.12}) are satisfied, we shall call $\Delta$ a \textit{difference operator of order $m-l$}. Acting with $\Delta$ on an arbitrary smooth function $f(x)$ we have
\bea \label{5.13}
\Delta f(x) &=& \frac{1}{\sigma} \sum_{k=l}^m a_k f(x + k \sigma) = \frac{1}{\sigma} \sum_{n=0}^{\infty} \frac{f^{(n)}(x)}{n!} \sigma^n \gamma_n, \\ \nonumber
\gamma_n &=& \sum_{k=l}^m  a_k k^n, \quad \gamma_0 = 0, ~ \gamma_1 = 1.
\eea
Thus we have
\be \label{5.14}
\Delta f(x) = \frac{\partial f}{\partial x} + \sum_{n=2}^{\infty} \frac{f^{(n})(x)}{n!} \sigma^{n-1} \gamma_n.
\ee
For $m-l \ge 2$ we can impose further conditions, namely
\be \label{5.15}
\gamma_n =0, \quad n = 2, 3, \ldots, m-l,
\ee
and obtain
\be \label{5.16}
\Delta = \frac{d}{dx} + O(\sigma^{m-l}).
\ee

From (\ref{5.8}) the operator $\beta$ conjugate to $\Delta$ is
\be \label{5.17}
\beta = ( \sum_{k=l}^m a_k k T_{\sigma}^k )^{-1}
\ee and the basic polynomials are
\be \label{5.18}
P_n(x) = (x \beta )^n 1.
\ee
It was shown in \cite{ref33} that the eq.(\ref{5.18}) yields a well defined polynomial of order $n$ in $x$.  For any differential operator $\Delta$ all coefficients in $P_n(x)$ are finite and each involves a finite number of positive powers of shifts in $\sigma$.

The simplest examples of difference operators and the related quantities are
\begin{enumerate}
\item 
\bea \label{5.19}
\Delta^+ & = & \frac{T - 1}{\sigma}, \quad \beta = T^{-1}, \\ \nonumber
P_n(x) & = & (x)_n = x ( x - \sigma ) ( x - 2 \sigma ) \ldots ( x - (n-1) \sigma ), \quad n \ge 1
\eea
Order $m-l = 1$.
\item
\bea \label{5.20}
\Delta^- & = & \frac{1 - T^{-1}}{\sigma}, \quad \beta = T,   \\ \nonumber
P_n(x) & = & (x)^n = x ( x + \sigma ) ( x + 2 \sigma ) \ldots ( x + (n-1) \sigma ), \quad n \ge 1
\eea
Order $m-l = 1$.
\item
\bea \label{5.21}
\Delta^s = \frac{T - T^{-1}}{ 2 \sigma}, \quad \beta = ( \frac{T + T^{-1}}{2})^{-1}, \\ \nonumber
P_{2n}(x) = x^2 ( x^2 - 4 \sigma^2 ) \ldots ( x^2 - ( 2 n - 2)^2 \sigma^2 ), \quad n \ge 1, \\ \nonumber
P_{2n+1}(x) = x ( x^2 - \sigma^2 ) \ldots ( x^2 - ( 2 n - 1)^2 \sigma^2 ), \quad n \ge 1.
\eea
Order $m-l = 2$.
\end{enumerate}
For any $\Delta$ we have $P_0 = 1$ and $P_1 = x$.

An important tool in the umbral calculus is the \textit{umbral correspondence}. This is a bijective mapping  $M$ between two delta operators $U_1$ and $U_2$.  This will induce a mapping between the corresponding operators $\beta_1$ and $\beta_2$, and also between the corresponding basic polynomials:
\be \label{5.22}
U_1 ~\overset{M}{\longleftrightarrow}~ U_2, \qquad \beta_1~ \overset{M}{ \longleftrightarrow} ~ \beta_2, \qquad P_n^{(1)}~ \overset{M}{\longleftrightarrow} ~P_n^{(2)}.
\ee

Let us now consider linear operators ${\hat L}(x \beta, U)$ that are polynomials, or formal power series in $x \beta$ and $U$. Since the umbral correspondence preserves the Heisenberg commutation relation (\ref{5.7}), it will also preserve commutation relations between the operators ${\hat L}(x \beta, U)$. In particular, we can take
\be \label{5.23}
U_1 = \partial_x, \quad \beta_1 = 1, \qquad U_2 = \Delta, \quad \beta_2 = \beta,
\ee
where $\Delta$ is any one of the difference operators (\ref{5.10}) and $\beta$ is as in (\ref{5.17}).

Let $A_1$ be a Lie algebra of vector fields of the form
\be \label{5.24}
{\widehat X}_{\alpha} = \sum_{j=1}^n a^j_{\alpha}(x_1, \cdots, x_n) \partial_{x_j}, \quad [ {\widehat X}_{\alpha}, {\widehat X}_{\beta}] = c^{\gamma}_{\alpha \beta} {\widehat X}_{\gamma}.
\ee
The umbral correspondence will map this algebra onto an isomorphic Lie algebra $A_2$ of difference operators
\be \label{5.25}
{\widehat X}_{\alpha}^u = \sum_{j=1}^n a^j_{\alpha}(x_1 \beta_1, \cdots, x_n \beta_n) \Delta_{x_j}.
\ee

\subsection{Umbral Calculus and Symmetries of Linear Difference Equations} \label{sec.n32}

Lie point symmetries of linear differential equations can be expressed in terms of commuting linear operators. Indeed, let us consider a linear differential equation
\be \label{5.26}
 L u = 0
 \ee
 where $L$ is some linear differential operator. The Lie point symmetry algebra of this equation can be realized by evolutionary vector fields of the form (\ref{1.16}), satisfying eq. (\ref{1.19}). If (\ref{5.26}) is an ordinary differential equation of order 3 or higher, or a partial differential equation of order 2 or higher, then the characteristic $Q$ of the vector field (\ref{1.16}) will have a specific form
 \be \label{5.27}
 Q = f(\overrightarrow{x}) + {\widehat X} u,
 \ee
 where $f(\overrightarrow{x})$ is the general solution of eq. (\ref{5.26}) and ${\widehat X}$ is a first order linear operator of the form \cite{ref55}
 \be \label{5.27a}
 {\widehat X} = \sum_{i=1}^n \xi_i(\overrightarrow{x}) \partial_{x_i} - \phi(\overrightarrow{x}),
 \ee
 satisfying
 \be \label{5.28}
 [ L , {\widehat X} ] u |_{L u = 0} = 0.
 \ee
 
 Using the umbral correspondence, we can carry this result over to a class of linear difference equations. Indeed, let us associate a difference operator $L_D$ with $L$ by the umbral correspondence $\partial_{x_i} \rightarrow \Delta_{x_i}$, $x_i \rightarrow x_i \beta_i$. Eq. (\ref{5.26}) is replaced by a difference equation
 \be \label{5.29}
 L_D u = 0.
 \ee
 The analogue of eq. (\ref{5.28}) will hold, namely
 \be \label{5.30}
 [ L_D, {\widehat X}_D ] u |_{L_D u = 0} = 0.
 \ee
 The difference operator ${\widehat X}_D$ will not generate point transformations taking solutions into solutions (they cannot be integrated to point transformation in general). They do, however, provide commuting flows, i.e. difference equations compatible with eq. (\ref{5.29}).
 
\subsection{Symmetries of the Time Dependent Schr\"odinger Equation on a Lattice} \label{sec.n33}

As an example let us apply the umbral correspondence to the free time - dependent Schr\"odinger equation. We obtain the difference equation
\be \label{5.31}
L_D \psi = 0, \qquad L_D = i\Delta_t + \frac{1}{2} \sum_{k=1}^n \Delta_{x_k x_k}.
\ee
The symmetries of eq. (\ref{5.31}) will be represented in terms of the "discrete" evolutionary vector fields
 \bea \label{5.32}
 {\widehat X}_D^E = Q_D \partial_{\psi} + Q_D^* \partial_{\psi^*},\quad
 Q_D =\eta_D - \tau_D \Delta_t \psi -\xi^k_D \Delta_{x_k} \psi,
 \eea
 where $\eta_D$, $\tau_D$ and $\xi_D^k$ are functions of $t \beta_t$, $x_j \beta_j$ and $\psi$, $\psi^*$ (the $*$ indicates complex conjugation).
 
 As in the continuous case, it can be shown \cite{ref33}, that the characteristic $Q$ will in this case have the form
 \be \label{5.33}
 Q_D = \chi(x_k \beta_k, t \beta_t) + i {\widehat X}_D \psi,
 \ee
 where $\chi$ is the general solution of eq. (\ref{5.31}). The first order difference operator ${\widehat X}_D$ satisfies
 \bea \label{5.34}
 && {\widehat X}_D = i [ \tau_D(t \beta_t) \Delta_t + \sum_{k=1}^n \xi_D^k \Delta_{x_k} - i \eta_D ], \\ \label{5.35}
 && [ L_D, {\widehat X}_D ] \psi |_{L_D \psi = 0} = 0
 \eea
Explicitating and solving eq. (\ref{5.35}), we obtain a difference realization of the Schr\"odinger algebra, first obtained in the continuous case by Niederer \cite{niederer}.
A basis for this algebra is given by the following operators:
\bea \label{5.36}
&& \hat P_0 = \Delta_t, \quad
\hat P_j = \Delta_{x_j}, \quad
\hat L_{j,k} = (x_j \beta_j ) \Delta_{x_k} - (x_k \beta_k ) \Delta_{x_j}, \\ \nonumber
&& \hat B_k = ( t \beta_t) \Delta_{x_k} - \frac{i}{2} x_k \beta_k, \quad
\hat D =  2 ( t \beta_t) \Delta_{t} - \sum_{k=1}^n (x_k \beta_k) \Delta_{x_k} + \frac{1}{2}, \\ \nonumber
&& \hat C = ( t \beta_t)^2 \Delta_{t} + \sum_{k=1}^n ( t \beta_t)(x_k \beta_k) \Delta_{x_k} + \frac{1}{2}  t \beta_t - \frac{i n}{4}  \sum_{k=1}^n (x_k \beta_k)^2,  \quad
\hat W = i.
\eea

Comparing with the continuous limit (or using the umbral correspondence), we see that $P_0$, $P_j$ correspond to time and space translation, $L_{j, k}$ to rotations, $B_k$ to Galilei boosts, $D$ and $C$ to dilations and projective transformations and $W$ to changes of phase of the wave function.

\subsection{Symmetries of the Discrete  Heat Equation} \label{sec.n34}

As a further example let us consider the discrete heat equation
\be \label{5.37}
\Delta_{xx} u(x,t) - \Delta_t u(x,t) = 0.
\ee
Eq. \eqref{5.37} is a linear partial difference equation on a two dimensional lattice. Floreanini, Negro, Nieto and Vinet showed in \cite{ref29} that \eqref{5.37} has a group of symmetries isomorphous to that of the continuous heat equation
\be \label{5.38}
u_{,xx}(x,t) - u_{,t}(x,t) = 0.
\ee 
This result can be easily recovered using the umbral calculus \cite{ref31,olmo1}.

Since \eqref{5.37} is linear, the symmetries are obtained by considering an evolutionary vector field of the form
\be \label{5.39}
{\hat X}_e = Q \partial_u = (\tau \Delta_t + \xi \Delta_x + f ) u \partial_u
\ee
and the determining equation is give by
\be \label{5.40}
\Delta_t Q - \Delta_{xx} Q |_{\Delta_{xx} u = \Delta_t u}
= 0
\ee
whose explicit expression is
\bea \label{5.41}
\Delta_t ( \xi \Delta_x u + \tau \Delta_t u +  f u ) -
 [ \Delta_{xx} ( \xi \Delta_x u + \tau \Delta_t u +  f u ) |_{\Delta_{xx} u = \Delta_t u}
= 0.
\eea
Defining $D_x f = [ \Delta, f ]$ it is immediate to see that $(D_x f ) 1 = (\Delta f) 1$ and for the operator $D$ the Leibnitz's  rule is satisfied: $D_x fg = (D_x f) g + f (D_x g)$.  In term of the operator $D$ we can split eq. \eqref{5.41} into the following overdetermined system of equations
\bea \label{5.42}
\begin{array}{cc} D_x \tau = 0, & D_t \tau - 2 D_x \xi = 0 \\
D_t \xi - D_{xx} \xi - 2 D_x f = 0, & D_t f - D_{xx} f = 0. \end{array}
\eea

From the umbral correspondence the solution of eq. \eqref{5.42} is
\bea \label{5.43}
\tau & = & \tau_2 (t \beta_t ) ^2 + \tau_1 t \beta_t  + \tau_0 \\ \nonumber
\xi & = & \frac{1}{2} ( \tau_1 + 2 \tau_2 t \beta_t  ) x \beta_x  + \xi_1 t \beta_t  + \xi_0 \\ \nonumber
 f & = &\tau_2 [  \frac{1}{4}  (x  \beta_x )^2 + \frac{1}{2} t  \beta_t  ] + \frac{1}{2} \xi_1 x \beta_x  + f_0
 \eea
 where $\tau_0$, $\tau_1$, $\tau_2$, $\xi_1$, $\xi_0$ and $f_0$ are arbitrary constants, functions of the lattice spacing and shifts. By a suitable
choice of these constants, we get the following representation of the symmetries   
\bea \label{5.44}
&& {\hat P}_0 = (\Delta_t u )\partial_u, \quad {\hat P}_1 =  (\Delta_x u )\partial_u \\ \nonumber
&& {\hat W} = u \partial_u, \quad {\hat B} = [ 2 t \beta_t  \Delta_x u + x \beta_x  u ] \partial_u \\ \nonumber
&& {\hat D} = [ 2 t \beta_t  \Delta_t u  + x \beta_x  \Delta_x u + \frac{1}{2}  u ] \partial_u
\\ \nonumber
&& {\hat K } = [ (t \beta_t )^2 \Delta_t u + t \beta_t x \beta_x  \Delta_x u + ( \frac{1}{4} 
( x \beta_x )^2 + \frac{1}{2} t \beta_t  ) u ] \partial_u.
\eea
For any choice of the $\Delta$ and the corresponding $\beta$ operators we have a different representation of the symmetry algebra of the heat equation. 

\section{Generalized Symmetries on Fixed Uniform Lattices} \label{sec.n4}

\subsection{Generalized symmetries of difference equations} \label{sec.n41}
We consider here the construction of generalized symmetries for 
difference equations and show the structure of  the infinite dimensional Lie 
algebra 
of point  and higher symmetries.

Generalized symmetries are symmetries whose infinitesimal generators 
depend on more than one point of the lattice and on derivatives with 
respect to the variables which vary in a continuous way. E. Noether was 
the first to notice in 1918 \cite{noe} that one can extend symmetries by including 
higher derivatives of the dependent variables in the transformation. 
They are more 
rare than point symmetries.  However, we can obtain an infinite number of 
generalized symmetries \cite{ref1} when the system is integrable \cite{cdI,AS,AC,N,ref59,FT,M}, i.e 
when the equations can be written as the compatibility condition for 
 an overdetermined system of linear 
equations (the Lax pair) and can be linearized either directly or via 
an Inverse Scattering Transform.

Lie symmetries, both point and generalized \cite{ref1}, can be 
obtained as flows 
commuting with the equation at study. We introduce here just the minimal 
notions on integrability necessary to be able to construct the commuting flows. 
For more on integrability see \cite{ref59,cdI,AS,AC,FT,N,M,1,1t}.

 A nonlinear partial differential equation for one function of two variables 
 \be \label{l1a}
 E(x, t, u(x,t), u_x(x,t), u_t(x,t),\ldots)=0
 \ee
  is said to be integrable if it 
can be characterized by a Lax pair \cite{lax}
\bea \label{l1}
L(u) \psi = \lambda \psi, \\ \label{l2} \psi_t = - M(u) \psi,
\eea
a system of  linear 
equations compatible only on the solution set of solutions of eq. (\ref{l1a}). In eqs. (\ref{l1}, \ref{l2}) 
$\lambda$ is an eigenvalue, $L$ and $M$ are two linear 
operators in x with coefficients
depending on $u$ but  not on $\lambda$. The function $\psi$, 
often called the wave or spectral function, depends on the 
independent variables ($x$,$t$) and on $\lambda$. If $\lambda_t = 0$ 
than $\lambda$ is an integral of motion, together with all the 
functions which depend only on it. 
In the particular case when eq. (\ref{l1}) represents the Schr\"odinger Spectral Problem
\be \label{e1}
\psi_{,xx} + [\lambda - u] \psi =0, \qquad \lambda = k^2,
\ee
for $u(x)$ vanishing at infinity, the solution of eq. (\ref{e1}) has the following asymptotic behaviour:
\bea \label{e2}
\psi(x, k) \rightarrow & b(k) e^{i k x} + a(k) e^{-i k x}, \quad & (x \rightarrow +\infty), \\ \label{e3}
\psi(x, k) \rightarrow &  e^{-i k x},  & (x \rightarrow -\infty).
\eea
It can be easily proved that, when $u(x,t)$ evolves according to the KdV equation \cite{cdI}, the function $a(k)$ is conserved, i.e. $\dot a(k) = 0$.
Thus, an infinite set of conservation laws 
can be obtained by expanding the transmission coefficient $1/a(\lambda)$
in powers of $\lambda$.  

The compatibility of eqs.(\ref{l1}, \ref{l2}) implies the Lax equation
\be \label{l3}
L_t(u) = [ L(u), M(u) ]
\ee
if $\lambda_t = 0$ or 
\be \label{l4}
L_t(u) = [ L(u), M(u) ] + f(L(u),t)
\ee
for $\lambda_t = f(\lambda,t)$. Here $f(z,t)$ is an entire function of its 
first argument. 

Eq. (\ref{l3}) or eq. (\ref{l4}) is equivalent to the 
nonlinear 
partial differential equation (\ref{l1a}).
In the first case we say that the equation (\ref{l1a}) is obtained 
as an isospectral deformation of (\ref{l1}) while in the second case by
a non-isospectral deformation. For isospectral deformations the existence 
of a Lax pair implies the existence of an infinite number of conserved 
quantities while
in the non-isospectral case these quantities will have a prescribed 
evolution in time.

The solution of the Cauchy problem 
for these systems is obtained by solving the Inverse Scattering Problem 
\cite{cdI,AC,AS}. The integrability of the isospectral 
systems has also been 
confirmed by the Painlev\'e test, by deriving the global properties 
of its solutions through singularity analysis \cite{kjh}. There is 
strong evidence, supported by many results, that all integrable 
equations have the Painlev\'e property, i.e. all solutions are 
single-valued 
around all movable singularities. In the case of ordinary differential 
equations this idea
was used by Kowalevski \cite{k1,k2} to identify integrable  rotating 
tops. For partial 
differential equations it lead to the Ablowitz, Ramani and Segur 
conjecture \cite{ars1,ars2} which states that {\it any ordinary differential equation 
which arises as a (symmetry) reduction of an integrable partial 
differential equation
possesses the Painlev\'e property, possibly after a transformation 
of variables.} This conjecture has never been proven but there are no known counterexamples.

Each integrable nonlinear 
partial differential equation (\ref{l1a}), be it isospectral, or non--isospectral,
 is characterized by an 
operator $M(u)$,  defined up to an arbitrary constant matrix. 
To each linear problem (\ref{l1}) we can associate a recursion operator
$\mathcal L$ and thus a hierarchy of nonlinear partial differential equations 
\cite{cdI}. The operators $M(u)$ appear in a sequence constructed in terms of 
$\mathcal L$.   The construction of $\mathcal L$ from the linear problem (\ref{l1}) 
is algorithmic \cite{akns,ca-re,BR,br}. 
The solution of the Cauchy problem for the integrable equation (\ref{l1a}) is obtained by performing the Spectral Transform and solving the Inverse Spectral 
Problem for the linear problem (\ref{l1}). Starting from a given initial 
potential $u(x,t=0)$ we can characterize the spectrum $S[u]$ of eq. (\ref{l1}). 
Solving the linear equation for the spectrum  
we can obtain the potential $u(x,t)$ for any $t > 0$ by solving the inverse problem.  
Each equation of the hierarchy 
\be \label{l4a}
E_j((x, t, u(x,t), u_x(x,t), u_t(x,t),\ldots)=0
\ee
is characterized by a different linear evolution of its spectrum $S_j[u)]$. 
At the level of the spectrum,  it is easy to prove that from the hierarchy 
of equations associated to a given linear equation (\ref{l1}) we can obtain 
a set of commuting flows. Thus every equation of the hierarchy (\ref{l4a})  can provide a set of evolutionary, in general generalized, symmetries \cite{Fokas1,Fu,Fua,f1,f2}. 
We have described here this procedure in the case of partial differential equations, where 
it was first introduced. This procedure has then been extended to the case of differential-difference 
and difference-difference equations \cite{discrete,ALJ,2,chiuLadik,16,16a,17,MF,N2,N1,S1,S2,SY}.

In the case of a differential--difference equation  
\be \label{l4b}
E(n, t, u_n(t), u_{n+1}(t), u_{n,t}(t),\ldots)=0 
\ee
the linear operators $L$ and $M$ that characterize it, depend on the shift operators in the discrete variable $n$. The Lax equations (\ref{l3}, \ref{l4}) are still valid. The recursion operator $\mathcal L$ and the Lax pair will depend on the shift operators, instead of $x$ derivatives. This implies that the higher equations of the hierarchy 
\be \label{l4c}
E_j(n, t, u_n(t), u_{n+1}(t), u_{n,t}(t),\ldots)=0
\ee
 and the higher symmetries will depend on points further away from the point $n$ instead of depending on higher derivatives.  The situation is slightly different in the case of difference--difference equations 
 \be \label{l4d}
  E(n, m, u_{n,m}, u_{n+1,m}, u_{n,m+1}, \ldots)=0.
  \ee
   The Lax pair in this case involves two linear functions $L_{n,m}$ and $M_{n,m}$ of the shift operator in $n$ with coefficients depending on $u_{n,m}$ The the linear equation (\ref{l2}) is replaced by
\be \label{e4}
\psi_{n,m+1} = -M_{n,m}(u)\psi_{n,m}.
\ee
The Lax equation in the isospectral regime ($\lambda_{m+1} = \lambda_m$) now reads:
\be \label{e5}
 L_{n,m+1} M_{n,m} = M_{n,m} L_{n,m}
 \ee 
 and in the nonisospectral case, when $\lambda_{m+1} = f(\lambda_m)$, with $f(.)$ an entire function of its argument, 
\be \label{e5b}
 L_{n,m+1} M_{n,m} = M_{n,m} f(L_{n,m}).
 \ee 
 Few results are known on generalized symmetries of difference--difference equations \cite{N3,N4,ref37}. In \cite{ref37}, using the techniques presented below, we computed the point and generalized symmetries for the discrete time Toda Lattice\eqref{a14}. The symmetries are provided by flows commuting with it. We are considering Lie symmetries, so these flows are given by differential difference equations. Indeed the group parameter is a continuous variable. So the symmetries of the discrete time Toda Lattice \eqref{a14} are given by the Toda Lattice hierarchy of nonlinear differential--difference equations \eqref{g2.1}, with the evolution not in the time variable, but in the group parameter. 
 
 The  Painlev\'e  property for integrable PDEs is not directly applicable. Several approaches have been proposed in the recent years \cite{ramani,ramani2,dp1,dp2,dp3,dp4,dp5,dp6,dp6b,dp7,dp8,dp9,dp10,dp11,dp12,dp13}. The first is the {\sl singularity confinement} proposed by Grammaticos et. al. \cite{ramani2}. iIn this approach one requires that any singularity of the solution of a discrete system be confined and should not propagate. This criterion is not sufficient, a system can satisfy the singularity confinement, but give rise to numerical chaos \cite{dp13}.
  More refined techniques have been proposed later. Among them let us mention that of Conte and Musette \cite{dp10} in which they propose a natural extension of the Painlev\'e property to discrete systems, that of Viallet and Hietarinta \cite{dp1,dp11,dp12,dp13} based on the study of growth and complexity properties and that of Ablowitz et. al.  \cite{dp3} based on the study of the asymptotic behaviour using the Nevanlinna theory of value distribution and growth of meromorphic functions.

In the following we will 
construct an infinite class of symmetries for the  Toda lattice. We will construct formally all the generalized symmetries 
and present explicitly the simplest examples. We will show the structure of the algebra of the generalized 
and point symmetries, using the one-to-one correspondence between the 
configuration space and the spectral space, where all equations are linear. In the continuous limit 
the Lie algebra of the point symmetries of the potential KdV (\ref{g4.4}) are recovered, as are the generalized symmetries. Finally, we present the 
B\"acklund transformations for the Toda lattice and show their relation to the symmetries.
For the corresponding results for the Volterra equation, the completely discrete
Toda lattice and for the Discrete 
Nonlinear Schr\"odinger equation we refer to the literature 
\cite{3,ref34,ref37,chiuLadik,hl,ref36}.

\subsection{The Toda system and its symmetries} \label{BB2}

The  Toda equation  (\ref{a16}) can be rewritten in the form of a system 
\begin{equation}\label{g1.1}
\dot{a}_n(t) = a_n(t) ( b_n(t) - b_{n+1}(t)), \qquad  
\dot{b}_n(t) = a_{n-1}(t) - a_n(t),
\end{equation}
where
\begin{equation}
b_n=\dot{v}_n,\quad a_n=\rm e^{v_n-v_{n+1}}.
\label{g2.8}
\end{equation}
It can be associated to the discrete Schr\"odinger spectral problem 
\begin{equation}
\psi(n-1,t;\lambda)+b_n \psi(n,t;\lambda)+ 
a_n\psi(n+1,t;\lambda)=\lambda \psi(n,t;\lambda).\label{g2.2} 
\end{equation}
The time evolution of the wave function $\psi(n,t;\lambda)$ is given by
\be \label{g2.2a}
\psi_t(n,t;\lambda) = - a_n(t) \psi(n+1,t;\lambda).
\ee

For the point symmetries of the Toda equation see Section \ref{sect3}. 

To the spectral problem (\ref{g2.2}) we can associate a  set of nonlinear differential 
difference equations 
(the Toda Lattice hierarchy)
\begin{eqnarray}
\begin{pmatrix}\dot{a}_n\cr \dot{b}_n\cr\end{pmatrix} = f_1({\cal L},t) 
\begin{pmatrix}a_n(b_n -b_{n+1})\cr a_{n-1}-a_n\cr\end{pmatrix}, \label{g2.1}
\end{eqnarray}
where $f_1({\cal L},t)$ is an entire function of its first argument and 
the recursion operator $\cal L$, is given by
\begin{eqnarray}
{\cal L}\begin{pmatrix}p_n\cr q_n\end{pmatrix}  =
\begin{pmatrix}p_n b_{n+1} + a_n(q_n+q_{n+1})+(b_n-b_{n+1})s_n\cr
b_n q_n + p_n +s_{n-1}-s_n\end{pmatrix}.\label{g2.3}
\end{eqnarray}
The operator (\ref{g2.3}) was firstly obtained by Dodd \cite{2} and by Bruschi, Levi and 
Ragnisco \cite{BLR-toda}.
In eq. (\ref{g2.3}) $s_n$ is a solution of the nonhomogeneous first order equation 
\begin{equation} 
s_{n+1} = {\frac{a_{n+1}} {a_n}} ( s_n - p_n ).\label{g2.3b} 
\end{equation}
For any equation of the hierarchy (\ref{g2.1}) we can write down an 
explicit evolution 
equation for the function $\psi(n,t;\lambda)$ \cite{BL,BR,BRL1} such 
that $\lambda$ does not 
evolve in time.
This is possible if the following boundary conditions  
\begin{equation}
\lim_{|n|\rightarrow \infty} a_n-1=
\lim_{|n|\rightarrow \infty} b_n= \lim_{|n|\rightarrow \infty} 
s_n=0,\label{g2.4} \end{equation}
are imposed on the fields $a_n$, $b_n$ and $s_n$.
The boundedness of the solutions of eqs.(\ref{g2.4}) was not required 
in the literature \cite{br,BLR,BRL1} but it is necessary to get a hierarchy of 
nonlinear differential-difference equations with well defined evolution 
of the spectra.
 We can than associate to 
eq.(\ref{g2.2}) a spectrum $S[u]$ \cite{BL,16,16a,17,1} 
defined in the complex plane of the variable $z$ ($\lambda = z
+  z^{-1}$):
\begin{equation}
S[u] = \{R(z,t),\; z\in {\bf C}_1; z_j,c_j(t),\; |z_j|<1, 
j=1,2,\ldots,N\},\label{g2.5}
\end{equation}
where $R(z,t)$ is the reflection coefficient, ${\bf C}_1$ is the unit 
circle in the complex 
$z$ plane, $z_j$ are isolated points inside the unit disk and $c_j$ 
are some complex 
functions of $t$ related to the residues of $R(z,t)$ at the poles $z_j$. 
When $a_n$, $b_n$ and $s_n$ 
satisfy the boundary conditions (\ref{g2.4}), the spectral data 
define the potentials in a 
unique way.
Thus, there is a one-to-one correspondence between the 
evolution of the potentials ($a_n$,$b_n$) of the discrete 
Schr\"odinger spectral problem (\ref{g2.2}), given by eq. (\ref{g2.1}) 
and that of the reflection coefficient $R(z,t)$, given by 
\begin{equation}
{\frac{d R(z,t)}{ d t}}=\mu f_1(\lambda,t)R(z,t),\quad 
\mu=z^{-1}-z. \label{g2.6}
\end{equation}
In eq. (\ref{g2.6}) and below, $\frac{d}{d t}$ denotes the total 
derivative with respect to $t$.

The Toda system is obtained from eq. (\ref{g2.1}) by choosing 
$f_1(\lambda,t) =1$ and thus the evolution equation of the reflection 
coefficient is given by
\begin{equation}
{\frac{d R(z,t)}{d t}}=\mu R(z,t). \label{g2.7}
\end{equation}

The symmetries for the Toda system (\ref{g1.1}), or the Toda equation (\ref{a16})  are provided by all flows commuting with the 
equation itself.  Let us introduce the following denumerable set of equations
\begin{equation}
\begin{pmatrix} a_{n,\epsilon_k} \cr b_{n,\epsilon_k} \end{pmatrix} = {\cal
 L}^k
\begin {pmatrix}a_n ( b_n
-b_{n+1}) \cr a_{n-1} - a_n \end{pmatrix}.\label{g2.12}
\end{equation}
Here $k$ is any positive integer and $\epsilon_k$ is a  
variable.
We can  associate to 
the equation (\ref{g2.12}) an evolution of 
the reflection coefficient
\begin{equation}
{\frac{d R}{d \epsilon_k}}=\mu \lambda^k R . \label{g2.14} 
\end{equation}
It can be verified directly that the flows (\ref{g2.7}) and (\ref{g2.14}) 
commute and hence the same must be true for the 
corresponding equations, i.e. for the Toda system (\ref{g1.1}) 
and the equations
(\ref{g2.12}). This implies that eqs. (\ref{g2.12}) are symmetries 
of the Toda system and consequently $\epsilon_k$ is a group parameter.
 From the point of view of the spectral problem (\ref{g2.2}), 
 eqs. (\ref{g2.12}) correspond to isospectral
deformations, i.e. we  have $\lambda_{\epsilon_k} =0$.
For any $\epsilon_k$, the solution of the Cauchy problem for 
eq. (\ref{g2.12}),  provides a 
solution of  the Toda system (\ref{g1.1}) 
[$a_n(t,\epsilon_k)$,$b_n(t,\epsilon_k)$]  in terms of 
the initial condition
[$a_n(t,\epsilon_k=0)$,$b_n(t,\epsilon_k=0)$]. The group transformation
corresponding to the group parameter 
$\epsilon_k$ can usually be written explicitly only for the lowest values of 
$k$.  In the case of the generalized symmetries, the group action is obtained in principle
by solving the Cauchy problem for the nonlinear characteristic equation starting 
from a generic solution of the Toda lattice. This often cannot be done. 
In all cases we can 
construct just a few classes of explicit group transformations corresponding to very specific 
solutions of the Toda lattice equation, namely the solitons, the rational 
solutions and the periodic solutions \cite{1}.
In all cases one can use the symmetries (\ref{g2.12}) 
to do, for example, symmetry reduction 
and to reduce the equation under consideration to an ordinary 
differential equation, or possibly a functional one. This is done by looking for fixed pints of the transformation, i.e. putting $a_{n,\epsilon_k} = 0$, $b_{n,\epsilon_k} = 0$,

We can extend the class of symmetries by considering nonisopectral 
deformations of the spectral problem (\ref{g2.2}) \cite{3,lz,Fokas1,Fu,Fua}. Thus for 
the Toda system we obtain
\begin{eqnarray}
\begin{pmatrix} a_{n,\epsilon_k}\cr b_{n,\epsilon_k} \end{pmatrix} = &
{\cal L}^{k+1} t
\begin{pmatrix} a_n( b_n
-b_{n+1})\cr a_{n-1}-a_n \end{pmatrix}
\nonumber\\
& +{\cal L}^k
\begin{pmatrix} a_n[(2n+3)b_{n+1}-(2n-1)b_n] \cr 
b_n^2-4+2[(n+1)a_n-(n-1)a_{n-1}] \end{pmatrix}.\label{g2.15}
\end{eqnarray}
In correspondence with eq.(\ref{g2.15}) we have the evolution of the 
reflection coefficient (\ref{g2.5}), given by
\begin{equation}
\frac{d R}{d \epsilon_k} = \mu \lambda^{k+1} t R, \qquad 
\lambda_{\epsilon_k} = \mu^2 \lambda^k, \label{g2.17}
\end{equation}
The proof that the eqs. (\ref{g2.15}) are symmetries is done by 
 showing that the flows (\ref{g2.17}) and (\ref{g2.7}) in the 
space of the 
reflection coefficients commute.

In addition to the above  two hierarchies of symmetries 
(\ref{g2.12}) and (\ref{g2.15}), we can 
construct two further cases, 
which, however, do not satisfy the asymptotic boundary conditions 
(\ref{g2.4}). They are:
\begin{eqnarray}
\begin{pmatrix} a_{n,\epsilon}\cr
b_{n,\epsilon} \end{pmatrix} =  \begin{pmatrix} 0 \cr 1 \end{pmatrix}, \quad
\begin{pmatrix} a_{n,\epsilon}\cr b_{n,\epsilon} \end{pmatrix} =
t  \begin{pmatrix} a_n(b_n -b_{n+1})\cr a_{n-1}-a_n \end{pmatrix}+
 \begin{pmatrix} 2 a_n \cr b_n \end{pmatrix}.\label{g2.19}
\end{eqnarray}
 As these exceptional symmetries do not 
satisfy the asymptotic boundary conditions (\ref{g2.4}), we cannot 
write a corresponding evolution equation for the reflection 
coefficient (\ref{g2.5}).

Let us now write down the lowest order symmetries for the  Toda system 
(\ref{g1.1}) that one can get from the hierarchies (\ref{g2.12}, \ref{g2.15}).
In the case of the Toda lattice (\ref{a16}) the symmetries are 
obtained from those of the Toda system (\ref{g1.1}) by using the transformation 
(\ref{g2.8}). The symmetries of the 
Toda lattice and the Toda system, corresponding to the isospectral and 
nonisospectral flows, will have the same evolution of the reflection 
coefficient. The transformation (\ref{g2.8}) involves an integration 
(to obtain $u_{n}$). The integration constant must be chosen so as to 
satisfy 
the following boundary conditions:
\begin{equation}
\lim_{|n|\rightarrow \infty} v_n=0.\label{g2.22}
\end{equation}
In the case of the exceptional symmetries such an integration will 
provide an additional symmetry. 

Taking $k=0,1,$ and $2$ in eq. (\ref{g2.12}) we obtain the first three 
isospectral symmetries for the Toda 
system, namely:
\begin{eqnarray}
a_{n,\epsilon_0} = a_n ( b_n - b_{n+1} ), \nonumber \\
b_{n,\epsilon_0} = a_{n-1} - a_n,
\label{g2.23}
\end{eqnarray}
\begin{eqnarray}
a_{n,\epsilon_1} = a_n ( b_n^2 - b_{n+1}^2 + a_{n-1} - a_{n+1} ), 
\nonumber \\
b_{n,\epsilon_1} = a_{n-1} ( b_n + b_{n-1} ) - a_n ( b_{n+1} + b_n ), 
\label{g2.24}
\end{eqnarray}
\begin{eqnarray}
a_{n,\epsilon_2} = a_n ( b_n^3 - b_{n+1}^3 + a_{n} b_{n} -2 a_{n+1} 
b_{n+1} + a_{n-1} b_{n-1} + 2 a_{n-1} b_{n} \nonumber \\
 - a_{n+1} b_{n+2} - a_{n} 
b_{n+1} - 2 b_{n} + 2 b_{n+1}), 
\nonumber \\
b_{n,\epsilon_2} = a_{n-1} ( b_n^{2} + b_{n-1}^{2} + 
b_{n}b_{n-1}+a_{n-1}+a_{n-2}-2 )  \nonumber \\
- a_n (  b_n^{2} + b_{n+1}^{2} + 
b_{n}b_{n+1}+a_{n+1}+a_{n}-2).
\label{g2.24b}
\end{eqnarray}
The lowest nonisospectral symmetry is obtained from eq. (\ref{g2.15}), 
taking $k=0$. It is:
\begin{eqnarray} \label{g2.25} 
a_{n,\nu} = a_n \{ t ( b_n^2 - b_{n+1}^2 + a_{n-1} - a_{n+1} 
) + (2 n + 3) b_{n+1} - (2 n - 1) b_n \}, \\ \nonumber 
b_{n,\nu} = t \{a_{n-1} ( b_n + b_{n-1} ) - a_n ( b_{n+1} + 
b_n ) \} +  b_n^2 - 4 + 2 [ (n+1) a_n - ( n - 1 ) a_{n-1} ]. 
\end{eqnarray}
The nonisospectral symmetries for $k>0$ are nonlocal.

The exceptional symmetries (\ref{g2.19}) are: 
\begin{eqnarray} \label{g2.26}
a_{n,\mu_0} &=& 0, \quad
b_{n,\mu_0} = 1, \\
a_{n,\mu_1} &=& 2 a_n  + t \dot a_{n}, \quad
b_{n,\mu_1} = b_n + t \dot b_{n}.
\label{g2.27}
\end{eqnarray}
The corresponding 
symmetries for the Toda lattice are
\begin{equation}
v_{n,\epsilon_0} =\dot  v_{n}
\label{g2.28}
\end{equation}
\begin{equation}
v_{n,\epsilon_1} = {\dot v_{n}}^2 + e^{v_{n-1}-v_n} + e^{v_n-v_{n+1}} 
-2 \label{g2.29}
\end{equation}
\begin{equation}
v_{n,\epsilon_2} = {\dot v_{n}}^3 - 2 {\dot v_{n}} + 
e^{v_{n-1}-v_n}({\dot v_{n-1}} + 
2 {\dot v_{n}}) + e^{v_n-v_{n+1}}({\dot v_{n+1}} + 2 {\dot v_{n}}) 
\label{g2.29b}
\end{equation}
\begin{equation}
 v_{n,\nu} = t \{{\dot v_{n}}^2 + e^{v_{n-1}-v_n} + 
e^{v_n-v_{n+1}} -2 \} - (2 n - 1 ) {\dot v_{n}} + w_n(t),
\label{g2.30}
\end{equation}
where $w_n(t)$ is defined by the following compatible system of equations:
\begin{equation}
 w_{n+1}(t)-w_n(t) = - 2 {\dot v}_{n+1}, \qquad {\dot w}_n(t) = 2 
(e^{v_n-v_{n+1}}-1).
\label{g2.30a}
\end{equation}
Under the assumption (\ref{g2.22}) we can integrate eqs. (\ref{g2.30a}) and 
obtain a formal solution. That is, we can write 
$w_n(t)$ in the form of an infinite sum; 
\begin{equation}
w_n(t) = 2 \sum_{j=n+1}^{\infty} {\dot v}_j + \alpha,
\label{g2.30b}
\end{equation}
where $\alpha$ is an arbitrary integration constant which can be interpreted 
as an additional symmetry.  The exceptional symmetries read:
\begin{equation}
v_{n,\mu_1} = t {\dot v_{n}} - 2 n
\label{g2.31}
\end{equation}
\begin{equation}
v_{n,\mu_0} = t
\label{g2.32}
\end{equation}
and the additional one, due to the integration \begin{equation}
v_{n,\mu_{-1}} = 1.
\label{g2.33}
\end{equation}

\subsubsection{The symmetry algebra for the Toda lattice} \label{B3}

To define the structure of the symmetry algebra for the Toda lattice
 we need to compute the commutation relations 
between the symmetries, i.e. between the flows commuting with the equations 
of the hierarchy. 
Using the one-to-one correspondence between the integrable equations 
and the evolution equations for the reflection coefficients, we 
 calculate the commutation relations between the 
symmetries and thus  analyze the structure of the obtained infinite 
dimensional Lie algebra.

If we define
\begin{eqnarray}
{\cal L}^k = \begin{pmatrix} {\cal L}_{11}^{(k)} \qquad {\cal 
L}_{12}^{(k)}\cr {\cal L}_{21}^{(k)} \qquad {\cal L}_{22}^{(k)} \end{pmatrix} 
,\label{g3.1}
\end{eqnarray}
we can write the generators for the isospectral symmetries as
\begin{eqnarray}
\hat X_k^T = & \{ {\cal L}_{11}^{(k)}[a_n ( b_n - b_{n+1} ) ]  +
{\cal  L}_{12}^{(k)} ( a_{n-1} - a_n ) \} \partial_{a_n}
\nonumber \\  & +\{  {\cal L}_{21}^{(k)}[a_n(b_n - b_{n+1})] +
{\cal L}_{22}^{(k)}(a_{n-1}  - a_n )\} \partial_{b_n}.
\label{g3.2}
\end{eqnarray}
The superscript on the generator $\hat X$, is there to indicate that this is the symmetry generator for the Toda system (\ref{g1.1}). We will indicate by the superscript $TL$ the symmetry generators for the Toda Lattice (\ref{a16}).
To these generators we can associate  symmetry 
generators in the space of the reflection coefficients. These 
generators are written as
\begin{equation}
{\hat {\cal X}}_k^T = \mu \lambda^k R {\partial_R}. 
\label{g3.6}
\end{equation}

In agreement with Lie theory, whenever $R$ is an analytic function of 
$\epsilon$, the corresponding flows are given by solving the 
equations \begin{equation}
\frac{ d \tilde R}{d \epsilon_k} =  \mu \lambda^k \tilde R,  
\qquad  \frac{ d \tilde \lambda}{d \epsilon_k} = 0,\qquad 
{\tilde R}(\epsilon_k=0) = R,
 \qquad \tilde \lambda 
(\epsilon_k=0)=\lambda. \label{g3.7}
\end{equation}

One can prove   that the isospectral symmetry generators (\ref{g3.2})   
commute amongst each
other
\begin{equation}
[\hat X_k^T, \hat X_m^T ] = 0,
\label{g3.4}
\end{equation}
by computing the corresponding commutation relation in the space of 
the reflection coefficients 
\begin{equation}
[\hat {\cal X}_k^T , \hat {\cal X}_m^T ] = [ \mu \lambda^k R \partial_R, 
\mu \lambda^m R \partial_R ] = 0.
\label{g3.8}
\end{equation}

So far, the use of the vector fields in the reflection coefficient 
space has just reexpressed a known result, namely that the commutation 
of the reflection coefficients is 
rewritten as eq. (\ref{g3.8}). We now extend the use of vector fields 
in the space of the spectral data to the case of the nonisospectral 
symmetries (\ref{g2.15}). Using the definition 
(\ref{g3.1}) we can introduce the generators of the nonisospectral 
symmetries for the Toda system. 
The symmetry vector fields  are:
\begin{eqnarray}
\hat Y_k^T = &\{ t [ {\cal L}_{11}^{(k+1)}[a_n (b_n -
b_{n+1}  )] + {\cal L}_{12}^{(k+1)}(a_{n-1} - a_n )]
\nonumber \\ &+{\cal  L}_{11}^{(k)}[a_n ( (2 n + 3)b_{n+1} - (2
n - 1 )b_n )] \nonumber 
\\ &+{\cal L}_{12}^{(k)}[b_n^2-4+2(n+1)a_n - 2 (n-1)a_{n-1} 
]\}\partial_{a_n} \nonumber \\ & +\{ t [ {\cal 
L}_{21}^{(k+1)}[a_n (b_n - b_{n+1} )] + {\cal 
L}_{22}^{(k+1)}(a_{n-1} - a_n )] \nonumber  \\ & +{\cal 
L}_{21}^{(k)}[a_n ( (2 n + 3)b_{n+1} - (2 n - 1 )b_n )] \nonumber 
\\ & +{\cal L}_{22}^{(k)}[b_n^2-4+2(n+1)a_n - 2 (n-1)a_{n-1} 
]\}\partial_{b_n}.
\label{g3.11}
\end{eqnarray}
Taking into account eq. (\ref{g2.17}), 
 we can define the symmetry generators 
(\ref{g3.11}) in the space of the spectral data, as
\begin{equation}
\hat  {\cal Y}_k^T = \mu \lambda^{k+1} t 
R\partial_R+\mu^2\lambda^k\partial_{\lambda} . 
\label{g3.13}
\end{equation}
Commuting $\hat  {\cal Y}_k^T$ with $\hat  {\cal Y}_m^T$ we have: 
\begin{equation}
[\hat  {\cal Y}_k^T,\hat  {\cal Y}_m^T]=(m-k)[\hat  {\cal 
Y}_{k+m+1}^T-4\hat  {\cal Y}_{k+m-1}^T].
\label{g3.15}
\end{equation}
From the isomorphism between the spectral space and the space of the 
solutions, we conclude that the vector fields representing the 
symmetries of the studied evolution equations, satisfy the same 
commutation relations
\begin{equation}
[\hat  { Y}_k^T,\hat  { Y}_m^T]=(m-k)[\hat  { Y}_{k+m+1}^T-4\hat  { 
Y}_{k+m-1}^T] ,
\label{g3.17}
\end{equation}
In a similar manner we can work out the commutation relations between 
the $\hat Y_k$ and $\hat X_m$ symmetry generators.
We get:
\begin{equation}
[\hat  {\cal X}_k^T,\hat  {\cal Y}_m^T]=-(1+k) \hat  {\cal 
X}_{k+m+1}^T+4k\hat  {\cal X}_{k+m-1}^T ,
\label{g3.19}
\end{equation}
and consequently
\begin{equation}
[\hat  { X}_k^T,\hat  { Y}_m^T]=-(1+k) \hat  { X}_{k+m+1}^T+4k\hat  { 
X}_{k+m-1}^T .
\label{g3.21}
\end{equation}
Relations like (\ref{g3.17}) and (\ref{g3.21}) can also be checked directly, but the use of vector field in the reflection coefficient space is much more efficient.

Let us now consider the commutation relations involving the 
exceptional symmetries (\ref{g2.19}).
We write them as:
\begin{equation}
\hat Z_0^T = \partial_{b_n}
\label{g3.23}
\end{equation}
\begin{equation}
\hat Z_1^T = [2 a_n + t {\dot a_n}]\partial_{a_n}+ [b_n+ t {\dot 
b_n}]\partial_{b_n}.
\label{g3.24}
\end{equation}
As  mentioned in Section \ref{sec.n41}, these symmetries do not satisfy the 
asymptotic conditions (\ref{g2.4}). Hence we cannot write down the 
commutation relations in all generality for all symmetries simultaneously.
We calculate explicitly the commutation relations involving just
$\hat Z_0^T$ and $\hat Z_1^T$, $\hat X_0^T$, $\hat X_1^T$ and $\hat Y_0^T$.
The non zero commutation relations  are:
\begin{eqnarray}
&&[{\hat X}_0^T,{\hat Z}_1^T] = 
- {\hat X}_0^T, \quad [{\hat Z}_0^T,{\hat Z}_1^T] = {\hat Z}_0^T 
\nonumber \\ 
&&\null [{\hat Y}_0^T,{\hat Z}_0^T  ] =-2 {\hat Z}_1^T,
\quad  [{\hat Y}_0^T,{\hat Z}_1^T] =
 - {\hat Y}_0^T -8 {\hat Z}_0^T, \quad [{\hat X}_1^T,{\hat Z}_0^T] = 
- 2{\hat X}_0^T, \nonumber \\ 
 &&\null [ {\hat X}_1^T,{\hat Z}_1^T  ] = - 2{\hat X}_1^T, 
[{\hat  X}_0^T,{\hat Y}_0^T] = - {\hat X}_1^T, [{\hat
X}_1^T,{\hat Y}_0^T] =  - 2{\hat X}_2^T + 4 {\hat X}_0^T. 
\label{g3.25}
\end{eqnarray}

In the case of the Toda lattice (\ref{a16}) we have (see eqs.(\ref{g2.32}, \ref{g2.31}))
\bea \label{g3.26}
&&\hat Z_0^{TL} =t \partial_{v_n} \\ \label{g3.27}
&&\hat Z_1^{TL} =[t \dot v_n -2 n] \partial_{v_n}
\eea
and
\begin{equation}
\hat Z_{-1}^{TL} =\partial_{v_n}
\label{g3.28}
\end{equation}
in correspondence with eq. (\ref{g2.33}).  As eqs. (\ref{a16}) and (\ref{g1.1}) are just two different representations of the same system, the symmetry generators in the space of the spectral data are the same. Consequently the commutation relations between $\hat X_n^{TL}$ and $\hat Y_m^{TL}$ are given by  eqs. (\ref{g3.8}, \ref{g3.17} and \ref{g3.21}).
The symmetries $\hat X_0^{TL}$, $\hat X_1^{TL}$  and $\hat Y_0^{TL}$, 
according to 
eqs.(\ref{g2.28}), (\ref{g2.29}), (\ref{g2.30}) are given by:
\begin{eqnarray}
 \hat X_0^{TL} &=&{\dot v_n} \partial_{v_n}, \quad \hat X_1^{TL}=[{\dot 
v_{n}}^2 +
e^{v_{n-1}-v_n} + e^{v_n-v_{n+1}}  -2]\partial_{v_n} \nonumber \\ 
 \hat Y_0^{TL} &=& 
\{t [v_{n,t}^2 + e^{v_{n-1}-v_n} + e^{v_n-v_{n+1}} -2 ] - (2 n - 1 ) 
v_{n,t} + w_n(t)\}\partial_{v_n} \nonumber \\
&& w_{n+1}(t)-w_n(t) = - 2 {\dot v}_{n+1}, \quad {\dot w}_n(t)
= 2  (e^{v_n-v_{n+1}}-1).
\label{g3.29}
\end{eqnarray}
The nonzero commutation relations are:
\begin{eqnarray}
&&[\hat X_0^{TL},\hat Z_0^{TL}] = - \hat Z_{-1}^{TL}, \quad [\hat 
X_0^{TL},\hat Z_1^{TL}] = - \hat X_0^{TL}, \nonumber \\ && \null
[\hat X_0^{TL},\hat Y_0^{TL}]  = - \hat 
X_1^{TL} + \omega \hat Z_{-1}^{TL}, \nonumber \\ 
&&\null [\hat 
X_1^{TL},\hat Z_0^{TL}] = - 2 \hat X_0^{TL},\quad [ \hat
X_1^{TL},\hat  Z_1^{TL}  ] = - 2 \hat X_{1}^{TL} - 
4 \hat Z_{-1}^{TL}, \nonumber \\
&&\null [\hat X_1^{TL},\hat Y_0^{TL}  ] = - 2 \hat X_2^{TL}
+ 4 
\hat X_0^{TL} + \sigma \hat Z_{-1}^{TL}, \nonumber \\
&&\null [ \hat Y_0^{TL},\hat Z_{-1}^{TL} ] = \beta \hat
Z_{-1}^{TL}, 
\quad
 [ \hat Y_0^{TL},\hat Z_0^{TL} ] =-2 \hat Z_1^{TL} +
\gamma \hat  Z_{-1}^{TL}, \nonumber \\ 
&&\null [\hat Y_0^{TL},\hat Z_1^{TL}  ] = - \hat Y_0^{TL} -8
\hat  Z_0^{TL} + \delta \hat Z_{-1}^{TL}, \quad [\hat
Z_0^{TL},\hat Z_1^{TL}] = \hat  Z_0^{TL}, \label{g3.30}
\end{eqnarray}
where $(\beta, \gamma, \delta, \omega, \sigma)$ are  integration constants.
The presence of these integration constants indicates that the 
symmetry algebra of the Toda equation is not completely specified. The 
constants appear whenever the symmetry $\hat Y_0^{TL}$ is involved. The 
ambiguity is related to the ambiguity in the definition of $\hat Y_0^{TL}$ 
itself, i.e. in the solution of eq. (\ref{g3.29}) for $w_n(t)$.
We fix these coefficients by requiring that one obtains the correct  continuous
limit, i.e. in the asymptotic limit, when $h$ goes to zero, a combination of the generators of the Toda Lattice (\ref{a16}) and Toda system (\ref{g1.1}) goes over to the symmetry algebra of the potential Korteweg - de Vries equation (see below).

The commutation relations obtained above determine the structure of 
the infinite dimensional Lie symmetry algebras.
The first symmetry generators are given in 
eqs.(\ref{g3.2}), (\ref{g3.11}),
(\ref{g3.23}), (\ref{g3.24}) and the corresponding commutation relations are 
given 
by
eqs.(\ref{g3.17}), (\ref{g3.21}), (\ref{g3.25}). As one can see, the symmetry 
operators 
$\hat Y_k^T$
and $\hat Z_k^T$ are linear in $t$ and the coefficient of $t$ is an isospectral 
symmetry operator $\hat X_k^T$.
Consequently, as the operators $\hat X_k^T$ commute amongst each other, the 
commutator of 
$\hat X_m^T$ with any of the $\hat Y_k^T$ or $\hat Z_k^T$ symmetries will not have 
any 
explicit time dependence and thus can be written in terms of
$\hat X_n^T$ only.  Thus the structure of the Lie algebra for the Toda
system can be written as:
\begin{equation}
L = L_0 \niplus L_1,\quad L_0=\{\hat h,\hat e,\hat f,\hat 
Y_1^T,\hat 
Y_2^T,\cdots\}, L_1=\{\hat X_0^T, \hat X_1^T,\cdots\}
\label{g3.39}
\end{equation}
where $\{ \hat h = \hat Z_1^T, \hat e = \hat Z_0^T, \hat f = \hat Y_0^T + 4 
\hat Z_0^T\}$ denotes a sl(2,{\bf R}) subalgebra with $[\hat h,
\hat e] = 
\hat e$, $[\hat h, \hat f] = - \hat f$, $[\hat e, \hat f]= 2 \hat h$. 
The algebra $L_{0}$ is perfect, i.e. we have $[L_{0},L_{0}] = L_{0}$.
It is worthwhile to notice that $\hat Z_0^T, \hat Z_1^T$ and $\hat X_0^T$ are point 
symmetries while all the others are generalized symmetries. 

For the Toda lattice equation the point transformations are $\hat X_{0}^{TL}, 
\hat Z_{0}^{TL}$ and 
$\hat Z_{1}^{TL}$, as for the Toda system, plus the additional $\hat 
Z_{-1}^{TL}$. Taking into account
eqs. (\ref{g3.26}--\ref{g3.30}), the  structure of the Lie
algebra is  the same as that of the Toda system with 
$L_{0}=\{\hat Z_{-1}^{TL}, \hat Z_{0}^{TL},\hat Z_{1}^{TL},\hat Y_{0}^{TL},\hat Y_1^{TL},\hat 
Y_2^{TL},\cdots\}$, ~$L_1=\{ \hat X_0^{TL}, \hat X_1^{TL},\hat 
X_{2}^{TL}, 
\cdots\}$. 

\subsubsection{Contraction of the symmetry algebras in the continuous limit} \label{B2}

It is well known \cite {1,5,6,BL,BRL1}, that the Toda 
equation has the potential Korteweg--de Vries equation as one of its possible  
continuous limits. In fact, by setting 
\begin{eqnarray}
v_{n}(t) = - {\frac {1}{2}} h u(x,\tau) \label{g4.1} \quad
x = (n - t) h \quad
\tau = - {\frac {1}{24}} h^{3} t 
\end{eqnarray}
we can write eq. (\ref{a16}) as
\begin{equation}
(u_{\tau} - u_{xxx} -3 u_{x}^2)_x = {\cal O}(h^{2}) \label 
{g4.4}
\end{equation}
i.e. the once differentiated potential Korteweg--de Vries
equation. Let us now rewrite the symmetry generators in the new
coordinate  system defined by (\ref{g4.1}) and
develop them for small 
$h$ in Taylor series. We have:
\begin{eqnarray}
{\hat X}_{0}^{TL} & = & \{ - u_{x}(x,\tau) h - {\frac {1}{24}} 
u_{\tau}(x,\tau) h^{3} \} {\partial_{u}} \label {g4.5} \\
\hat{X}_{1}^{TL} & = &  \{ - 2 u_{x}(x,\tau) h - {\frac {1}{3}} 
u_{\tau}(x,\tau) h^{3}  + {\cal O}(h^{5}) \} \partial_{u}
\label {g4.6} 
\\ {\hat X}_{2}^{TL} & = &  \{ - 4 u_{x}(x,\tau) h - {\frac 
{7}{6}} u_{\tau}(x,\tau) h^{3}  + {\cal O}(h^5) \} \partial_{u} 
\label {g4.7} \\ 
{\hat Y}_{0}^{TL} & = &  \{  2 [ u(x,\tau) + x 
u_{x}(x,\tau)  + 3 \tau u_{\tau}(x,\tau) ] + {\cal O}(h) \} 
{\partial_{u}} \label {g4.8} \\ 
{\hat Z}_{-1}^{TL} & = & - {\frac {2}{h}} 
{\partial_{u}}  \label {g4.9}, \qquad
{\hat Z}_{0}^{TL}  =  {\frac 
{48}{h^{4}}} \tau {\partial_{u}}   \\
{\hat Z}_{1}^{TL} & = & \{- {\frac {96}{h^{4}}}\tau + {\frac {4}{h^{2}}} 
[x + 6 \tau u_{x}(x,\tau)] + {\cal O}(1)  \}{\partial_{u}}. 
 \label {g4.11} 
\end{eqnarray}
To obtain eqs. (\ref{g4.6}--\ref{g4.8}) we have used the following 
evolution for $u$:
\begin{equation} 
u_{\tau} = u_{xxx} + 3 u_{x}^2 \label{pkdv}.
\end{equation}
The point symmetry generators written in the evolutionary
form, for  the potential Korteweg--de Vries equation
(\ref{pkdv}) read:
\begin{eqnarray}
{\hat P}_{0} & = &  u_{\tau} {\partial_{u}}, \label{g4.12} \quad
{\hat P}_{1}  =  u_{x} {\partial_{u}}, \quad
{\hat B}  =  [x +6 \tau u_{x}] {\partial_{u}}, \\ 
{\hat D} & = & [ u + x 
u_{x}  + 3 \tau u_{\tau} ]{\partial_{u}}, \label{g4.16} \quad
{\hat \Gamma}  =   {\partial_{u}},
\end{eqnarray}
and their commutation table is:
\begin{equation}
\begin{tabular}{c|ccccc} 
 & $\hat P_{0}$ &$\hat P_{1}$ &$\hat B$   & $\hat D$  & $\hat
\Gamma$  \\
\hline
$\vphantom{\hat{\tilde{T}}}\hat P_{0}$  &$0$          &
$0$       & $-6 \hat P_{1}$ & $-3 \hat P_{0}$  &$0$\\
$\hat P_{1}$  &            &   $0$       &   $- {\hat \Gamma}$   & $- \hat 
P_{1}$ 
&$0$ \\
$\hat B$      &            &           &   $0$           & $2 \hat B$ &$0$ \\
$\hat D$      &            &           &               &   $0$ & $- {\hat 
\Gamma}$ 
\\ 
$\hat \Gamma$ & & & & &  $0$ \\
\end{tabular} \label{g4.17}
\end{equation}

Let us now consider the continuous limit $h \rightarrow 0$ of the symmetry algebra of the Toda equation. We can write the simplest symmetry generators of the Toda
equation  as a linear combination of the generators 
(\ref{g4.5}--\ref{g4.11}), so that in the 
continuous limit they go over to the generators of the point symmetries of the 
potential Korteweg--de 
Vries  equation (\ref{g4.12}, \ref{g4.16}):
\begin{eqnarray}
{\tilde P}_{0} &  =&  {\frac{4}{h^{3}}} ( 2\hat X_{0}^{TL}-\hat X_{1}^{TL}),  
\label{g4.18} \quad
{\tilde P}_{1}  =  -{\frac{1}{h}} \hat X_{0}^{TL}, \quad {\tilde D}  =  {\frac{1}{2}}  \hat Y_{0}^{TL}, \\ 
{\tilde B} & = & {\frac{h^2}{4}} ( 2 \hat Z_{0}^{TL}+\hat Z_{1}^{TL}),  
\quad 
{\tilde \Gamma}  =   -{\frac{h}{2}} \hat Z_{-1}^{TL}. \label{g4.22}
\end{eqnarray}
Taking into account the commutation table between the generators 
$\hat X_{0}^{TL}$, $\hat X_{1}^{TL}$, $\hat Z_{-1}^{TL}$, $\hat Z_{0}^{TL}$, 
$\hat Z_{1}^{TL}$ and $\hat Y_{0}^{TL}$, given by (\ref{g3.30})
and the continuous limit of $\hat X_{2}^{TL}$  given by
eq. (\ref{g4.7}),  we get:
\begin{equation}
\begin{tabular}{c|ccccc} 
 & $\tilde P_{0}$ &$\tilde P_{1}$ &$\tilde B$   & $\tilde D$  & $\tilde 
\Gamma$  \\
\hline
 $\vphantom{\hat{\tilde{T}}}\tilde P_{0}$  &$0$          &
$0$       & $-6 \tilde P_{1} + {\mathcal   O}(h^2)$ & $-3
\tilde P_{0} +{\mathcal  O}(h^2)$ &$0$\\
$\tilde P_{1}$  &            &   $0$       &   $-\tilde \Gamma + {\mathcal  
O}(h^2)$   & $- \tilde P_{1} + {\mathcal  O}(h^2)$ &$0$ \\
$\tilde B$      &            &           &   $0$           & $2 \tilde B + 
{\mathcal  O}(h^2)$ &$0$ \\
$\tilde D$      &            &           &               &   $0$ & $-\tilde 
\Gamma$ \\ 
$\tilde \Gamma$ & & & & &  $0$ \\
\end{tabular} \label{g4.23}
\end{equation}
The results contained in table (\ref{g4.23}) are obtained by setting 
 $\beta=-2$, $2\gamma+\delta=0$, and
$\omega=\sigma=0$. 
\
Thus, we  have reobtained
in the continuous limit, all point symmetries of the potential 
KdV equation. The limit partially fixes the  previously
undetermined constants in eq. (\ref{g3.30}). 
To get all point
symmetries of the  potential KdV equation we needed not only the
point symmetries $\hat X_0^{TL}$, 
$\hat Z_0^{TL}$, $\hat Z_{-1}^{TL}$ and $\hat Z_1^{TL}$ of the
Toda equation,  but also the higher symmetries $\hat X_1^{TL}$,
$\hat Y_0^{TL}$.

This procedure can be viewed as a new application of the concept of Lie algebra contractions. Lie algebra contractions were first introduced by In\"on\"u and Wigner \cite{12} in order to relate the group theoretical foundations of relativistic and nonrelativistic physics. The speed of light $c$ was introduced as a parameter into the commutation relations of the Lorentz group. For $c \rightarrow \infty$ the Lorentz group "contracted" to the Galilei group. Lie algebra contractions thus relate different Lie algebras of the same dimension, but of different isomorphism classes. A systematic study of contractions, relating large families of nonisomorphic Lie algebras of the same dimension, based on Lie algebra grading, was initiated by Moody and Patera \cite{13}.

In general Lie algebra and Lie group contractions are extremely useful when describing the mathematical relation between different theories. The contraction parameter can be the Planck constant, when relating quantum systems to classical ones. It can be the curvature $k$ of a space of constant curvature, which for $k \rightarrow 0$ goes to a flat space. The contraction will then relate special functions defined e.g. on spheres, to those defined in a Euclidean space \cite{15}.

In our case the contraction parameter is the lattice spacing $h$. Some novel features appear. First of all, we are contracting an infinite dimensional Lie algebra of generalized symmetries, that of the Toda lattice. The contraction leads to an infinite dimensional Lie algebra, not isomorphic to the first one. This "target algebra" is the Lie algebra of point and generalized symmetries of the potential KdV equation. A particularly interesting feature is that the five dimensional Lie algebra of point symmetries of the potential KdV is obtained from a subset of point and generalized symmetries of the Toda equation. This 5 dimensional subset is not an algebra (it is not closed under commutations). It does contract into a Lie algebra in the continuous limit. 

\subsubsection{B\"acklund transformations for the Toda equation} \label{B1}

In addition to symmetry transformations presented in Section (\ref{B3}), the Toda system admits B\"acklund
transformations \cite{dl2,dl1,lr,ref35,BLR,br,cdI,2}.  
They are discrete transformations (i.e.~mappings)
that starting from a  solution,  produce a new
solution.  B\"acklund transformations commute amongst each other, 
allowing the definition of a soliton superposition
formula that endows the evolution equation with an integrability
feature.  Using the spectral transform~\bref{cdI} we can
 write down families of B\"acklund transformations. 
 They are obtained by requiring the existence of two essentially different solutions to the Lax equations (\ref{l1}, \ref{l2}), $\psi(x,t;\lambda)$ and $\tilde \psi(x,t;\lambda)$. These two solutions will be associated to two different solutions to the nonlinear partial differential equation (\ref{l1a}), $u(x,t)$ and $\tilde u(x,t)$ and consequently two different Lax pairs ($L(u)$, $M(u)$) and ($ L(\tilde u)$, $ M(\tilde u)$). If a transformation exists between $u$ and $\tilde u$ than there must exist a transformation between $\psi(x,t;\lambda)$ and $\tilde \psi(x,t;\lambda)$ and between 
 ($L(u)$, $M(u)$) and ($ L(\tilde u)$, $ M(\tilde u)$). This implies that there will exist an operator $D(u,\tilde u)$, often called the {\sl Darboux operator} which will relate $\tilde \psi(x,t;\lambda)$ and $\psi(x,t;\lambda)$, i.e.
 \be \label{d1}
 \tilde \psi(x,t;\lambda) = D(u,\tilde u) \psi(x,t;\lambda).
 \ee
 Taking into account the Lax equations for $\psi(x,t;\lambda)$ and $\tilde \psi(x,t;\lambda)$, we get from (\ref{d1}) the following operator equations for $D$:
 \bea \label{d2}
& & \tilde L(\tilde u) D(u,\tilde u)  = D(u,\tilde u)  L(u), \\ \label{d3}
& & D_t(u,\tilde u) = D(u,\tilde u) M(u) - M(\tilde u) D(u,\tilde u).
 \eea
 From eqs. (\ref{d2}, \ref{d3}) we get a class of B\"acklund transformations, which we will symbolically write as $B_j(u(x,t), \tilde u(x,t), \ldots)$ characterized by a recursion operator $\Lambda$.
 
   In the discrete case, the
technique is basically the same, and in the case of the matrix
discrete Schr\"odinger spectral problem, a generalization of the
scalar problem~\fref{g2.2}, the appropriate developments are found
in~\bref{br}. 
 Specializing to the scalar case, the class of B\"acklund transformations associated to the Toda system (\ref{g1.1}) is given by
\begin{equation}\label{eqBack}
    \gamma(\Lambda)\begin{pmatrix}\tilde{a}(n)-a(n)\\\tilde{b}(n)-b(n)
       \end{pmatrix}=
    \delta(\Lambda)\begin{pmatrix}\tilde{\Pi}(n)\Pi^{-1}(n{+}1)(\tilde{b}(n)-b(n{+}1))\\
          \tilde{\Pi}(n{-}1)\Pi^{-1}(n)-\tilde{\Pi}(n)\Pi^{-1}(n{+}1),
       \end{pmatrix}
\end{equation}
where $\gamma(z)$ and $\delta(z)$ are entire functions of their 
argument and we 
have denoted
\begin{gather}
\Pi(n)=\prod_{j=n}^\infty a(j),\quad \tilde{\Pi}(n)=\prod_{j=n}^\infty 
\tilde{a}(j).
\end{gather}
Above, $\Lambda$ is the recursion operator
\begin{gather}
    \Lambda\left[\hspace{-2mm}\begin{array}{c}
    p(n)\\q(n)\end{array}\hspace{-2mm}\right]=
    \left[\hspace{-1mm}\begin{array}{l}
    \begin{aligned}
        p(n)b(n{+}1)+\tilde{a}(n)[q(n)+q(n{+}1)] + 
    \Sigma(n)&[\tilde{b}(n)-b(n{+}1)]  \\
        +&[a(n)-\tilde{a}(n)]{\ds\sum_{j=n}^\infty} p(j)
    \end{aligned}\\
        p(n)+\tilde{b}(n)q(n)-\Sigma(n)+\Sigma(n{-}1)+ 
        [b(n)-\tilde{b}(n)]{\ds\sum_{j=n}^\infty}q(j)
    \end{array}\hspace{-1mm}\right]
\end{gather}
and
\begin{gather}
\Sigma(n)=\tilde{\Pi}(n)\left[\sum_{j=n}^\infty\tilde{\Pi}(j)^{-1}p(j)\Pi(j{+}1)\right]
\Pi^{-1}(n{+}1).
\end{gather}
In~\bref{BRL1} it is proven that  whenever 
$(a_n,b_n)$ and $(\tilde{a}_n,\tilde{b}_n)$ satisfy the asymptotic 
conditions~\fref{g2.4} and the 
B\"acklund transformations~\fref{eqBack}, the reflection 
coefficient satisfies  the  equation
\begin{equation}\label{Backlund1}
    \tilde{R}(\lambda)=\frac{\gamma(\lambda)-\delta(\lambda)z}
    {\gamma(\lambda)-\delta(\lambda)\big/z}\,R(\lambda).
\end{equation}

For the Toda lattice~\fref{a16} the one-soliton 
B\"acklund transformation, when the functions~$\gamma(\lambda)$ 
and $\delta(\lambda)$  are constant, reads:
\begin{equation}\label{oneSolT}
    \dot{u}(n)-\dot{u}(n{+}1)=\beta \{{\rm e}^{u(n{+}1)-\tilde{u}(n+1)}-{\rm e}^{\tilde{u}(n-1)-u(n)}
    -{\rm e}^{u(n)-\tilde{u}(n)}+{\rm e}^{\tilde{u}(n)-u(n+1)} \}
\end{equation}
Equation~\fref{oneSolT}  can be
interpreted as a (nonlinear) three point difference equation
for~$\tilde{u}$, where~$u$ is some chosen solution of~\fref{a16}. 

The 
formulas~(\ref{eqBack}--\ref{Backlund1}) provide also much more general
transformations, i.e.~higher order B\"acklund transformations. 
If~the
arbitrary functions~$\gamma(\lambda)$ and~$\delta(\lambda)$ are
polynomials, then we have a finite order B\"acklund transformation
that can be interpreted as a composition of a finite number of
one-soliton transformations.  In more general cases, when  
$\gamma(\lambda)$ and~$\delta(\lambda)$ are entire functions, we face an
infinite-order B\"acklund transformations.

In the following section we discuss how B\"acklund
transformations are related to continuous symmetry transformations,
allowing, albeit formally, an integration of the latter.

\subsubsection{Relation between B\"acklund transformations and higher 
symmetries}

A general isospectral higher symmetry of the Toda equation is given by
\begin{equation}\label{gen1}
\begin{pmatrix}a_{n,\epsilon}\\ b_{n,\epsilon}\end{pmatrix}=\phi({\cal L})
\begin{pmatrix}a_n(b_n-b_{n+1})\\a_{n-1}-a_n
\end{pmatrix},
\end{equation}
with the spectum evolution
\begin{equation}
\frac{d R(\lambda,\epsilon)}{d \epsilon}=\mu \phi(\lambda) R(\lambda,\epsilon).
\label{symSpec} 
\end{equation}
These equations generalize eq. (\ref{g2.12}), used above.
In eqs. (\ref{gen1}, \ref{symSpec}) the function $\phi$ is an entire function of its argument.
Eq.
~\fref{symSpec} can be formally
integrated in the spectral parameter ($\lambda$) space, giving:
\begin{equation}\label{intSym}
    R(\lambda,\epsilon)= \exp(\mu\phi(\lambda))R(\lambda,0).
\end{equation}
Taking into account the following definitions of $\lambda$ and $\mu$ in terms of $z$ 
\begin{gather}\lambda=\frac1{z}+z,\quad\mu=\frac1{z}-z,\quad 
    \mu^2=\lambda^2-4\label{eqmu},\\
    z=\frac{\lambda-\mu}2,\quad\frac1{z}=\frac{\lambda+\mu}2,\label{eqz}
\end{gather}
we can rewrite the general B\"acklund transformation~\fref{Backlund1} 
as
$$\tilde{R}(\lambda)=\frac{2-(\lambda-\mu)\beta(\lambda)} 
{2-(\lambda+\mu)\beta(\lambda)}
\,R(\lambda),\quad
\beta(\lambda)=\frac{\delta(\lambda)}{\gamma(\lambda)}$$
In order to identify a general symmetry transformation with a 
B\"acklund transformation, and vice versa, we 
equate~$R(\lambda,\epsilon)=\tilde{R}(\lambda)$ 
\begin{equation}\label{eqexp}
    \exp(\mu\phi(\lambda))=\frac{2-(\lambda-\mu)\beta(\lambda)} 
{2-(\lambda+\mu)\beta(\lambda)}
\end{equation}
and find that~$\phi(\lambda)$ in eq.~\fref{intSym} is given by
\begin{equation}\label{eqPhi}
    \phi(\lambda)=\frac1\mu\ln\frac{2-(\lambda-\mu)\beta(\lambda)} 
{2-(\lambda+\mu)\beta(\lambda)}.
\end{equation}
The rhs of eq. (\ref{eqPhi}) must not depend on~$\mu$.
Relations~\fref{eqmu} allow us to separate 
the exponential in~\fref{eqexp} into two entire components~$E_0(\lambda)$ 
and~$E_1(\lambda)$ satisfying
\begin{equation}\label{defEs}
    \exp(\mu\phi(\lambda))=
    \cosh(\mu\phi(\lambda))+
    \mu\frac{\sinh(\mu\phi(\lambda))}{\mu}=E_0(\lambda)+\mu E_1(\lambda).
\end{equation}
 Noticing that the rhs of eq. (\ref{eqPhi}) is rational in~$\mu$, developing~$\mu^2$ 
and identifying powers (0th and 1st) of~$\mu$, we get a system of two compatible equations
\begin{align}
  -(2-\lambda\beta)E_0+(\lambda^2-4)\beta E_1&=-(2-\lambda\beta),\label{sysE}\\
  -\beta E_0+(2-\lambda\beta)E_1&=\beta.\label{eqDelta}
\end{align}
 Eqs. (\ref{sysE}, \ref{eqDelta}) provide us with explicit 
formulas relating a given general higher symmetry (characterized by~$\phi$, and thus $E_0$, 
$E_1$) with a general B\"acklund transformation 
(characterized by $\gamma$ and $\delta$, and thus by~$\beta$):
\begin{equation}
    \beta(\lambda) = \frac{\delta(\lambda)}{\gamma(\lambda)}=
    \frac{2E_1}{E_0+\lambda E_1+1}=\frac{2\sinh(\mu\phi)\big/\mu}
    {\cosh(\mu\phi)+\lambda\sinh(\mu\phi)\big/\mu+1}.\label{rel1}
\end{equation}
From this equation we see that whatsoever be the symmetry, we find a 
B\"acklund transformation, i.e.~for an arbitrary function~$\phi$ we 
obtain the two entire functions~$\gamma$ and~$\delta$.
Vice versa, given a general B\"acklund transformation, we can find 
a corresponding generalized symmetry
\begin{equation}
   E_{0} = - \frac{2 (\beta^{2}-1) +\lambda \beta ( 2 - \lambda \beta)}{2 (\beta^{2} - \lambda \beta + 1)}, \qquad E_1=
    -\frac{(\lambda\beta-2)\beta}{2(\beta^2-\lambda\beta+1)}\label{rel3},
\end{equation}
or more explicitly,
\begin{equation}
    \phi(\lambda)=\frac1\mu\sinh^{-1}\left[\ds-\mu\frac{(\lambda\beta-2)\beta}
{2(\beta^2-\lambda\beta+1)}\right].
\end{equation}

In the case of a one-soliton B\"acklund transformation 
with $\beta=1$, we have:
$$E_0=-\frac{\lambda}{2},\quad E_1= \frac12.$$
 and we can write~$\phi(\lambda)$ as
\begin{equation}\phi(\lambda) = 
\frac{\sinh^{-1}({\sqrt{\lambda^2-4}}\big/2)}{\sqrt{\lambda^2-4}}.
\end{equation}
In this simple case  we
can write the symmetry in closed form as an infinite sequence of elementary
symmetry transformations:
\begin{equation}\label{symgenfun}
    \phi(\lambda)=\sum_{k=0}^\infty\left[
    \frac{(2k)!\pi}{k!(k{-}1)!2^{4k+2}}\lambda^{2k} + 
    \frac12\frac{k!(k{+}1)!}{(2k{+}2)!} 
    \lambda^{2k+1}\right].
\end{equation}
In this way, the existence of a one-soliton transformation 
implies the existence of an infinite-order generalized symmetry.

Let us consider the time shift symmetry given by $\phi(\lambda)=1$. Then 
eq.~\fref{defEs} implies that $E_0=\cosh\mu$ and~$E_1=\sinh\mu\big/\mu$.  According 
to~\fref{rel1} the
corresponding higher B\"acklund transformation is
\begin{align}
    \delta(\lambda)&=2\sinh\mu\big/\mu \\
    \gamma(\lambda)&=\cosh\mu+\lambda\sinh\mu\big/\mu+1
\end{align}
This B\"acklund transformation, corresponding to the 
point symmetry studied, is of infinite order.

\section{Point Symmetries Transforming Solutions and Lattices} \label{sec.n2}

\subsection{Lie Point Symmetries of Ordinary Difference Schemes}\label{sec2}

\subsubsection{Ordinary Difference Schemes}\label{subsec2.1}
In this section we take a complementary view to the previous ones. 
We shall restrict ourselves to  continuous point symmetries only. On the 
other hand we shall consider flexible difference schemes, rather than just 
difference equations on fixed lattices. For brevity and simplicity of 
exposition, we limit ourselves to ordinary difference schemes 
(O$\Delta$S). For partial difference schemes, involving more than one 
independent variable, see Ref. \cite{ref42,ref49}.

An O$\Delta$S involves two objects, a difference
equation and a lattice. We shall specify an O$\Delta$S by a system of two
equations, both involving two continuous variables $x$ and $u(x)$, evaluated at
a discrete set of points $\{x_n\}$.

Thus, a difference scheme of order $K$ will have the form
\begin{equation}\label{2.2}
\begin{array}{c}
E_a(\{x_k\}^{n+N}_{k=n+M}, \{u_k\}^{n+N}_{k=n+M}) = 0, ~a = 1, 2\\[\jot]
K= N  - M + 1, \quad n, M, N \in {\mathbb Z}, \quad N \ge M,  \quad u_k \equiv u(x_k).
\end{array}
\end{equation}
At this stage we are not imposing any boundary conditions, so the reference
point $x_n$ can be arbitrarily shifted to the left, or to the right. The order
$K$ of the system is the number of points involved in the scheme (\ref{2.2})
and it is assumed to be finite. We also assume that if the values of $x_k$ and
$u_k$ are specified at $(N-M)$ neighbouring point, we can calculate their
values at the point to the right, or to the left of the given set, using
equations (\ref{2.2}) i.e., the Jacobians satisfy
\be \label{2.3}
\frac{\partial(E_1,E_2)}{\partial(x_{n+N}, u_{n+N})} \ne 0, \qquad \frac{\partial(E_1,E_2)}{\partial(x_{n+M}, u_{n+M})} \ne 0.
\ee

A continuous limit, when the spacings between all neighbouring points go to
zero, if it exists, will take one of the equations (\ref{2.2}) into a
differential equation of order $K' \leq K$,  the 
other into an identity (like $0 =
0$).

When taking the continuous limit it is convenient to introduce different
quantities, namely differences between neighbouring points and discrete
derivatives like
\begin{eqnarray}
h_+ (x_n) & = & x_{n+1}-x_n, \quad h_-(x_n) =x_n - x_{n-1},\nonumber\\
u_{x} & = & \frac{u_{n+1}-u_n}{x_{n+1}-x_n}, \quad u_{\underline 
x}= \frac{u_n-u_{n-1}}{x_n-x_{n-1}},\label{2.3a}\\
u_{x\underline x} & = &
2\frac{u_{,x}-u_{,\underline x}}{x_{n+1}-x_{n-1}},\dots\nonumber
\end{eqnarray}

In the continuous limit, we have
\[
h_+ \to 0, \quad h_- \to 0, \quad u_{,{x}} \to u ^{\prime},\quad
u_{{\underline x}} \to u ^{\prime}, \quad u_{x\underline x} \to
u ^{\prime\prime}.
\]

As a clarifying example of the meaning of the difference scheme (\ref{2.2}),
let us consider a three-point scheme that will approximate a second-order
linear difference equation:
\begin{eqnarray}
E_1 & = & \frac{u_{n+1}-2u_n+u_{n-1}}{(x_{n+1}-x_n)^2} - u_n = 0,\label{2.4}\\
E_2 & = & x_{n+1} - 2x_n + x_{n-1} = 0.\label{2.5} 
\end{eqnarray}
The solution of eq. $E_2=0$, determines a uniform lattice
\begin{equation}\label{2.6}
x_n = hn + x_0.
\end{equation}
The scale $h$ and the origin $x_0$ in eq.~(\ref{2.6}) are not fixed by
eq.~(\ref{2.5}), instead they appear as integration constants, {\it
  i.e.}, they are {\it a
priori} arbitrary. Once they are chosen, eq.~(\ref{2.4}) reduces to a linear
difference equation with constant coefficients, since we have $x_{n+1} - x_n =
h$. Thus, a solution of eq.~(\ref{2.4}) will have the form
\begin{equation}\label{2.7}
u_n = \lambda^{x_n}.
\end{equation}
Substituting (\ref{2.7}) into (\ref{2.4}) we obtain the general solution of the
difference scheme (\ref{2.4}), (\ref{2.5}),
\begin{eqnarray}
u(x_n) & = & c_1\lambda^{x_n}_1+c_2\lambda^{x_n}_2, \quad x_n = hn +
x_0,\label{2.8}\\
\lambda_{1, 2} & = & \left(\frac{2+h^2\pm h\sqrt{4+h^2}}{2}\right)^{1/2}.
\nonumber
\end{eqnarray}
The solution (\ref{2.8}) of system (\ref{2.4})--(\ref{2.5}) depends on 4
arbitrary constants $c_1$, $c_2$, $h$ and $x_0$.

Now let us consider a general three-point scheme of the form
\begin{equation}\label{2.9}
E_a(x_{n-1}, x_n, x_{n+1}, u_{n-1}, u_n, u_{n+1}) = 0, \quad a = 1, 2,
\end{equation}
satisfying (\ref{2.3}) with $N = 1$, $M = -1$
(possibly after an up or down shifting).
The two conditions on the Jacobians (\ref{2.3}) are sufficient to allow us to
calculate $(x_{n+1}, u_{n+1})$ if $(x_{n-1}, u_{n-1}, x_n, u_n)$ are known.
Similarly, $(x_{n-1}, u_{n-1})$ can be calculated if $(x_n, u_n, x_{n+1},
u_{n+1})$ are known. The general solution of the scheme (\ref{2.9}) will hence
depend on 4 arbitrary constants and will have the form
\begin{eqnarray}
u_n & = & f(x_n, c_1, c_2, c_3, c_4)\label{2.11}\\
x_n & = & \phi(n, c_1, c_2, c_3, c_4).\label{2.12}
\end{eqnarray}

A more standard approach to difference equations would be to consider a fixed
equally spaced lattice, {\it e.g.}, with spacing $h = 1$. We can then identify the
continuous variable $x$, sampled at discrete points $x_n$, with the discrete
variable $n$,
\begin{equation}\label{2.13}
x_n = n.
\end{equation}
Instead of a difference scheme we then have a difference equation, lke the one we considered in Section \ref{sec.n2}
\begin{equation}\label{2.14}
E(\{u_k\}^{n+N}_{k=n+M}) = 0,
\end{equation}
involving $K = N - M + 1$ points. Its general solution has the form
\begin{equation}\label{2.15}
u_n = f(n, c_1, c_2, \dots c_{N-M}) \ ,
\end{equation}
i.e., it depends on $N - M$ constants.

Below, when studying point symmetries of discrete equations we will see the
advantage of considering difference systems like system (\ref{2.2}). A treasury of information on difference equations and their solutions can 
be found in the classical book by Milne--Thompson \cite{ML}.

\subsubsection{Point Symmetries of Ordinary Difference Schemes}\label{subsec2.2}

In this section we shall follow rather closely the article \cite{ref48}. We
shall define the symmetry group of an ordinary difference scheme in the same
manner as for ODEs: it is, a group of continuous local point transformations
of the form (\ref{1.2}) taking solutions of the O$\Delta$S (\ref{2.2}) into
solutions of the same scheme. The transformations considered being continuous,
we will adopt an infinitesimal approach, as in eq.~(\ref{1.3}). We drop the
labels $i$ and $\alpha$, since we are considering the case of one independent
and one dependent variable only.

As in the case of differential equations, our basic tool will be vector fields
of the form (\ref{1.4}). In the case of O$\Delta$S they will have the form
\begin{equation}\label{2.16}
X = \xi(x, u)\partial_x + \phi(x, u)\partial_u,
\end{equation}
with
\[
x \equiv x_n, \quad u \equiv u_n = u(x_n).
\]
Because we are considering point transformations, $\xi$ and $\phi$ in
(\ref{2.16}) depend on $x$ and $u$ at one point only.

The prolongation of the vector field $X$ is different from that of the case
of ODEs. Instead of prolonging to derivatives, we prolong to all points of the
lattice figuring in scheme (\ref{2.2}). Thus we set
\begin{equation}\label{2.17}
\pr X = \sum^{n+N}_{k=n+M} \xi(x_k, u_k)\partial_{x_k} + \sum^{n+N}_{k=n+M}
\phi(x_k, u_k)\partial_{u_k}.
\end{equation}
In these terms the requirement that the transformed function $\tilde u(\tilde
x)$ should satisfy the same O$\Delta$S as the original $u(x)$ is expressed by
the requirement
\begin{equation}\label{2.18}
\pr X E_a\mid_{E_1=E_2=0} = 0, \quad a = 1, 2.
\end{equation}
Since we must respect both the difference equation and the lattice, we have two
conditions (\ref{2.18}) from which to determine $\xi(x, u)$ and $\phi(x, u)$.
Since each of these functions depends on a single point  $(x, u)$ and the
prolongation (\ref{2.17}) introduces $N - M + 1$ points in space $X \times U$,
equation (\ref{2.18}) will imply a system of determining equations for
$\xi$ and $\phi$. Moreover, in general this will be an overdetermined system of
linear functional equations that we transform into an overdetermined system of
linear differential equations \cite{ref57,ref58}.

Let us first of all check that the prolongations (\ref{2.17}) have the continuous limit. We consider the first prolongation in the discrete case
\bea \label{2.19}
pr^1 X =&& \xi(x,u) \partial_x + \phi(x,u) \partial_u +\xi(x_+, u_+) \partial_{x_+} + \phi(x_+, u_+) \partial_{u_+}, \\ \nonumber
&&x_+ \equiv x_{n+1},\qquad u_+ \equiv u_{n+1},
\eea
and apply it to a function of the variables  
$x, ~u, ~h_+, ~u_x$ (\ref{2.3a}) .
The continuous limit is recovered by taking $h \rightarrow 0$ and putting
\bea \label{2.21}
u_+ & = & u(x_+) = u(x) + h u'(x) + \ldots , \\ \nonumber
\xi(x_+, u_+) & = & \xi(x,u) + \xi_{,x} h + \xi_{,u} u' h + \ldots, \\ \nonumber
\phi(x_+, u_+) & = & \phi(x,u) + \phi_{,x} h + \phi_{,u} u' h + \ldots .
\eea
where $u'$ is the (continuous) derivative of $u(x)$. We have
\bea \nonumber
pr^1_D X F(x, u, h, u_x) = \{ \xi \partial_x + \phi \partial_u + [ \xi_{,x} + \xi_{,u} u' ] h \partial_h + \\ \nonumber
+ [ - u' ( \xi_{,x} + \xi_{,u} u') + \phi_{,x} + \phi_{,u} u' ] \partial_{u_x} \} F .
\eea
In the limit when $h \rightarrow 0$ we obtain 
\bea \label{2.22}
\lim_{h \rightarrow 0} pr^1_D X = \xi(x,u) \partial_x + \phi(x,u) \partial_u + 
\phi^x(x,u,u') \partial_{u'}, \\ \label{2.23}
\phi^x = \phi_{,x} + (\phi_{,u} - \xi_{,x} ) u' - \xi_{,u} u'^2,
\eea
which is the correct expression for the first prolongation in the continuous case \cite{ref1}.
The proof that the $n$-th prolongation (\ref{2.17}) has the correct continuous limit can be performed by induction.

\subsection{Examples}\label{subsec2.3}

1. 
\textit{Power nonlinearity on a uniform lattice.}

Let us  consider a difference scheme that is a discretization of the ODE
\begin{equation}\label{2.45}
u ^{\prime\prime} - u^N = 0, \quad N \neq 0, 1.
\end{equation}
For $N \neq - 3$, eq.~(\ref{2.45}) is invariant under a two-dimensional Lie
group whose Lie algebra is given by
\begin{equation}\label{2.46}
X_1 = \partial_x, \quad X_2 = (N-1)x\partial_x - 2u\partial_u
\end{equation}
(translations and dilations).
For $N = -3$ the symmetry algebra is three-dimensional, isomorphic to $\Sl(2,
\mathbb R)$, i.e., it contains a third element in addition to (\ref{2.46}). A
convenient basis for the symmetry algebra of the equation
\begin{equation}\label{2.47}
u ^{\prime\prime} - u^{-3} = 0
\end{equation}
is
\begin{equation}\label{2.48}
X_1 = \partial_x, \quad X_2 = 2x\partial_x + u\partial_u, \quad X_3 = x(x
\partial_x+u\partial_u).
\end{equation}

A very natural O$\Delta$S that has (\ref{2.45}) as its continuous limit is
\begin{eqnarray}
E_1 & = & \frac{u_{n+1}-2u_n+u_{n-1}}{(x_{n+1}-x_n)^2} - u^N_n =  0 \quad N
\neq 0, 1\label{2.49a}\\
E_2 & = & x_{n+1} - 2x_n +  x_{n-1} = 0.\label{2.49b}
\end{eqnarray}

Let us now apply the symmetry algorithm described in Section \ref{subsec2.2} to
system (\ref{2.49a})--(\ref{2.49b}). To illustrate the method, we shall
present all calculations in detail.

First, we choose two variables that will be substituted in
eq.~(\ref{2.18}), once the prolonged vector field (\ref{2.17}) is applied to
system (\ref{2.49a})--(\ref{2.49b}), namely
\begin{eqnarray}
x_{n+1} & = & 2x_n - x_{n-1}\label{2.50}\\
u_{n+1} & = & (x_n-x_{n-1})^2u^N_n + 2u_n - u_{n-1}\nonumber
\end{eqnarray}
We apply $\pr X$ of (\ref{2.17}) to eq.~(\ref{2.49b}) and obtain
\begin{equation}\label{2.51}
\xi(x_{n+1}, u_{n+1}) - 2\xi(x_n, u_n) + \xi(x_{n-1}, u_{n-1}) = 0,
\end{equation}
where, $x_n$, $u_n$ $x_{n-1}$, $u_{n-1}$ are independent, but $x_{n+1}$,
$u_{n+1}$ are expressed in terms of these quantities, as in eq.~(\ref{2.50}).
Taking this into acccount, we differentiate (\ref{2.51}) first with respect to
$u_{n-1}$, then with respect to $u_n$. We obtain successively
\begin{eqnarray}
- \xi_{,u_{n+1}}(x_{n+1}, u_{n+1}) + \xi_{,u_{n-1}}(x_{n-1}, u_{n-1}) & = &0
\label{2.52}\\
(N(x_n-x_{n-1})^2 u^{N-1}_n +2) \xi_{,u_{n+1}u_{n+1}} (x_{n+1}, u_{n+1}) 
& = & 0.\label{2.53}
\end{eqnarray}
Eq.~(\ref{2.53}) is the desired one-term equation. It implies that 
\begin{equation}\label{2.54}
\xi(x, u) = a(x)u + b(x).
\end{equation}
Substituting (\ref{2.54}) into (\ref{2.52}) we obtain
\begin{equation}\label{2.55}
- a(2x_n-x_{n-1}) + a(x_{n-1}) = 0.
\end{equation}
Differentiating with respect to $x_n$, we obtain $a = a_0 =\mbox{const}$. Finally,
we substitute (\ref{2.54}) with $a = a_0$ into (\ref{2.51}) and obtain
\begin{equation}\label{2.56}
a = a_0 = 0, \quad b(2x_n-x_{n-1}) - 2b(x_n) + b(x_{n-1}) = 0,
\end{equation}
and hence
\begin{equation}\label{2.57}
\xi = b = b_1x + b_0,
\end{equation}
where $b_0$ and $b_1$ are constants. To obtain the function $\phi(x_n, u_n)$,
we apply $\pr  X$ to eq.~(\ref{2.49a}) and obtain
\begin{eqnarray}
\lefteqn{\phi(x_{n+1}, u_{n+1}) - 2\phi(x_n, u_n) + \phi(x_{n-1}, u_{n-1})}
\nonumber\\
&&\hspace{1in}-(x_n-x_{n-1})^2 (N\phi(x_n, u_n)u^{N-1}_n + 2b_1u^N_n) =
0.\label{2.58}
\end{eqnarray}
Differentiating successively with respect to $u_{n-1}$ and $u_n$ (taking
(\ref{2.50}) into account), we obtain
\begin{eqnarray}
- \phi_{,u_{n+1}}(x_{n+1}, u_{n+1}) + \phi_{,u_{n-1}}(x_{n-1}, u_{n-1}) & = &
0\label{2.59}\\
(N(x_n-x_{n-1})^2u^N_n +2) \phi_{,u_{n+1}u_{n+1}} & = & 0,\label{2.60}
\end{eqnarray}
and hence
\begin{equation}\label{2.61}
\phi = \phi_1u + \phi_0(x), \quad \phi_1 = const.
\end{equation}
Eq.~(\ref{2.58}) now reduces to
\begin{eqnarray}
\lefteqn{\phi_0(2x_n-x_{n-1}) - 2\phi_0(x_n) + \phi_0(x_{n-1})}\nonumber\\
&&\hspace{0.50in} -(x_n-x_{n-1})^2 ((N-1)\phi_1 + 2b_1)u^N_n\nonumber\\
&&\hspace{0.50in} - N(x_n-x_{n-1})^2\phi_0u^{N-1}_n = 0.\label{2.62}
\end{eqnarray}
We have $N \neq 0, 1$ and hence (\ref{2.62}) implies that
\begin{equation}\label{2.63}
\phi_0 = 0, \quad (N-1)\phi_1 + 2b_1 = 0.
\end{equation}
We have thus proven that the symmetry algebra of the O$\Delta$S 
(\ref{2.49a})--(\ref{2.49b}) 
is the same as that of the ODE (\ref{2.45}), namely the
algebra (\ref{2.46}).

We observe that the value $N = -3$ is not distinguished here and that system
(\ref{2.49a})--(\ref{2.49b}) 
is not invariant under $SL(2, \mathbb R)$ for
$N = -3$. Actually, a difference scheme invariant under $SL(2, \mathbb R)$ does
exist and it will have eq.~(\ref{2.47}) as its continuous limit. It will
not however have the form (\ref{2.49a})--(\ref{2.49b}),
and the lattice will not be uniform
\cite{ref45,ref47}. The corresponding SL(2,R) invariant scheme is presented below in Section \ref{sec3.3}.

Had we taken a two-point lattice, $x_{n+1}-x_n = h$ with $h$ fixed, instead of
$E_2 = 0$ as in (\ref{2.49b}), we would only have obtained translational
invariance for the equation (\ref{2.49a}) and lost the dilational invariance
represented by $X_2$ of eq.~(\ref{2.46}).

2. \textit{An O$\Delta$S involving an 
arbitrary function on a uniform lattice}

We consider
\begin{eqnarray}
E_1 & = & \frac{u_{n+1}-2u_n+u_{n-1}}{(x_{n+1}-x_n)^2} - f\left(\frac{u_n-
u_{n-1}}{x_n-x_{n-1}}\right) = 0,\label{2.64a}\\
E_2 & = & x_{n+1} -2x_n + x_{n-1} = 0,\label{2.64b}
\end{eqnarray}
where $f(z)$ is some sufficiently smooth function satisfying
\begin{equation}\label{2.65}
f ^{\prime\prime}(z) \neq 0.
\end{equation}
The continuous limit of eq.~(\ref{2.64a}) and (\ref{2.64b}) is
\begin{equation}\label{2.66}
u ^{\prime\prime} - f(u ^\prime) = 0,
\end{equation}
and it is invariant under a two-dimensional group with Lie algebra,
\begin{equation}\label{2.67}
X_1 = \partial_x, \quad X_2 = \partial_u,
\end{equation}
for any function $f(u ^\prime)$. For certain functions $f$ the
symmetry group is three-dimensional, where the additional basis element of the
Lie algebra is
\begin{equation}\label{2.68}
X_3 = (ax+bu)\partial_x + (cx+du)\partial_u.
\end{equation}

Now let  us consider the discrete system (\ref{2.64a})--(\ref{2.64b}).
Before applying $\pr X$ to this system we choose two variables to substitute in
eq.~(\ref{2.18}), namely
\begin{eqnarray}
x_{n+1} & = & 2x_n - x_{n-1}\label{2.70}\\
u_{n+1} & = & 2u_n - u_{n-1} + (x_n-x_{n-1})^2 f\left(\frac{u_n-u_{n-1}}
{x_n-x_{n-1}}\right).\nonumber
\end{eqnarray}
Applying $\pr X$ to eq.~(\ref{2.64b}) and performing the same kind of passages necessary to solve the determining equation as we did for eq. (\ref{2.50}) we get 
\begin{equation}\label{2.72}
\xi = \alpha x + \beta,
\end{equation}
with $\alpha = \mbox{const}$, $\beta = \mbox{const}$. 
Now let us apply $\pr X$ to  eq.~(\ref{2.64a}). We get the following equation for $\phi$
\begin{eqnarray}
\lefteqn{\phi\big(x+h, u+hz+h^2f(z)\big) - 2\phi(x, u) + \phi(x-h,
u-hz)}\nonumber\\
&&\hspace{.25in}=2\alpha h^2f(z)
+ h^2f'(z)\biggl(\frac{\phi(x, u)-\phi(x-h, u-hz)}{h}-\alpha
z\biggr).\label{2.76}
\end{eqnarray}
where we have defined
\begin{equation}
z = \frac{u_n-u_{n-1}}{x_n-x_{n-1}}, \quad h = x_{n+1} - x_n.
\end{equation}
In general, eq.  (\ref{2.76}) is quite difficult to solve and for most functions 
$f(z)$ it has no further solutions.

   However, if we make the choice
\be \label{2.77}
f(z) = e^{-z}
\ee
we find that eq. (\ref{2.76}) is solved by putting
\be \label{2.78}
\alpha = 1, \quad \phi = x + u.
\ee
Finally, we find that the system
\be \label{2.79}
\frac{u_{n+1} - 2 u_n + u_{n-1}}{(x_{n+1} -x_n)^2} = e^{\frac{u_n - u_{n-1}}{x_n - x_{n-1}}}, \quad x_{n+1} - 2 x_n + x_{n-1} = 0
\ee
is left invariant by a three-dimensional transformation group, generated by the solvable Lie algebra with basis
\be \label{2.80}
X_1 = \partial_x; \qquad X_2 = \partial_u; \qquad X_3 = x \partial_x + ( x + u ) \partial_u.
\ee 
For further examples see the original article \cite{ref48} and the lectures \cite{158}.

\subsection{Symmetry Preserving Discretization of Ordinary Differential Equations} \label{sec3}

\subsubsection{General Comments} \label{sec3.1}

In Section \ref{sec2} we assumed that an O$\Delta$S (\ref{2.2}) is given and we showed how to determine its symmetries.

Here we will discuss a different problem, namely the construction of O$\Delta$S with a priori given symmetry groups. More specifically, we start from a given ODE
\be \label{3.1}
E(x,y,{\dot y}, {\ddot y}, \ldots ) = 0
\ee
and its symmetry algebra, realized by vector fields of the form (\ref{2.16}).  We now wish to construct an O$\Delta$S (\ref{2.2}), approximating the ODE (\ref{3.1}) and having the same symmetry algebra (and the same symmetry group).

This can be done systematically, once the order of the ODE (\ref{3.1}) is fixed. In general, the motivation for such a study is multifold. In physical applications the symmetry may actually be more important than the equation itself. A discrete scheme with the correct symmetries has a good chance of describing the physics correctly. This is specially true if the underlying phenomena really are discrete and the differential equations come from a continuum approximation. Furthermore, the existence of point symmetries for differential and difference equations makes it possible to obtain explicit analytical solutions.  Finally, a discretization respecting point symmetries should provide improved numerical methods.

Let us first outline the general method of discretization. If the ODE (\ref{3.1}) is of order $N$ we need a O$\Delta$S involving at least $N+1$ points, i.e. $N+1$ pairs 
\be \label{3.2}
\{ x_i, y_i; i = 1, \ldots, N+1 \}.
\ee

The procedure is as follows
\begin{enumerate}
\item  Take the Lie algebra \textgoth{g} of the symmetry group \textgoth{G} of the ODE (\ref{3.1}) and prolong the (known) vector fields $\{ X_1, \ldots, X_n \}$ to all $N+1$ points (\ref{3.2}), as in eq. (\ref{2.17}).
\item Find a basis for all invariants of the (prolonged) Lie algebra \textgoth{g} in the space (\ref{3.2}) of independent and dependent variables. Such a basis will consist of $K$ functionally independent invariants
\be \label{3.3}
I_a = I_a ( x_1, \ldots, x_{N+1}, y_1, \ldots, y_{N+1} ), \quad 1 \le a \le K .
\ee
They are determined by solving the differential equations 
\be \label{3.4}
pr X_i I_a ( x_1, \ldots, x_{N+1}, y_1, \ldots, y_{N+1} ) = 0, \quad  i = 1, \ldots, n.
\ee
The actual number $K$ satisfies
\be \label{3.5}
K = 2 N + 2 - ( \mbox{dim} \textgoth{g} - \mbox{dim} \textgoth{g}_0 )
\ee
where $\textgoth{g}_0$ is the Lie algebra of the group $\textgoth{G}_0 \subset \textgoth{G}$, stabilizing the $N+1$ points (\ref{3.2}).

We need at least two independent invariants of the form (\ref{3.3}) to write an invariant difference scheme.
\item If the number of invariants is not sufficient, we can make use of invariant manifolds. To find them, we first write out the matrix of coefficients of the prolonged vector fields $\{ X_1, \ldots, X_n \}$ :
\be \label{3.6}
M =  \begin{pmatrix} \xi_{11} & \xi_{12} & \ldots & \xi_{1N+1} & \phi_{11} & \phi_{12} & \ldots & \phi_{1N+1} \\
\vdots & \vdots &  & \vdots & \vdots & \vdots & & \vdots & \\
 \xi_{n1} & \xi_{n2} & \ldots & \xi_{nN+1} & \phi_{n1} & \phi_{n2} & \ldots & \phi_{nN+1} \end{pmatrix}
 \ee
and determine the manifolds on which the rank of $M$, $\mbox{Rank}(M)$, satisfies
\be \label{3.7}
\mbox{Rank}(M) < \mbox{min}(n, 2N+2),
\ee
i.e.  is less than maximal. The invariant manifolds are then obtained by requiring that eq. (\ref{3.4}) be satisfied on the manifold satisfying eq. (\ref{3.7}). 
\end{enumerate}

\subsubsection{Symmetries of Second Order ODEs} \label{sec3.2}

Let us now restrict to the case of a second order ODE
\be \label{3.8}
{\ddot u} = F(x, u, {\dot u}).
\ee
Sophus Lie gave a symmetry classification of second order ODE's (over the field of complex numbers $\C$ ) \cite{Lie1,Lie2}. A similar classification over $\R$ is much more recent \cite{Mal1,Mal2}.

The main classification results can be summed up as follows.
\begin{enumerate}
\item  The dimension $n = \mbox{dim}\textgoth{g}$ of the symmetry algebra of eq. (\ref{3.8}) can be $\mbox{dim}\textgoth{g} = 0,1,2,3$ or $8$.
\item If we have $\mbox{dim}\textgoth{g} = 1$ we can decrease the order of eq. (\ref{3.8}) by one. If the dimension is $\mbox{dim}\textgoth{g} \geq 2$ we can integrate by quadratures.
\item If we have $\mbox{dim}\textgoth{g} = 8$, then the symmetry algebra is $sl(3,\C)$, or $sl(3,\R)$, respectively. The equation can be transformed into ${\ddot y} = 0$ by a point transformation.
\end{enumerate}
Further symmetry results are due to E. Noether \cite{noe} and Bessel--Hagen \cite{ebh}. Every ODE (\ref{3.8}) can be interpreted as an Euler--Lagrange equation for some Lagrangian density
\be \label{3.9}
{\cal L }= {\cal L }(x, u, \dot u).
\ee
The equation is
\be \label{3.10}
\frac{\partial {\cal L}}{\partial u} - D (\frac{\partial {\cal L}}{\partial {\dot u}}) = 0, \quad 
D = \frac{\partial}{\partial x} + {\dot u} \frac{\partial}{\partial u} + {\ddot u} \frac{\partial}{\partial {\dot u}} + \ldots .
\ee
An {\it infinitesimal divergence symmetry}, or a Lagrangian symmetry is a vector field $X$ (\ref{2.16}) satisfying
\be \label{3.11}
pr X ({\cal L}) + {\cal L} D(\xi) = D(V), \quad V = V(x,u),
\ee
where $V$ is some function of $x$ and $u$. A symmetry of the Lagrangian $\cal L$ is always a symmetry of the Euler--Lagrange equation (\ref{3.10}), however equation (\ref{3.10}) may have additional, non-Lagrangian symmetries.

A relevant symmetry result is that if we have $\mbox{dim}\textgoth{g} = 1$, or $\mbox{dim}\textgoth{g}= 2$ for eq. (\ref{3.8}), then there always exists a Lagrangian having the same symmetry. For $\mbox{dim} {\cal L} = 3$, at least a two-dimensional subalgebra of the Lagrangian symmetries exists. For $\mbox{dim}\textgoth{g} = 8$ a four-dimensional solvable subalgebra of Lagrangian symmetries exists.

\subsubsection{Symmetries of the Three-point Difference Schemes} \label{sec3.3}

A symmetry classification of three-point difference schemes was performed quite recently \cite{ref45,ref47}. It is similar to Lie's classification of second order ODE's and goes over into this classification in the continuous limit. We shall now review the main results of the classification following the method outlined in Section \ref{sec3.1} above.

Sophus Lie \cite{Lie3} gave a classification of all finite dimensional Lie algebras that can be realized by vector fields of the form (\ref{2.16}). This was done over the field $\C$ and thus amounts to a classification of finite dimensional subalgebras of $\mbox{diff}(2,\C)$, the Lie algebra of the group of diffeomorphisms of the complex plane $\C^2$. A similar classification of finite dimensional subalgebras of $\mbox{diff}(2,\R)$ exists \cite{g-l}, but we restrict ourselves to the simpler complex case. 

We use the following notation for 3 neighbouring points on the lattice:
\be \label{3.12}
x_- = x_{n-1}, ~x = x_n, ~x_+ = x_{n+1}, ~u_- = u_{n-1}, ~u = u_n, ~u_+ = u_{n+1}.
\ee

Let us now proceed by dimension of the symmetry algebras.
\begin{description}
\item[$\mbox{dim} \textgoth{g} = 1$ ]  A single vector field can always be rectified into the form
\be \label{3.14}
A_{1,1} : \qquad X_1 = \frac{\partial}{\partial u}
\ee
The invariant ODE is 
\be \label{3.15}
{\ddot u} = F(x, {\dot u}).
\ee
Putting ${\dot u} = y$ we obtain a first order ODE.

The difference invariants of $X_1$ are
\be \label{3.17}
x, ~h_+ = x_+ - x, ~h_- = x - x_-,~\eta_+ = ~ u_+ - u, ~\eta_- = ~ u - u_-.
\ee
Using these invariants and the notation (\ref{2.3a}), we can write a difference scheme
\be \label{3.18}
y_{x \underline{ x}} = F( x, \frac{u_x + u_{\underline{ x}}}{2}, h_- ), \quad h_+ = h_- G( x, \frac{u_x + u_{\underline{ x}}}{2}, h_- ).
\ee
This scheme goes into eq. (\ref{3.15}) if we  require that the otherwise arbitrary functions $F$ and $G$ satisfy
\be \label{3.19}
\lim_{h_- \rightarrow 0} F( x, \frac{u_x + u_{\underline{ x}}}{2}, h_- ) = F( x, {\dot u}), \quad \lim_{h_- \rightarrow 0} G( x, \frac{u_x + u_{\underline{ x}}}{2}, h_- )  < \infty.
\ee
\item[$\mbox{dim}\textgoth{g} = 2$ ] Precisely four equivalence classes of two-dimensional subalgebras of $\mbox{diff}(2,\C)$ exist. Let us consider them separately.
\begin{description}
\item[$A_{2,1}$ ] 
\be \label{3.20}
X_1 = \partial_x, \quad X_2 = \partial_u
\ee
The algebra $A_{2,1}$ is Abelian, the elements $X_1$ and $X_2$ are linearly nonconnected (linearly independent in any point $(x,y)$). The invariant ODE is
\be \label{3.21}
{\ddot u} = F({\dot u}),
\ee
and can be immediately integrated.

An invariant difference scheme is given by any two relations between the invariants
$h_+, ~h_-, ~\eta_+, ~\eta_-$ of eq. (\ref{3.17}), 
for instance
\be \label{3.23}
u_{x {\underline{ x}}} = F(\frac{u_x + u_{\underline{ x}}}{2}, h_-), ~ h_+ = h_- G(\frac{u_x + u_{\underline{ x}}}{2}, h_-), 
\ee
with conditions (\ref{3.19}) imposed on the functions $F$ and $G$.
\item[$A_{2,2}$ ] 
\be \label{3.24}
X_1 = \partial_u, \quad X_2 =x \partial_x + u  \partial_u
\ee
This Lie algebra in non-Abelian, the two elements are linearly nonconnected. The invariant ODE is
\be \label{3.25}
{\ddot u} = \frac{1}{x} F({\dot u}).
\ee
A basis for the difference invariance is
\be \label{3.26}
\{ x u_{x \underline{ x}},~ u_x + u_{\underline{ x}}, ~\frac{h_+}{h_-},~ \frac{h_-}{x} \}
\ee
so a possible invariant difference scheme is
\be \label{3.27}
u_{x \underline{ x}} = \frac{1}{x} F(\frac{u_x + u_{\underline{ x}}}{2}, \frac{h_-}{x}), ~ h_+ = h_- G(\frac{u_x + u_{\underline{ x}}}{2}, \frac{h_-}{x}).
\ee
\item[$A_{2,3}$ ] 
\be \label{3.28}
X_1 = \partial_u, \quad X_2 =x  \partial_u
\ee
The algebra is Abelian, the elements $X_1$ and $X_2$ are linearly connected. The invariant ODE is
\be \label{3.29}
{\ddot u} = F(x).
\ee
This equation is linear and hence has an eight dimensional symmetry algebra (of which $A_{2,3}$ is just a subalgebra).

The difference invariants are
\be \label{3.30}
\{ u_{x {\bar x}},~ x, ~h_+, ~h_- \}
\ee
so the invariant difference scheme will also be linear (at least in the dependent variable $u$).
\item[$A_{2,4}$ ] 
\be \label{3.31}
X_1 = \partial_u, \quad X_2 = u  \partial_u
\ee
The algebra is non-Abelian and isomorphic to $A_{2,2}$, but with linearly connected elements. The invariant ODE is again linear,
\be \label{3.32}
{\ddot u} = F(x) {\dot u},
\ee
as is the invariant difference scheme. Eq. (\ref{3.32}) is invariant under the group $SL(3,\C)$. Difference invariants are
\be \label{3.33}
\{ \xi =  2 \frac{ u_{x \underline{ x}}}{u_x + u_{\underline{ x}} }, \quad x, ~h_+, ~h_- \}
\ee
and a possible invariant O$\Delta$S is
\be \label{3.34}
2 \frac{ u_{x \underline{ x}}}{u_x + u_{\underline{ x}} } = F(x, h_-), \quad G(x, h_+, h_-) = 0.
\ee
\end{description}

\item[$\mbox{dim}\textgoth{g} = 3$ ]  We now turn to difference schemes invariant under three-dimensional symmetry groups. We will restrict ourselves to the case when the corresponding ODE is nonlinear. Hence we will omit all algebras that contain $A_{2,3}$ or $A_{2,4}$ subalgebras (they were considered in \cite{ref45}).
\begin{description}
\item[$A_{3,1}$ ] 
\be \label{3.33a}
X_1 = \partial_x, \quad X_2 =   \partial_u, \quad X_3 = x \partial_x + k u \partial_u, ~ k \ne 0, \frac{1}{2},1,2
\ee
The invariant ODE is 
\be \label{3.34a}
{\ddot u} = {\dot u}^{\frac{k-2}{k-1}}.
\ee
For $k=1$ there is no invariant second order equation; for $k=2$ the equation is linear, for $k=\frac{1}{2}$ it is trasformable into a linear equation and has a larger symmetry group.

Difference invariants are
\be \label{3.35}
I_1 = \frac{h_+}{h_-}, \quad I_2 = u_x h_+^{1-k}, \quad I_3 = u_{\underline{ x}} h_-^{1-k}.
\ee

A simple invariant difference scheme is
\be \label{3.36}
u_{x \underline{ x}} = ( \frac{  u_x + u_{\underline{ x}} }{2} )^[\frac{k-2}{k-1} f( \frac{  u_x + u_{\underline{ x}} }{2} h_-^{1-k} ), \quad h_+ = h_- g( \frac{  u_x + u_{\underline{ x}} }{2} h_-^{1-k} ).
\ee

We shall see below that other invariant schemes may be more convenient.
\item[$A_{3,2}$ ] 
\be \label{3.37}
X_1 = \partial_x, \quad X_2 =   \partial_u, \quad X_3 = x \partial_x + (x + u) \partial_u, 
\ee
The invariant ODE is
\be \label{3.38}
{\ddot u} = e^{-{\dot u}}.
\ee
Difference invariants in this case are
\be \label{3.39}
I_1 = \frac{h_+}{h_-}, \quad I_2 = h_+ e^{-u_x}, \quad I_3 = h_- e^{-{u_{\underline{ x}}}}.
\ee
A possible invariant scheme is 
\be \label{3.40}
u_{x \underline{ x}} = e^{-  \frac{  u_x + u_{\underline{ x}} }{2}} f(\sqrt{h_- h_+} e^{-  \frac{  u_x + u_{\underline{ x}} }{2}} ), \quad  h_+ = h_- g(\sqrt{h_- h_+} e^{-  \frac{  u_x + u_{\underline{ x}} }{2}} ).
\ee

No further solvable three-dimensional subalgebras of $\mbox{diff}(2,\C)$ exist (though there is another family for $\mbox{diff}(2,\R)$ \cite{ref45}).
\end{description}

Two inequivalent realizations of $sl(2,\C)$ exist. Let us consider them separately.
\begin{description}
\item[$A_{3,3}$ ] 
\be \label{3.41}
X_1 = \partial_x, \quad X_2 = 2 x \partial_x + u \partial_u, \quad X_3 = x^2 \partial_x + x u \partial_u, 
\ee
The corresponding invariant ODE is
\be \label{3.42}
{\ddot u} = u^{-3},
\ee
and its general solution is
\be \label{3.43}
u^2 = A (x - x_0)^2 + \frac{1}{A}, \qquad A \ne 0.
\ee

A convenient set of difference invariants is
\bea \label{3.44}
I_1 & = & \frac{h_+}{u u_+}, \quad I_2  =  \frac{h_-}{h_+ + h_-} \frac{u_+}{u}, \\ \nonumber
I_3 & = & \frac{h_+}{h_+ + h_-} \frac{u_-}{u}, \quad I_4  =  \frac{h_-}{u u_-}.
\eea
Any three of these are independent; the four satisfy the identity
\be \label{3.45}
I_1 I_2 = I_3 I_4.
\ee

An invariant difference scheme can be written as
\be \label{3.46}
2 ( I_2 + I_3 -1 ) = I_1^2 I_2 \frac{I_2 + I_3}{I_3} f(I_1 I_2 ), \quad I_1 + I_4 = 4 I_1 I_2 g( I_1 I_2 ),
\ee
i.e.
\bea \label{3.47}
&& u_{x {\bar x}}  =  \frac{1}{h_+ + h_-} \frac{1}{u^2} (\frac{h_+}{u_+} + \frac{h_-}{u_-}) f( \frac{1}{u^2} \frac{h_+ h_-}{h_+ + h_-}), \\ \nonumber
&& \frac{h_+}{u_+} + \frac{h_-}{u_-}  =  \frac{4}{u} \frac{h_+ h_-}{h_+ + h_-} g( \frac{1}{u^2} \frac{h_+ h_-}{h_+ + h_-}).
\eea
For $f = g = 1$ this scheme approximates the ODE (\ref{3.42}).
\begin{description}
\item[$A_{3,4}$ ] 
\be \label{3.48}
X_1 = \partial_u, \quad X_2 = x \partial_x + u  \partial_u, \quad X_3 = x^2 \partial_x + ( -x^2 + u^2) \partial_u.
\ee
This algebra is again $sl(2,\C)$ and can be transformed into
\be \label{3.49}
Y_1 = \partial_x + \partial_u, \quad Y_2 = x \partial_x + u \partial_u, \quad Y_3 = x^2 \partial_x + u^2 \partial_u.
\ee
The realization (\ref{3.49}) (and hence also (\ref{3.48})) is imprimitive; (\ref{3.37}) is primitive. Hence $A_{3.4}$ and $A_{3.3}$ are not equivalent. The invariant ODE for the algebra (\ref{3.48}) is
\be \label{3.50}
x {\ddot u} = C ( 1 + {\dot u}^2 )^{\frac{3}{2}} + {\dot u} ( 1 + {\dot u}^2 ),
\ee
where $C$ is a constant.
The general integral of eq. (\ref{3.50}) can be written as
\bea \label{3.51}
(x - x_0)^2 + (u - u_0)^2 & = & (\frac{x_0}{C})^2, \quad C \ne 0,
\\ \label{3.52}
x^2  +  (u - u_0)^2 & = & x_0^2, C = 0,
\eea
where $x_0$ and $u_0$ are integration constants.

The difference invariants corresponding to the algebra (\ref{3.49}) are
\bea \label{3.53}
I_1  =  \frac{x_+ - x}{x_+ x} ( 1 + u_x^2 ), \quad I_2  =  \frac{x - x_-}{x_- x} ( 1 + u_{\underline{ x}}^2 ), \\ \nonumber
I_3  =  - \frac{(x_+ - x) ( x - x_-)}{2 x x_+ x_-} \{ ( h_+ u_x^2 + x_+ + x) u_{\underline{ x}} + \\ \nonumber + ( h_- u_{\underline{ x}}^2 - x_- - x) u_x \}.
\eea
An invariant scheme representing the ODE (\ref{3.50}) can be written as
\be \label{3.54}
I_3 = C ( \frac{I_1 + I_2}{2} )^{\frac{3}{2}}, \quad I_1 = I_2,
\ee
(this is not the most general such scheme).
\end{description}
\end{description}
\subsubsection{Lagrangian formalism and solutions of three-point O$\Delta$S} \label{sec3.4}
In Section \ref{sec3.2} we presented a Lagrangian formalism for the integration of second order ODE's. Let us now adapt it to O$\Delta$S \cite{ref47}.

The Lagrangian density (\ref{3.9}) will now be a two-point function
\be \label{3.55}
{\cal L} = {\cal L}(x, u, x_+, u_+ ).
\ee
Instead of the Euler-Lagrange equation (\ref{3.10}) we have two \textit{quasiextremal equations} \cite{ref40,ref43,ref47} corresponding to "discrete variational derivatives" of $\cal L$ with respect to $x$ and $y$ independently
\bea \label{3.56}
\frac{\delta {\cal L}}{\delta x} & = & h_+ \frac{\partial {\cal L}}{\partial x} +  h_- \frac{\partial {\cal L}^-}{\partial x} + {\cal L}^- - {\cal L} = 0, \\ \nonumber
\frac{\delta {\cal L}}{\delta u} & =&  h_+ \frac{\partial {\cal L}}{\partial u} +  h_- \frac{\partial {\cal L}^-}{\partial u}  = 0
\eea
where ${\cal L}^-$ is obtained by downshifting $\cal L$ (replacing $n$ by $n-1$ everywhere). In the continuous limit both quasiextremal equations reduce to the same Euler-Lagrange equation.  Thus, the two quasiextremal equations together can be viewed as an O$\Delta$S, where e.g. the difference between them defines the lattice.

The Lagrangian density (\ref{3.55}) will be divergence invariant under the transformation generated by vector field $X$, if it satisfies
\be \label{3.57}
pr X ({\cal L}) + {\cal L} D_+(\xi) = D_+(V)
\ee
for some function $V(x,u)$ where $D_+(f)$ is the discrete total derivative
\be \label{3.58}
D_+ f(x,u) = \frac{f(x+h,u(x+h)) - f(x,u)}{h}
\ee
Each infinitesimal Lagrangian divergence symmetry operator $X$ will provide one first integral of the quasiextremal equation
\be \label{3.59}
h_- \phi \frac{\partial {\cal L}^-}{\partial u} + h_- \xi \frac{\partial {\cal L}^-}{\partial x}  + \xi {\cal L}^- - V = K
\ee
\cite{ref47}. These first integrals will have the form
\be \label{3.60}
f_a(x, x_+, u, u_+) = K_a, \quad a = 1, \ldots .
\ee
Thus, if we have two first integrals, we are left with a two point O$\Delta$S to solve. If we have three first integrals, then the quasiextremal equations reduce to a single two point difference equation, e.g. involving just $x_n$ and $x_{n+1}$. This can often be solved explicitly \cite{ML}.

This procedure has been systematically applied to three point O$\Delta$S in the original article \cite{ref47}. For brevity we will just consider some examples here.

Let us first consider a two-dimensional Abelian Lie algebra and the corresponding invariant second order ODE:
\be \label{3.61}
X_1 = \partial_x, ~ X_2 = \partial_u, \qquad {\ddot u} = F({\dot u}).
\ee
This equation is the Euler-Lagrange equation for the Lagrangian
\be \label{3.62}
{\cal L} = u + G({\dot u}), ~ {\ddot G} = \frac{1}{F},
\ee
and both symmetries are Lagrangian ones
\be \label{3.63}
pr X_1 {\cal L} + {\cal L} D(\xi_1) = 0, ~   pr X_2 {\cal L} + {\cal L} D(\xi_2) = 1 = D(x).
\ee
The corresponding two first integrals are
\be \label{3.64}
J_1 = u + G({\dot u}) - {\dot u} {\dot G}({\dot u}), ~ J_2 = {\dot G}({\dot u}) - x
\ee
Introducing $H$ as the inverse function of $\dot G$ we have
\be \label{3.65}
{\dot y} = H(J_2 + x ), ~ H(J_2 + x) = [ {\dot G}]^{-1}(J_2 + x).
\ee
Substituting into the first equation in eq. (\ref{3.64}), we obtain the general solution of (\ref{3.61}) as
\be \label{3.66}
y(x) = J_1 - G[H(J_2 + x ) ] + (J_2 + x) H(J_2 + x).
\ee

Now let us consider the discrete case. We introduce the discrete analogue of (\ref{3.62}) as
\be \label{3.67}
{\cal L} = \frac{u + u_+}{2} + G(u_x)
\ee
for some smooth function $G$. Equations (\ref{3.63}) hold (with $D$ interpreted as the discrete total derivative $D_+$). The two quasiextremal equations are
\bea \label{3.68}
\frac{x_+ - x_-}{2} - {\dot G}(u_x) + {\dot G}(u_{\underline{ x}}) = 0, \\ \nonumber
u_x {\dot G}(u_x) - u_{\underline{ x}} {\dot G}(u_{\underline{ x}}) -  G(u_x) + G(u_{\underline{ x}}) - \frac{u_+ - u_-}{2} = 0
\eea
The two first integrals obtained using Noether's theorem in this case can be written as
\bea \label{3.69}
{\dot G}(u_x) - \frac{x + x_+}{2} = B \\ \label{3.70}
-u_x {\dot G}(u_x) + G(u_x) + u + \frac{1}{2} (x_+ - x) u_x = A.
\eea

In principle, these two integrals can be solved to obtain
\be \label{3.71}
u_x = H[B + \frac{1}{2} (x_+ + x)], \quad u = \Phi(A, B, x, x_+ ),
\ee 
where $H[z] = [{\dot G}]^{-1}(z)$ and $\Phi$ is obtained by solving eq. (\ref{3.70}), once $u_x = H$ is substituted into this equation. A three point difference equation for $x_{n+2}$, $x_{n+1}$ and $x_n$, not involving $u$ is obtained from the consistency condition $u_x = \frac{u_{n+1} - u_n}{x_{n+1} - x_n}$. In general this equation is difficult to solve. We shall follow a different procedure which is less general, but works well when the considered O$\Delta$S has a three dimensional solvable symmetry algebra with $\{ \partial_x, \partial_u \}$ as a subalgebra. We add a third equation to the system (\ref{3.69}, \ref{3.70}), namely 
\be \label{3.72}
\frac{x_+ - x}{x - x_-} = 1 + \epsilon.
\ee
The general solution of eq. (\ref{3.72}) is
\be \label{3.73}
x_n = (x_0 + B)(1 + \epsilon)^n - B
\ee
where $x_0$ and $B$ are integration constants. We will identify $B$ with the constant in eq. (\ref{3.69}), but leave $\epsilon$ as an arbitrary constant. Eq. (\ref{3.73}) defines an exponential lattice (for $\epsilon \ne 0$). Using (\ref{3.73}) together with (\ref{3.69}) and (\ref{3.70}), we find
\bea \label{3.74}
u_x = H[(x_n + B) (1 + \frac{\epsilon}{2})], \qquad H[z] = [{\dot G}]^{-1}(z) \\ \label{3.75}
u_n = A + (x_n + B) H[(x_n + B) (1 + \frac{\epsilon}{2})] - G[H(x_n + B) (1 + \frac{\epsilon}{2})]
\eea
There is no guarantee that equation (\ref{3.74}) and (\ref{3.75}) are compatible. However, let us consider the two special cases with three-dimensional solvable symmetry algebras, namely algebras $A_{3,1}$ and $A_{3,2}$ of Section \ref{sec3.3}.
\begin{description}
\item[Algebra $A_{3,1}$.] We choose $G(u_x)$ to be 
\be \label{3.76}
G(u_x) = \frac{(k-1)^2}{k} u_x^{\frac{k}{k-1}},\quad k \ne 0,1
\ee
From eq. (\ref{3.74}) and (\ref{3.75}) we obtain
\bea \label{3.77}
u_x = (\frac{1}{k-1})^{k-1} x_n^{k-1} (1 + \frac{\epsilon}{2})^{k-1} \\ \label{3.78}
u_n = \frac{1}{k}   (\frac{1}{k-1})^{k-1} (x_n + B)^k (1 + \frac{\epsilon}{2})^{k-1} [1 + (1-k)\frac{\epsilon}{2} ]
\eea

The consistency condition (for $u_x$ to be the discrete derivative of $u_n$) provides us with a transcedental equation for $\epsilon$:
\be \label{3.79}
[ (1 + \epsilon )^k -1 ] [ 1+ (1-k) \frac{\epsilon}{2} ] = k \epsilon.
\ee
In the continuous limit we take $\epsilon \rightarrow 0$ and $u_n$ given by eq. (\ref{3.78}) goes to the general solution of the ODE (\ref{3.34}). In eq. (\ref{3.79}) terms of order $\epsilon^0$, $\epsilon$, and $\epsilon^2$ cancel. The solution $u_n$ coincides with the continuous limit up to terms of order $\epsilon^2$.

We mention that in the special case $k = -1$ all three symmetries of the O$\Delta$S are Lagrangian ones and in this case eq. (\ref{3.79}) is identically satisfied for any $\epsilon$.
\item[Algebra $A_{3,2}$.] We choose $G(u_x)$ to be
\be \label{3.80}
G(u_x) = e^{u_x}
\ee
and obtain
\bea \label{3.81}
u_x &=& \ln(x_n + B)(1 + \frac{\epsilon}{2}), \\ \label{3.82}
u_n &=& (x_n + B) \ln(x_n + B) + A +  \\ \nonumber&& + (x_n + B) [\ln(1 + \frac{\epsilon}{2}) - (1 + \frac{\epsilon}{2})].
\eea
The expressions (\ref{3.81}) and (\ref{3.82}) are consistent if $\epsilon$ satisfies
\be \label{3.83}
\epsilon (1 + \frac{\epsilon}{2}) - (1 + \epsilon ) \ln (1 + \epsilon ) = 0
\ee
Again (\ref{3.82}) coincides with its continuous limit up to terms of order $\epsilon^2$ and in (\ref{3.83}) terms of order $\epsilon^0$, $\epsilon^1$ and $\epsilon^2$ cancel.
\end{description}
For the $sl(3,\R)$ algebra $A_{3,3}$ all three symmetry operators $X_1$, $X_2$ and $X_3$ correspond to Lagrangian symmetries. The corresponding O$\Delta$S is integrated in \cite{ref47}.
\end{description}

\section{Conclusions and Open Problems} \label{co}

Let us compare the symmetry approach for difference equations with that
for differential ones. In both cases one is interested in transformations
taking solutions into solutions and in both cases, choices have to be
made. Lie's choice of point symmetries of differential equations is so
natural, that it is often forgotten that it is also just an Ansatz : the
vector fields should depend on the independent and dependent variables
only. Generally speaking, this Ansatz has turned out to be the most useful
one. Contact symmetries have much fewer applications than point ones.
Generalized symmetries are mainly of use for identifying integrable
nonlinear partial differential equations.

    In this review we have stressed the fact that for difference equations
this choice of pure point transformations must be modified. Without
significant modifications it remains mainly fruitful for differential--difference
equations rather than for purely difference ones (see Section \ref{sec.nn2}).

    Which modifications are needed for difference equations depends on the
application that we have in mind. For differential equations there are two
main types of applications in physics. In the first, the equations are
already known and group theory is used to solve them. In the second, the
symmetries of the physical problem are known and are used in building up
the theoretical model, i.e. the symmetries precede the equations. These
two aspects are also present in the case of difference equations, but
there are new features. First of all, the physical processes that are
being described may be discrete and the lattices involved may be real
physical objects.  If we are considering linear theories, like 
quantum mechanics, or quantum field theory on a lattice, then the 
generalized point symmetries of Section \ref{sec.n3} are extremely promising. 
A mathematical tool, umbral calculus, is ready to be used, both 
to solve equations and also to formulate models.
 For nonlinear theories on given fixed lattices the most appropriate 
symmetry  approach involves generalized symmetries, as reviewed in Section 
\ref{sec.n4}. Their main application is as in the continuous case: to identify 
integrable systems on lattices. Moreover it can be used to get new interesting solutions. An interesting feature is that some 
point symmetries of differential equations, in particular dilations, 
appear as generalized symmetries for difference equations.

   The second type of application of difference equations in physics is
more practical and in a certain sense, less fundamental. We have in mind
the situation when the physical processes are really continuous and are
described by differential equations. These are then discretized in order
to solve them. The lattices used are then our choice and they can be
chosen in a symmetry adapted way. Moreover, as shown in Section \ref{sec.n2}, the
difference equation and lattice are both part of a "difference scheme" and
the actual lattice is part of the solution of this scheme. We can then
restrict ourselves to point transformations, but they act simultaneously on
the solutions and on the lattice.

 In an attempt to keep this review reasonably short, we have left out many interesting and important topics. Among them we have not included a complete discussion of partial difference equations on transforming symmetry--adapted lattices \cite{ref49,ref44,ref42} and the use of Lie point symmetries to get conditions for the linearizability of difference equations \cite{ref18,ref27,ref28,ref28b,ref28c,ref28d,ref28e}. Also left out is the vast area of numerical methods of solving differential equations, making use of their symmetry properties \cite{Proc.WNM} or the treatment of asymptotic symmetries for difference equations \cite{glm2005}.  A very active area of research, not covered in the present review, is the use of higher symmetries to identify integrable lattice equations \cite{b12,SY,asy,ly1999,ly2000,ly2001,ly2004,S2}.

   Since we are reviewing a relatively new area of research, many open 
questions remain. Thus, for purely difference equations on fixed lattices,
the role of discrete symmetries has not been fully explored. Basically, 
what is needed, specially for partial difference equations, is to apply 
results from crystallography, to characterize discrete or finite 
transformations taking a lattice into itself. For differential-difference 
equations with three independent variables, one or two of them discrete,
a classification of equations with Kac-Moody-Virasoro symmetries should 
help to identify new integrable lattice equations. Applications to 
genuine physical systems would be of great interest. The umbral approach of
Section \ref{sec.n3} has so far been applied in a rather formal manner. The question 
of the convergence of formal power series solutions must be addressed.
The main  question is whether one can develop a complete convergent 
quantum  theory on a lattice. The generalized symmetries of Section \ref{sec.n4} are 
an integral part of the theory of integrability on a lattice. There the 
greatest challenge lies in the field of applications, i.e in applying the 
techniques of integrability to the real world of discrete phenomena. 
Finally, the greatest challenge in the direction of symmetry adapted 
discretizations is to establish whether they provide improved numerical 
methods, in particular for partial differential equations, or higher 
order ordinary ones.

\section*{Acknowledgments}

The research of P.W. was partly supported by a research grant from NSERC
of Canada. Both authors benefitted from the NATO collaborative grant
PST.CLG. 978431.
 D.L.  was partially supported by  PRIN Project "SINTESI-2000" of the  Italian Minister for  Education and Scientific Research and from  the Projects {\sl Sistemi dinamici nonlineari discreti: 
simmetrie ed integrabilit\'a} and {\it Simmetria e riduzione di equazioni differenziali di interesse fisico-matematico} of GNFM--INdAM. 



\bibliographystyle{amsplain}

 \end{document}